\documentclass{iopjournal}

\usepackage{amsmath}
\usepackage{amsfonts}
\usepackage{amssymb}
\usepackage{graphicx}

\usepackage{xcolor}
\usepackage{float}
\usepackage{booktabs}
\usepackage{subfig}
\usepackage{mathtools}
\usepackage{mathrsfs}
\usepackage{orcidlink}
\usepackage{hyperref}
\usepackage[noabbrev,capitalise]{cleveref}
\usepackage[numbers,sort&compress]{natbib}

\newenvironment{ruledtabular}{\begin{center}}{\end{center}}

\renewcommand{\articletype}[1]{}     % removes "Journal Name" / Crossmark / RECEIVED-REVISED box

\providecommand{\etal}{\textit{et al}}

\begin{document}
\articletype{Paper}  
\title{Survey of Novel Deep Learning Architectures for Denoising Gravitational-wave Signals}

\author{Rohan Raha\orcidlink{0009-0002-2354-2884}$^{1}$ and Prayush Kumar\orcidlink{0000-0001-5523-4603}$^{2}$}\\
\affil{$^{1}$Department of Physics, Indian Institute of Science, Bangalore 560012, Karnataka, India}\\
\affil{$^{2}$International Centre for Theoretical Sciences, Tata Institute of Fundamental Research, Bangalore 560089, India}
\email{raharohan@iisc.ac.in}\\

\begin{abstract}

Gravitational wave astronomy demands signal-extraction tools capable of
handling the full diversity of spinning, precessing binaries that dominate
observed catalogs. Recovering the true waveform underpins black-hole
parameter estimation, strong-field tests of general relativity, and
studies of compact-binary formation and evolution. Classical matched
filtering achieves this at a computational cost that will become
prohibitive as next-generation detectors push event rates beyond current
capabilities. Deep learning denoising offers a path to real-time
reconstruction, yet existing methods are developed on narrow parameter
spaces, precluding principled comparison and reliable deployment. We
present the first controlled comparison of five neural-network
architectures for gravitational-wave denoising, trained identically
across the complete range of astrophysically-motivated spinning
binary-black-hole configurations. A unifying principle emerges: matching
network structure to the spectral anatomy of a coalescence --- inspiral,
merger, ringdown --- outperforms brute-force model scaling. Our
Multi-Scale Frequency-Aware architecture embodies this through dedicated
parallel branches per frequency regime, achieving the best reconstruction
fidelity of all five while using fewer parameters than larger competing
models. The framework generalizes from simulated training to real
LIGO--Virgo--KAGRA observations without retraining, recovering merger
morphology with high fidelity across confirmed events spanning three
observing runs and both detectors, despite training on a single
detector's simulated noise. We construct population-level uncertainty
bands from denoising residuals and validate their calibration, and
stress-test the framework beyond its training distribution on extended
mass ratios and a spin population resembling hierarchical-merger
remnants. Applied to 24 hours of real detector noise containing no known
signal, the network suppresses its output almost everywhere, with rare
exceptions traced to real noise transients rather than a general weakness
on stationary noise, indicating the output statistic can discriminate
signal from noise and motivating a future detection study. Publicly
released model weights provide a deployable denoising tool and
reproducible benchmark for future extensions to neutron-star systems and
next-generation detectors.

\end{abstract}

\section{\label{sec:level1}Introduction}
Machine learning approaches to gravitational wave denoising have matured considerably over the past decade, transitioning from initial feasibility studies to practical tools that now contribute to routine astrophysical discoveries \citep{Benedetto:2023jwn, Koloniari:2024kww, Stergioulas:2024jgk}.

Early work by Powell \etal \cite{Powell:2015ona} addressed a critical challenge in the field: distinguishing between instrumental noise transients and genuine astrophysical signals. They introduced Principal Component Analysis for Transients (PCAT) alongside machine learning-enhanced wavelet detection methods \citep{Powell:2015ona, Capote:2024rmo}, providing the first systematic framework for automated glitch classification using machine learning. This work proved particularly timely, appearing during LIGO's 
first observing run, as the community recognized that manual 
inspection of noise artifacts would not scale with increasing detection rates \citep{Cuoco2025}.

In the past decade, the field witnessed a rapid adoption of deep learning architectures specifically tailored for gravitational wave denoising. The pioneering efforts of George and Huerta \cite{George:2017vlv} marked a significant paradigm shift by demonstrating that convolutional neural networks could achieve real-time gravitational wave detection and parameter estimation directly from LIGO strain data. This breakthrough, coupled with the concurrent work of Shen \etal \cite{Shen:2017jkj} on recurrent denoising autoencoders, established deep learning as a viable alternative to traditional matched filtering techniques for both signal extraction and noise mitigation.

The early success of these autoencoder-based approaches led to more sophisticated implementations, notably Wei and Huerta's \cite{Wei:2019zlc} specialized framework for binary black hole merger denoising. Their work demonstrated that deep learning methods can be used to learn the properties of the noise, and can be generalized across different source types. More recently, Wang \etal \cite{Wang:2022quo} introduced transformer-based architectures to gravitational wave denoising through their WaveFormer model, leveraging attention mechanisms to capture long-range temporal dependencies that traditional convolutional approaches might miss.

Despite these significant advances, several fundamental limitations constrain the practical deployment of existing machine learning denoising methods in operational gravitational wave analysis pipelines. Most critically, current approaches suffer from substantial domain gaps between their training environments and real detector conditions \citep{Wei:2019zlc, Wang:2022quo}. While George and Huerta \cite{George:2017vlv} demonstrated impressive performance on simulated LIGO noise, the translation to non-stationary, non-Gaussian detector artifacts remains problematic. This limitation is particularly evident in Shen \etal \cite{Shen:2017jkj}'s work, where training exclusively on white Gaussian noise led to suboptimal performance when applied to realistic detector glitches and environmental disturbances.

The scope of astrophysical signals addressed by existing methods remains unnecessarily narrow, with critical gaps in both mass parameter coverage and spin configurations. George and Huerta's \cite{George:2017vlv} Deep Filtering approach was explicitly trained on non-spinning binaries on quasi-circular orbits, representing only a 2-dimensional subset of the full $8+$ dimensional gravitational wave parameter space. More significantly, Wei and Huerta \cite{Wei:2019zlc} trained their WaveNet-based denoiser exclusively on non-spinning binary black hole mergers, despite overwhelming observational evidence that astrophysical black holes possess substantial spins \citep{KAGRA:2021duu}. While their work demonstrated some generalization to spinning systems in post-hoc testing, achieving lower overlaps for spin-precessing binaries, this represents a concerning performance degradation for signals that constitute the majority of detected events.

The binary masses and mass ratio limitations are equally restrictive across existing frameworks. Shen \etal \cite{Shen:2017jkj} constrained their analysis to mass ratios $q =m_1/m_2 \leq 10$ with total masses $M \in [5, 75]\,M_{\odot}$, while Wei and Huerta \cite{Wei:2019zlc} imposed similar restrictions with mass ratios $q \leq 10$ and component masses $m_{1,2} \in [5\,M_{\odot}, 75\,M_{\odot}]$. Even the more recent Wang \etal \cite{Wang:2022quo} transformer approach, while extending to broader mass ranges with $m_{1,2} \in [5, 150]\,M_{\odot}$ 
and mass ratios $q \in [1,8]$, demonstrated a notable bias against low-mass systems, with samples achieving low overlaps for signals with lower chirp masses.
The treatment of spin degrees of freedom presents an equally critical limitation. While Wang \etal \cite{Wang:2022quo} incorporated spin magnitudes $a_{1,2} \in [0,0.99]$ in their training, most existing approaches either neglect spins entirely or restrict to aligned-spin configurations. Wei and Huerta \cite{Wei:2019zlc} trained exclusively on non-spinning systems, demonstrating only limited generalization to spinning binaries in post-hoc testing, with notably degraded overlaps for spin-precessing systems. This restriction is particularly problematic given that observational evidence indicates a significant fraction of astrophysical binary black holes possess measurable component spins \citep{KAGRA:2021duu}, and hierarchical formation scenarios predict that second-generation mergers---where remnants from previous coalescences merge again---will exhibit substantial spins approaching the Kerr limit. Most critically, when component spins are misaligned with the orbital angular momentum, the resulting spin-induced precession produces characteristic amplitude and phase modulations that encode crucial astrophysical information about formation channels, yet these precessing signals represent precisely the morphologies for which current denoising methods demonstrate the poorest reconstruction fidelity.

Signal duration constraints imposed by computational architectures further limit the applicability of current methods. George and Huerta \cite{George:2017vlv} and Wei and Huerta \cite{Wei:2019zlc} restricted their analysis to $1$-second windows.  These durations, while suitable for high-mass binary black hole mergers, preclude effective denoising of longer-duration sources such as neutron star inspirals, eccentric binaries, or hierarchical mergers that may dominate future observation scenarios and provide unique tests of general relativity in strong-field regimes. Wang \etal \cite{Wang:2022quo}, however, extended the signal duration to $8.0625$ seconds.

In this work, we address many of these fundamental limitations through a comprehensive framework that bridges the gap between controlled laboratory demonstrations and operational deployment in gravitational wave astronomy. We treat denoising throughout as a post-detection reconstruction problem: the network is applied to data segments already identified as containing a candidate signal — whether by a matched-filter search pipeline or, for the real-event validation in Section~\ref{sec:real_events}, by confirmed LVK detections — rather than performing detection or significance assessment itself.
First, we implement signal windows of 2 seconds, doubling the 
duration of several existing approaches~\cite{George:2017vlv, 
Wei:2019zlc} that used 1-second windows, while remaining computationally efficient relative to the 8-second windows of Wang \etal \cite{Wang:2022quo}. This extension captures the complete inspiral-merger-ringdown evolution, particularly crucial for lower mass ratio configurations where previous methods truncated critical information. 
We systematically incorporate generically spinning BBH waveforms throughout our training framework, addressing the non-spinning restriction in Wei and Huerta's \cite{Wei:2019zlc} work and limited validation in George and Huerta \cite{George:2017vlv}. Our training encompasses the full range of astrophysically motivated spin configurations, including both aligned and precessing systems that dominate observational catalogs.
Our 14-dimensional parameter space described in \cref{sec:parameter} substantially broadens beyond the narrow mass ranges and ratios of previous studies. This addresses the demonstrated bias against extreme mass ratio and low-mass systems in current approaches.

Rather than pursuing ever-larger models like recent transformer approaches~\cite{Wang:2022quo}, we optimize for efficiency, achieving comparable performance with significantly fewer parameters, enabling real-time processing while maintaining acceleration options for high-throughput applications. Most importantly, we provide the first systematic comparison of neural network architectures—convolutional, recurrent, transformer-based, and hybrid approaches—under controlled conditions using identical training data and validation protocols. This reveals architecture-specific strengths and limitations previously obscured in single-method studies, providing crucial guidance for practitioners. We additionally address the absence of calibrated uncertainty quantification in other deep-learning-based literature on gravitational-wave denoising. We calculate population-level error bands constructed from the residual distribution of an independent evaluation ensemble, which are validated with probability-probability (P-P) calibration diagnostics. Thus, we verify that a stated $n$-sigma uncertainty band encompasses the true signal an $n$-sigma fraction of the time. We further test generalization beyond the training distribution using an out-of-sample (OOS) out-of-distribution (OOD) evaluation sets , 
complementing a related reconstruction-uncertainty and OOD robustness 
study \citep{Chatterjee:2024mdv},--- binaries with mass ratios extrapolated beyond the trained range, and a spin population drawn from a distribution shape absent from training --- providing a systematic extrapolation study for this class of denoiser. In addition, we characterize the network's behavior on an extended stretch of real, non-simulated detector noise containing no known gravitational-wave signal, and find that the rare cases in which its output remains non-negligible are driven almost exclusively by real noise transients (glitches) absent from the training distribution rather than by a general weakness on stationary detector noise -- a result that motivates a dedicated future study of detection performance while reinforcing the reconstruction-focused, post-detection scope of the present framework. Finally, to promote reproducibility and accelerate community progress, we publicly release our trained model weights and complete model implementation---a resource not provided by prior studies. This enables researchers to directly apply our denoising framework without requiring extensive computational resources for retraining, and provides a foundation for developing improved architectures and exploring transfer learning approaches to other gravitational wave source classes. The immediate availability of production-ready models lowers the barrier to entry for groups without access to large-scale computing infrastructure, democratizing access to state-of-the-art denoising capabilities for the broader gravitational wave community.

% ==========================================================
%  SECTION II:  METHODS
% ==========================================================

\section{\label{sec:summary}Methods}

This section summarizes the simulated dataset, preprocessing
pipeline, neural-network architectures, and training protocol
underlying our comparison. Complete specifications --- including
sampling algorithms, waveform morphology studies, layer-by-layer
architecture definitions, and optimization details --- are provided
in \cref{sec:methods,sec:preprocessing,sec:architectures,sec:training}.

\subsection{\label{sec:summary_data}Training Data and
Parameter-Space Coverage}

Our training and testing datasets each comprise 20{,}000 simulated
BBH signals, generated with the effective-one-body approximants
\texttt{SEOBNRv4\_opt} (non-spinning and aligned-spin systems) and
\texttt{SEOBNRv4P} (precessing systems)~\cite{Bohe:2016gbl,
Husa:2015iqa} using PyCBC~\cite{Usman:2015kfa}. Each signal spans
2\,s at a sampling rate of 4096\,Hz, with the merger placed 0.1\,s
before the end of the segment, capturing the complete
inspiral--merger--ringdown evolution. The datasets follow a 2:1:1
composition --- 50\% precessing-spin, 25\% aligned-spin, and 25\%
non-spinning systems --- reflecting the expectation that generic
spin orientations dominate the astrophysical
population~\cite{Vitale:2016avz}.

Component masses span $m_{1,2} \in [5, 100]\,M_\odot$ with mass
ratio $q \geq 1/6$, sampled uniformly in the symmetric mass ratio
$\eta$ to prevent overrepresentation of near-equal-mass systems.
Spin vectors are drawn isotropically in direction with magnitudes
$|\vec{s}_{1,2}| \leq 0.99$, and extrinsic parameters (sky
location, inclination, polarization, coalescence phase) are sampled
isotropically; all orbits are quasi-circular. Each waveform is
projected onto the LIGO Hanford (H1) detector response and injected
into an independent realization of colored Gaussian noise generated
from the aLIGO Zero-Detuned High Power design-sensitivity
PSD~\cite{Aasi_2015}. Luminosity distances are drawn from a
cosmologically weighted distribution, $p(d_L) \propto
d_L^2/(1+z)^3$, over $d_L \in [1, 1000]$\,Mpc, yielding a
realistic optimal-SNR distribution peaking at $\mathrm{SNR}
\sim 10$--$30$ with $\sim$98.7\% of samples below
$\mathrm{SNR} = 100$. The full parameter-space construction,
detector projection formalism, and waveform morphology studies are
detailed in \cref{sec:methods}, with the resulting dataset
statistics presented in \cref{sec:dataset_statistics}.

\subsection{\label{sec:summary_preprocessing}Preprocessing}

Each noisy data segment $d(t) = h(t; d_L) + n(t)$ and its clean
target $h(t)$ are whitened in the frequency domain by
$1/\sqrt{S_n(f)}$, equalizing the noise variance across the
detector band so that the network is not biased toward
high-sensitivity frequencies. A bandpass filter restricts the
content to $[20, 2000]$\,Hz, encompassing the inspiral, merger, and
ringdown frequencies of the full mass range considered. After
conversion back to the time domain, both the noisy input and clean
target are normalized by the standard deviation of the noisy
series; dividing both by the same factor scales the data to
unit variance while preserving the relative signal-to-noise
structure of each injection. The whitened, bandpassed, normalized
noisy segment serves as the network input, with the identically
processed clean waveform as the supervised target. The complete
preprocessing pipeline, including the distance-scaling and
whitening implementations, is described in
\cref{sec:preprocessing}.

\subsection{\label{sec:summary_architectures}Neural-Network
Architectures}

We implement and compare five architectures representing
fundamentally different processing paradigms, summarized in
Table~\ref{tab:architecture_comparison}:
\begin{enumerate}
    \item  \textbf{baseline LSTM encoder-decoder}
    ($\sim$243K parameters), establishing the performance floor of
    purely recurrent processing;
    \item  \textbf{CNN-LSTM hybrid} ($\sim$26.5M parameters),
    prepending a pyramid of convolutional kernels
    ($k = 512 \to 48$) that extract multi-scale local features
    before recurrent temporal integration;
    \item \textbf{ACRED-Net} ($\sim$22.0M parameters), 
    transformer-style encoder-decoder combining multi-head self-
    and cross-attention with convolutional and bidirectional LSTM
    processing;
    \item  \textbf{U-Net with attention gates}
    ($\sim$172M parameters), whose gated skip connections preserve
    fine-grained features across a 5-level multi-resolution
    hierarchy; and
    \item  \textbf{Multi-Scale Frequency-Aware
    (Multi-Frequency) architecture} ($\sim$59.1M parameters),
    which routes the input through seven parallel branches
    explicitly specialized for different frequency bands --- three
    large-kernel branches targeting the low-frequency inspiral and
    four small-kernel branches targeting the merger and ringdown
    --- followed by cross-frequency integration and recurrent
    reconstruction.
\end{enumerate}
This sequence represents a deliberate progression of inductive
biases: from generic temporal memory, through multi-scale and
attention-based feature extraction, to an explicit decomposition
aligned with the spectral anatomy of a compact binary coalescence.
Layer-by-layer specifications, mathematical formulations, and
schematic diagrams for all five architectures are provided in
\cref{sec:architectures}.

% ----------------------------------------------------------
% Table I (moved here from the appendix; delete the appendix
% copy in Step 3)
% ----------------------------------------------------------
\begin{table*}[t]
\caption{Comprehensive comparison of neural network architectures for gravitational wave denoising. All models process input sequences of length $T=8192$ samples (2 seconds at 4096 Hz sampling rate). Training time is relative to the baseline LSTM encoder-decoder. The receptive field indicates the maximum temporal span that can influence a single output time step.}
\label{tab:architecture_comparison}
\begin{ruledtabular}
\resizebox{\textwidth}{!}{%
\begin{tabular}{lcccccc}
\textbf{Architecture} & \textbf{Parameters} & \textbf{Layers} & \textbf{Key Components} & \textbf{Receptive Field} & \textbf{Attention} & \textbf{Training Time} \\
\hline
LSTM & 242,657 & 10 & 4 LSTM (128,64$\to$64,128) & Global & No & 1.0$\times$ \\
Encoder-Decoder & & & + Dense bottleneck (32) & (recurrent memory) & & (baseline) \\
 & & & + LayerNorm + Dropout & & & \\
\hline
CNN-LSTM & 26,470,737 & 34 & 8 Conv1D (k: 512$\to$48) & 512 samples & No & 8$\times$ \\
Hybrid & & & + 4 LSTM layers & ($\sim$125 ms) & & \\
 & & & + BatchNorm (16$\times$) & & & \\
\hline
ACRED-Net & 22,007,525 & 28 & 6 Conv1D + 4 BiLSTM & Global + 512 & Yes & 15$\times$ \\
(Transformer) & & & + Multi-head attn (1 head, key dim 4) & samples & (self+cross) & \\
 & & & + Positional encoding & & & \\
\hline
U-Net with & 172,022,593 & 42 & 20 Conv1D blocks & 512 samples & Yes & 15$\times$ \\
Attention Gates & & & + 4-level hierarchy & ($\sim$125 ms) & (gated skip) & \\
 & & & + Skip connections & Multi-scale (5) & & \\
\hline
Multi-Scale & 59,125,384 & 25 & 7 parallel branches & 512 samples & Limited & 30$\times$ \\
Frequency-Aware & & & + 14 LSTM (8 Bi + 6 Uni) & ($\sim$125 ms) & (self-attn) & \\
 & & & + Frequency fusion & Multi-scale (7) & & \\
\end{tabular}%
}
\end{ruledtabular}
\end{table*}

\subsection{\label{sec:summary_training}Training Protocol}

All five architectures are trained under identical conditions: the
mean-squared-error loss between predicted and clean whitened
waveforms, the Adam optimizer~\cite{Kingma:2014vow} with initial
learning rate $10^{-3}$ and \texttt{ReduceLROnPlateau} scheduling,
early stopping with a patience of 5 epochs (maximum 50), and model
checkpointing at each validation-loss minimum. To suppress
spurious signal hallucination on noise-dominated inputs, the
training set is augmented with pure-noise samples whose target
output is identically zero; the augmentation fraction of 30\%,
adopted uniformly across all architectures, is established by the
ablation study in \cref{sec:ablation}. All computations use double
precision on a single NVIDIA A40 GPU. Full optimization,
regularization, and infrastructure details are given in
\cref{sec:training}.

With the dataset, preprocessing, loss function, and optimization
procedure held fixed across all models, any performance
differences reported in the following section are attributable to
architectural design alone --- the controlled comparison that
constitutes the central methodological contribution of this work.

% ========================= END SECTION II =========================

\section{\label{sec:results}Results: Simulated Data}

We organize our quantitative evaluation into six parts. We first determine
the optimal pure noise augmentation fraction through an ablation study on
the baseline LSTM encoder-decoder (Section~\ref{sec:ablation}), then
compare training convergence across all five architectures under uniform
conditions (Section~\ref{sec:convergence}). We next present a systematic
comparative analysis of the temporal evolution of reconstruction quality
across all architectures (Section~\ref{sec:temporal_metrics}). We
follow with a qualitative assessment of waveform reconstruction fidelity
for the best-performing Multi-Frequency architecture across all three spin
configurations, together with population-level uncertainty bands
constructed from the residual distribution of an independent evaluation
ensemble (Section~\ref{sec:qualitative}), validate the statistical
calibration of those bands with population-wide P-P diagnostics
(Section~\ref{sec:calibration}), and present population-level performance
statistics for the Multi-Frequency model, including two evaluation sets
extending beyond the training distribution
(Section~\ref{sec:population}). Denoising of real gravitational wave events is presented in Section~\ref{sec:real_events}. Finally, we characterize the network's behavior on an extended stretch of real detector noise containing no known gravitational-wave signal, examining both its typical suppression behavior and the physical origin of rare high-amplitude outputs (Section~\ref{sec:pure_noise_real}).

% ===========================================================
\subsection{\label{sec:ablation}Ablation Study: Pure Noise Augmentation
Fraction}
% ===========================================================

\begin{figure}[htbp]
\centering
\includegraphics[width=0.7\columnwidth]{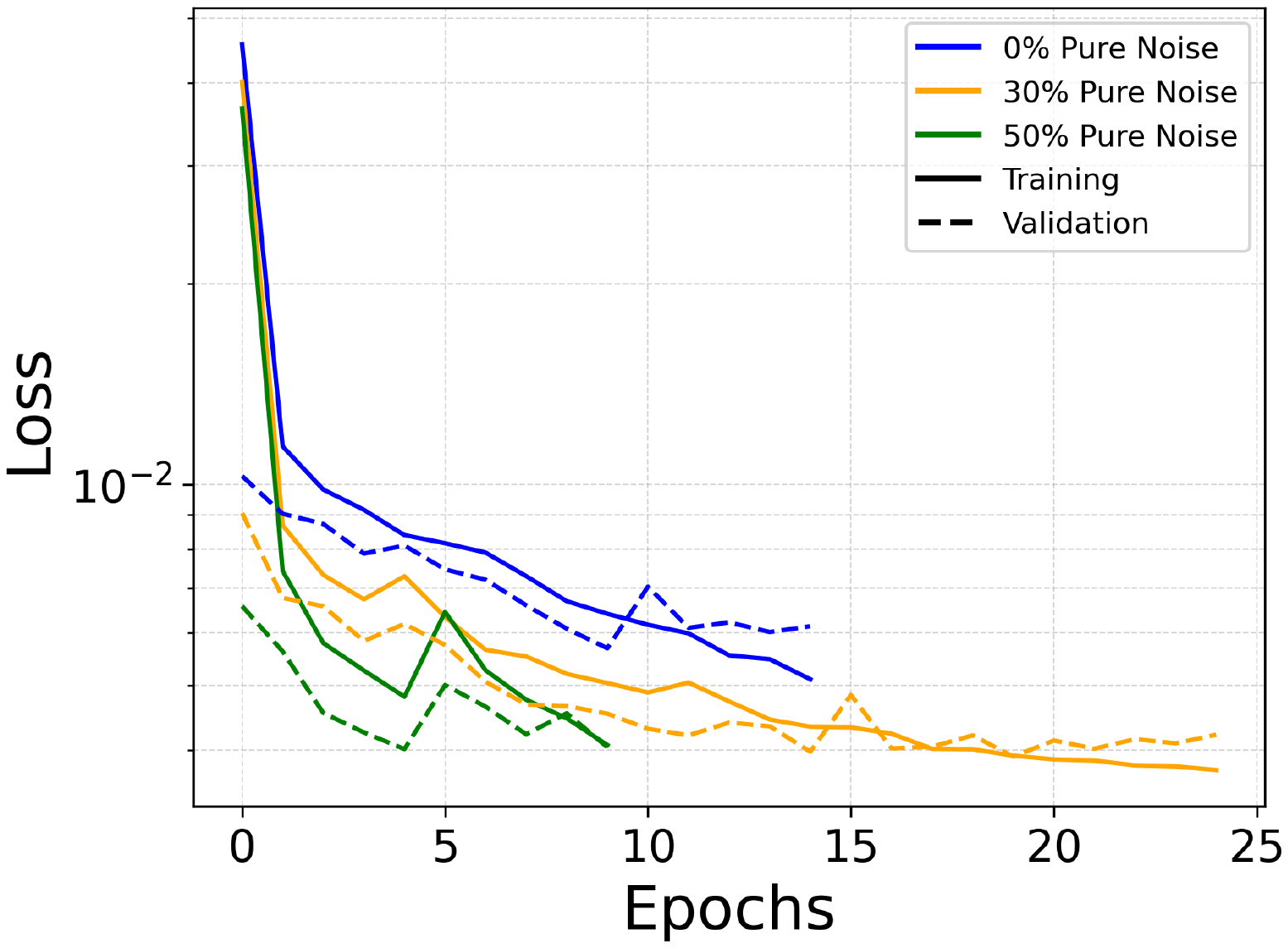}
\caption{Training (solid) and validation (dashed) MSE loss curves for the
baseline LSTM encoder-decoder trained with three pure noise fractions:
$0\%$ (blue), $30\%$ (orange), and $50\%$ (green). All models employ
identical optimizer configuration, early stopping patience of 5 epochs, and
the same training dataset described in Section~\cref{sec:methods}.}
\label{fig:ed2_loss}
\end{figure}

To determine the optimal fraction of pure noise samples to include during
training, we train three variants of the baseline LSTM encoder-decoder
(~\cref{sec:arch_lstm}) with pure noise fractions of $0\%$, $30\%$,
and $50\%$, holding all other hyperparameters fixed. As described in
Section~\cref{sec:training}, pure noise training samples provide
zero-signal targets that teach the network to suppress its output in the
absence of a gravitational wave signal, preventing spurious hallucination
on noise-dominated inputs. Table~\ref{tab:ablation} summarizes the
convergence statistics for all three variants; the corresponding loss
curves are shown in Figure~\ref{fig:ed2_loss}.

\begin{table}[htbp]
\centering
\caption{Training convergence statistics for the baseline LSTM
encoder-decoder trained with three pure noise augmentation fractions.
$\mathcal{L}^{*}_{\rm val}$ is the best validation MSE across all training
epochs; $e^{*}$ is the epoch at which it was achieved; $e_{\rm total}$ is
the total number of epochs completed before early stopping.}
\label{tab:ablation}
\begin{ruledtabular}
\begin{tabular}{lcccc}
\textbf{Noise fraction} & $\mathcal{L}^{*}_{\rm val}$
& $e^{*}$ & $e_{\rm total}$ \\
\hline
$0\%$  & $5.68 \times 10^{-3}$ & 10 & 15 \\
$30\%$ & $3.91 \times 10^{-3}$ & 20 & 25 \\
$50\%$ & $4.01 \times 10^{-3}$ &  5 & 10 \\
\end{tabular}
\end{ruledtabular}
\end{table}

Introducing $30\%$ pure noise samples produces a substantial improvement,
achieving the best validation MSE of all three variants and benefiting
from training across the greatest number of epochs before early stopping.
The validation curve of the $30\%$ model lies consistently below both
alternatives at every epoch, confirming a systematically superior
solution rather than a stochastic fluctuation.

The $50\%$ configuration terminates the earliest of the three variants
and converges to a minimum inferior to the $30\%$ model, despite the
higher noise fraction. We attribute this to a training budget imbalance:
with half of all mini-batches containing no signal, the network
encounters insufficient morphological diversity of gravitational wave
templates per epoch and settles into a local minimum associated with
aggressive noise suppression at the cost of waveform reconstruction
fidelity. The $0\%$ configuration converges to the worst minimum. Without training with pure noise, the network retains a residual bias toward reconstructing signal-like features even from noise-only inputs.

Based on these results, we adopt $30\%$ pure noise augmentation as the
standard training protocol for all subsequent architectures.

% ===========================================================
\subsection{\label{sec:convergence}Architecture Comparison: Training
Convergence}
% ===========================================================

\begin{figure}[htbp]
\centering
\includegraphics[width=0.7\columnwidth]{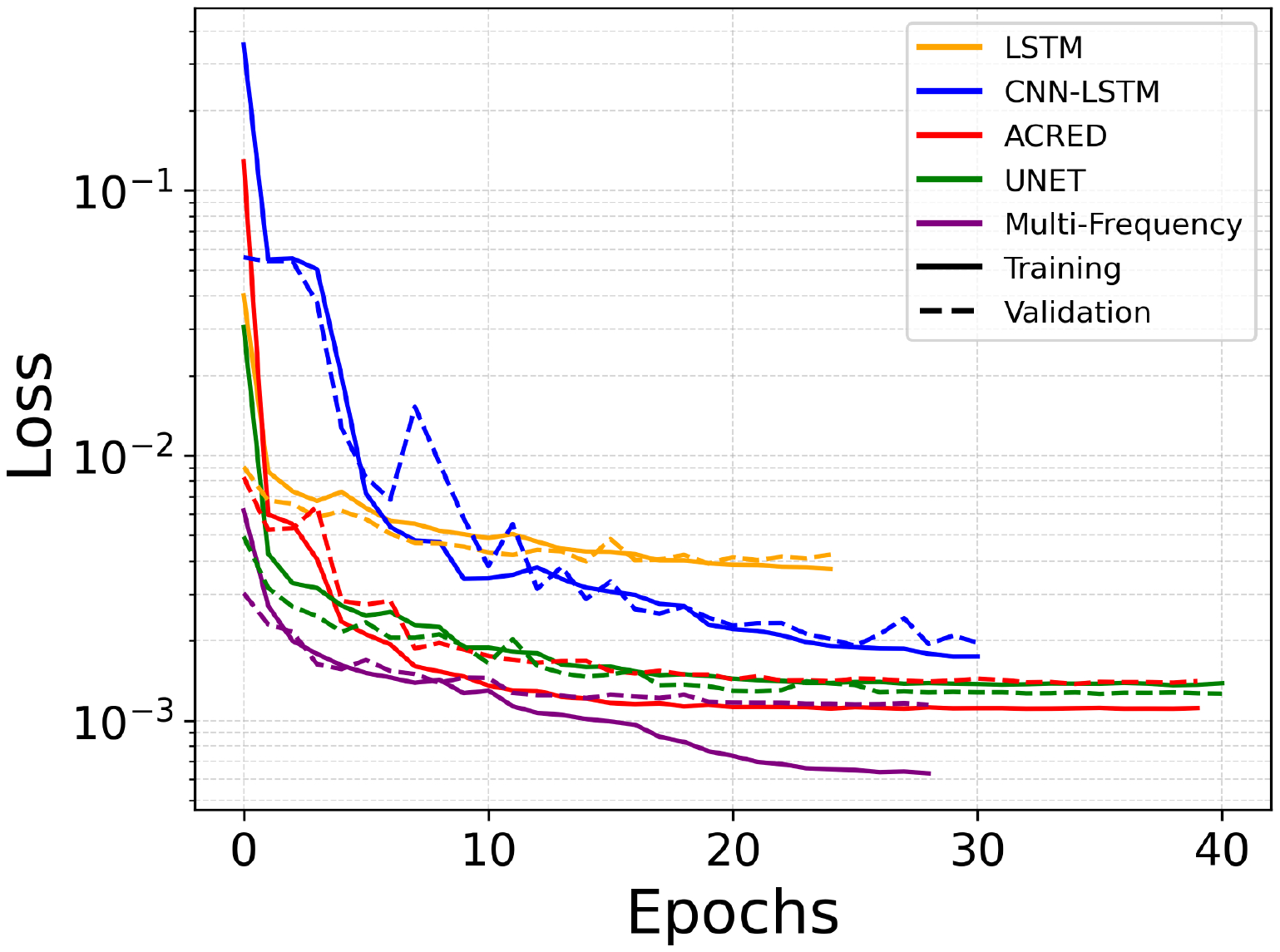}
\caption{Training (solid) and validation (dashed) MSE loss curves for all
five architectures, each trained with $30\%$ pure noise augmentation. Ranked by final validation MSE (lower is better), 
the hierarchy from worst to best is: LSTM $>$ CNN-LSTM 
$>$ ACRED-Net $>$ U-Net $>$ Multi-Frequency, with Multi-Frequency 
achieving the lowest best validation MSE of 
$1.15\times10^{-3}$.
}
\label{fig:arch_loss}
\end{figure}

Having established the optimal pure noise fraction, we train all five
architectures under uniform conditions with $30\%$ pure noise augmentation.
Figure~\ref{fig:arch_loss} shows the resulting loss curves, and
Table~\ref{tab:convergence} summarizes the convergence statistics.

\begin{table}[htbp]
\centering
\caption{Training convergence statistics for all neural network
architectures trained with $30\%$ pure noise augmentation.
$\mathcal{L}^{*}_{\rm val}$ is the best validation MSE across all training
epochs; $e^{*}$ is the epoch at which it was achieved; $E_{\rm total}$ is
the total epochs completed before early stopping.}
\label{tab:convergence}
\begin{ruledtabular}
\begin{tabular}{lcccc}
\textbf{Architecture} & \textbf{Parameters} & $\mathcal{L}^{*}_{\rm val}$
& $e^{*}$ & $e_{\rm total}$ \\
\hline
LSTM ED    & 243K   & $3.91 \times 10^{-3}$ & 20         & 25         \\
CNN-LSTM   & 26.5M  & $1.91 \times 10^{-3}$ & 26         & 31         \\
ACRED-Net  & 22.0M  & $1.37 \times 10^{-3}$ & 35         & 40         \\
U-Net      & 172M   & $1.26 \times 10^{-3}$ & 36         & 41         \\
Multi-Frequency & 59.1M  & $1.15 \times 10^{-3}$ & $24$ & $29$ \\
\end{tabular}
\end{ruledtabular}
\end{table}

The Multi-Frequency architecture achieves the lowest best
validation MSE, outperforming U-Net by $8.7\%$, ACRED-Net by $16.1\%$,
CNN-LSTM by $39.8\%$, and the LSTM baseline by $70.5\%$. Crucially, it
does so with $3\times$ fewer parameters than U-Net,
demonstrating that physically motivated frequency decomposition is more
parameter-efficient than model scale for this task.

The single largest performance gain occurs at the transition from 
LSTM to CNN-LSTM: a $51.1\%$ improvement in best validation MSE 
($3.91 \to 1.91 \times 10^{-3}$) from adding multi-scale 
convolutional preprocessing. The subsequent $28.2\%$ gain from 
CNN-LSTM to ACRED-Net ($1.91 \to 1.37 \times 10^{-3}$) reflects 
the benefit of self-attention in modeling arbitrary temporal 
dependencies. U-Net improves upon ACRED-Net by $8.0\%$ 
($1.37 \to 1.26 \times 10^{-3}$), with skip connections preserving 
fine-grained signal structure lost in the bottleneck compression. 
The final $8.7\%$ gain from U-Net to Multi-Frequency 
($1.26 \to 1.15 \times 10^{-3}$) reflects the benefit of explicit 
frequency decomposition over generic multi-scale feature extraction as the fundamental signal-processing challenge shifts from
representation capacity to spectral specialization.

% ===========================================================
\subsection{\label{sec:temporal_metrics}Temporal Evolution of
Reconstruction Quality}
% ===========================================================

To compare how reconstruction quality accumulates throughout the
inspiral-merger sequence across all five architectures, we compute two
cumulative metrics as a function of time for the same representative test
signals used in Section~\ref{sec:qualitative}. At each time step $t$
within the evaluation window $[t_{\rm merger} - 0.9\,\mathrm{s},\
t_{\rm merger} + 0.1\,\mathrm{s}]$, metrics are computed over the growing
sub-window from the window start to $t$, ensuring that at early times the
metrics reflect only the early inspiral, while at $t = 0$ they capture
the integrated performance over the full sequence. We impose a minimum
sub-window of $0.1$\,s to ensure numerical stability of frequency-domain
operations.

The first metric is the Pearson correlation coefficient
(Equation~\ref{eq:pearson}), which measures morphological fidelity of the
reconstruction independent of amplitude scaling. Perfect reconstruction
corresponds to $r = 1$.

The second is the noise-weighted match (maximized overlap),
\begin{equation}
\mathcal{M} =
\max_{t_c,\,\phi_c}
\frac{|\langle \hat{h}(t + t_c)\,e^{i\phi_c}\,|\,h\rangle|}
{\sqrt{\langle \hat{h}\,|\,\hat{h}\rangle\,
\langle h\,|\,h\rangle}},
\label{eq:match}
\end{equation}
where the maximization is performed over time-shift $t_c$ and overall
phase $\phi_c$, and the noise-weighted inner product is the matched-filter
statistic~\citep{Allen:2005fk},
\begin{equation}
\langle a\,|\,b\rangle = 4\,\mathfrak{R}
\int_{f_{\rm low}}^{f_{\rm high}}
\frac{\tilde{a}^{*}(f)\,\tilde{b}(f)}{S_{n}(f)}\,df,
\label{eq:inner_product}
\end{equation}
with $f_{\rm low} = 20$\,Hz and $f_{\rm high} = 2000$\,Hz.
$\mathcal{M}$ weights frequency components by the inverse detector noise
PSD, emphasizing the $100$--$500$\,Hz band where aLIGO sensitivity is
highest and where the late inspiral and merger carry the greatest spectral
power. We report the mismatch,
\begin{equation}
\textit{Mismatch} = 1 - \mathcal{M},
\label{eq:mismatch}
\end{equation}
on a logarithmic scale so that inter-model differences are visible
across the full dynamic range from early inspiral to merger. Perfect
reconstruction corresponds to a mismatch of zero.

At each time step $t$, $\mathcal{M}$ is evaluated between the clean
and denoised waveforms over the growing sub-window
$[t_{\rm merger} - 0.9\,\mathrm{s},\, t]$, so that the plotted
mismatch $1 - \mathcal{M}(t)$ reflects the cumulative
noise-weighted reconstruction quality from the start of the
inspiral window up to time $t$.

\begin{figure*}[htbp]
\centering
  \includegraphics[width=\columnwidth]{%
    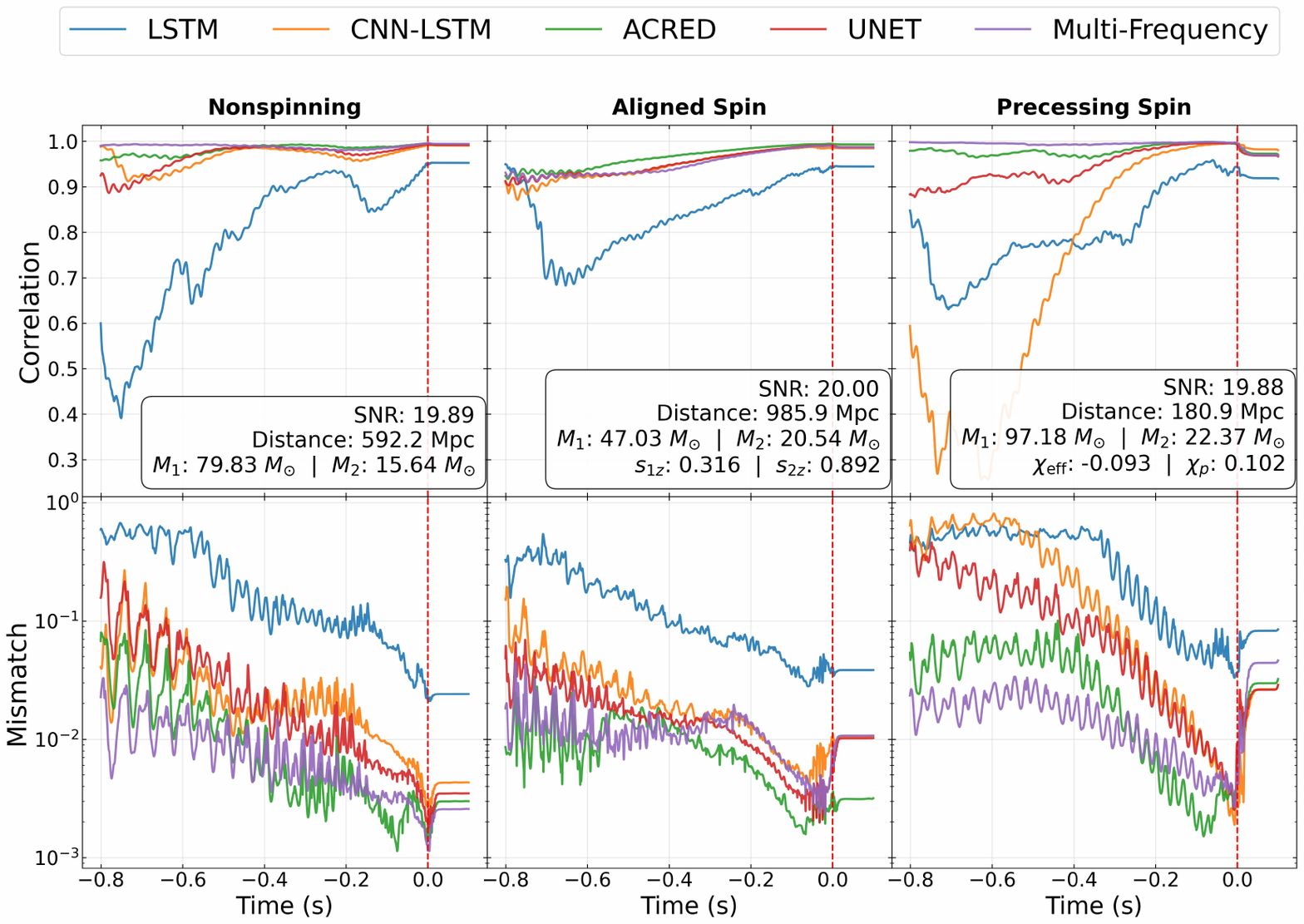}
\caption{
  Cumulative Pearson correlation $r$ (top panels) and noise-weighted
  mismatch $1-\mathcal{M}$ (bottom panels, logarithmic scale) as a
  function of time relative to merger for all five architectures, evaluated
  over a growing sub-window starting at $t = -0.9$\,s. The vertical red
  dashed line marks merger at $t = 0$. Metrics are computed on the same
  test signals as Figure~\ref{fig:denoised_waveforms}, all at
  $\mathrm{SNR} \approx 20$. Lower mismatch and higher correlation both
  indicate better reconstruction. Colors follow the legend at top:
  LSTM (blue), CNN-LSTM (orange), ACRED (green), U-Net (red),
  Multi-Frequency (purple). The Multi-Frequency and ACRED architectures
  maintain the lowest mismatch throughout all three configurations.
  The LSTM baseline exhibits the highest mismatch across all cases.
  The most distinctive architectural failure is visible in the precessing
  spin case, where CNN-LSTM shows a sharp correlation drop and mismatch
  spike during the mid-inspiral, a feature absent for the other four
  architectures.
}
\label{fig:temporal_metrics}
\end{figure*}

Figure~\ref{fig:temporal_metrics} presents both metrics for all five
architectures across all three spin configurations.

The performance hierarchy established by the training loss
curves---Multi-Frequency $\gtrsim$ U-Net $\gtrsim$ ACRED $\gtrsim$ CNN-LSTM 
$\gg$ LSTM---is reproduced at every time step across all three spin
configurations in both metrics, with the largest inter-model separation
appearing during the early inspiral ($t \lesssim -0.4$\,s), where the
signal-to-noise per cycle is lowest, and narrowing progressively toward
merger as the dominant high-amplitude signal accumulates into the growing
cumulative window. The Multi-Frequency architecture maintains the lowest
mismatch and highest correlation throughout all cases. The LSTM baseline
exhibits the largest mismatch values and sharpest correlation minima
across all configurations, reflecting its inability to extract
low-frequency signal content through recurrent processing alone.

The oscillatory structure prominent in the mismatch panels arises from
the beating of the growing cumulative window with the inspiral's
instantaneous frequency: as the window expands, periodic shifts in
PSD-weighted spectral content cause the noise-weighted match to fluctuate,
with the amplitude of these oscillations scaling inversely with
reconstruction quality. This behavior is most pronounced for LSTM and, in
the precessing case, CNN-LSTM, and is strongly suppressed for
Multi-Frequency and ACRED throughout all configurations.

The inter-model separation is most compressed for the nonspinning case,
where the absence of spin-induced amplitude and frequency modulations
simplifies the reconstruction task for all architectures.  The precessing case exposes the most
distinctive architectural failure: CNN-LSTM exhibits a pronounced
mid-inspiral correlation drop and accompanying mismatch spike, absent for
all other architectures, where spin-precession amplitude modulations and
the evolving chirp frequency are simultaneously active. The Multi-Frequency
and ACRED architectures are unaffected by this feature, indicating that
parallel frequency-decomposed processing is better suited to separating
the slower precession envelope from the underlying orbital phase evolution
than the fixed-kernel convolutional preprocessing used by CNN-LSTM.
U-Net, despite its encoder-decoder structure, shows elevated and more
oscillatory mismatch in the precessing case relative to ACRED,
suggesting that attention-gated skip connections alone are insufficient
to disentangle the precession modulation without explicit
frequency-domain decomposition.

Across all three cases, the noise-weighted mismatch consistently reveals larger inter-model separation than the Pearson
correlation $r$, particularly during the early-to-mid inspiral. This
amplification occurs because the PSD weighting emphasizes the
$100$--$500$\,Hz band, where frequency-specific processing capabilities
most directly determine reconstruction quality. The Multi-Frequency
architecture's advantage is most pronounced during the early inspiral
rather than at merger: all models benefit from the large-amplitude
merger signal near $t = 0$, but only those with dedicated
low-frequency processing extract reliable signal content from the quiet
early inspiral where the signal-to-noise per cycle is smallest.

% ===========================================================
\subsection{\label{sec:qualitative}Case Studies: Denoising Across Spin Configurations}
% ===========================================================

We examine representative denoised waveforms from the Multi-Frequency architecture for each spin configuration at a target
$\mathrm{SNR} \approx 20$, evaluated over a $1$-second window spanning
$0.9$\,s before to $0.1$\,s after merger which is also used for all further evaluations of performance metrics in the subsequent subsections.  We use this 1-second evaluation window, rather than the full 2-second preprocessed segment, because the 
dominant signal power for the mass range considered 
($m_{1,2} \in [5, 100]\,M_\odot$) is concentrated in 
the final second before and after merger; extending the 
window into the early inspiral beyond $t = -0.9$\,s 
introduces additional noise-dominated samples that 
inflate MSE without reflecting the network's ability 
to reconstruct the astrophysically informative portion 
of the signal. Reconstruction quality is
quantified using two complementary metrics: the energy-normalized mean
squared error
\begin{equation}
\mathrm{MSE}_{\mathrm{norm}} =
\frac{\displaystyle\sum_{t}
\left[h(t) - \hat{h}(t)\right]^{2}}
{\displaystyle\sum_{t} h^{2}(t)},
\label{eq:mse_norm}
\end{equation}
which measures the fractional residual reconstruction error relative to the signal energy, and the Pearson correlation coefficient $r$ defined as 
\begin{equation}
r = \frac{\displaystyle\sum_{t}
\bigl[h(t) - \bar{h}\bigr]
\bigl[\hat{h}(t) - \bar{\hat{h}}\bigr]}
{\sqrt{\displaystyle\sum_{t}
\bigl[h(t) - \bar{h}\bigr]^{2}
\;\cdot\;
\sum_{t}\bigl[\hat{h}(t) - \bar{\hat{h}}\bigr]^{2}}},
\label{eq:pearson}
\end{equation}
where overbars denote time averages over the evaluation window.
It measures the correlation of the phase of the signal
independent of amplitude. Perfect reconstruction corresponds to
$\mathrm{MSE}_{\rm norm} = 0$ and $r = 1$.

\begin{figure*}[htbp]
\centering

  \includegraphics[width=\columnwidth]{%
    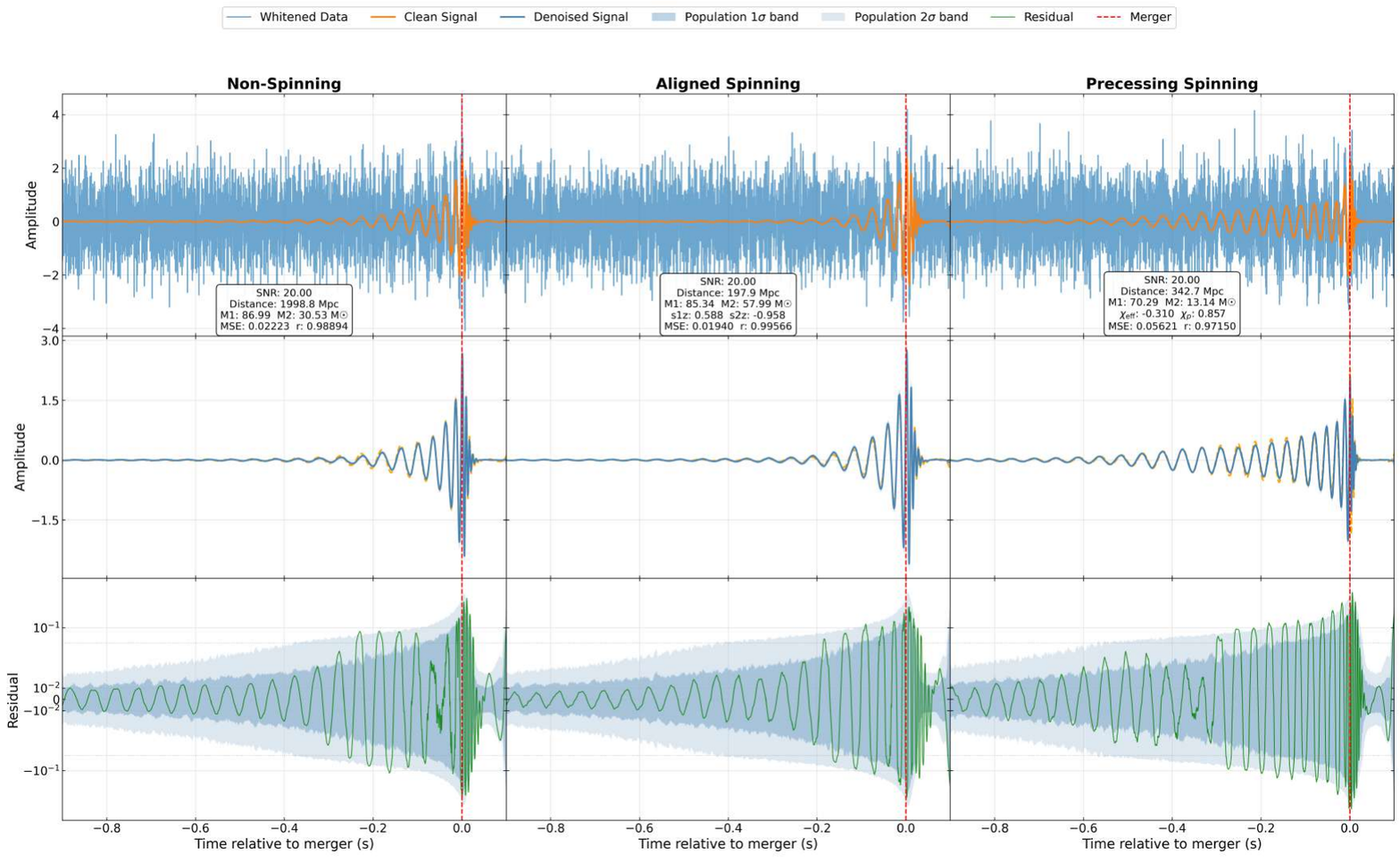}%

\caption{Denoised waveform outputs from the Multi-Frequency architecture
for representative test signals at $\mathrm{SNR} \approx 20$, spanning the
three spin configurations in our dataset, augmented with population-level
uncertainty bands. Each column contains three panels: the top panel shows
the whitened noisy input (light blue) with the whitened clean signal
(orange) embedded; the middle panel overlays the denoised output (dark
blue) on the clean target (orange dashed); the bottom panel shows the residual
$r(t) = \hat{h}(t) - h(t)$ (green) on a symmetric logarithmic scale,
together with the population $1\sigma$ and $2\sigma$ bands (shaded)
centered on zero. The $1\sigma$ and $2\sigma$ bands are broad enough to encompass the raw residual. The vertical red dashed line marks merger at $t=0$.
Source parameters, $\mathrm{MSE}_{\rm norm}$, and Pearson correlation $r$
for each example are annotated in the top panel. The population bands are
constructed independently of these three specific examples, from the
per-time-sample residual distribution of an $N=100$-per-category
evaluation ensemble held at fixed $\mathrm{SNR}=20$; their statistical
calibration is validated in Section~\ref{sec:calibration}.
}
\label{fig:denoised_waveforms}
\end{figure*}

Figure~\ref{fig:denoised_waveforms} (top and middle panel) presents the results for all three
spin configurations.
Across all three cases, the merger peak and exponentially damped ringdown are accurately reconstructed, with the denoised trace tracking the clean
signal most closely during the high-amplitude late inspiral
($t \gtrsim -0.4$\,s). Differences emerge in the early inspiral, where
the signal-to-noise per cycle is lowest: the non-spinning waveform is
recovered faithfully from the earliest visible cycles, while the
aligned-spin case shows a modest increase in $\mathrm{MSE}_{\rm norm}$
attributable to spin-orbit coupling modulations superimposed on the
underlying chirp~\citep{Ajith:2009bn}. The precessing case exhibits the
largest residual error, concentrated in the early inspiral where the
time-varying orbital plane orientation generates quasi-periodic amplitude
modulations~\citep{Schmidt:2014iyl} that are partially smoothed in the
denoised output.

Across the three cases, $\mathrm{MSE}_{\rm norm}$ increases monotonically
from non-spinning to aligned-spin to precessing, while $r$ decreases in
the same order---reflecting the physical hierarchy of waveform complexity.
Critically, even for the most morphologically complex precessing case,
$r > 0.97$, confirming effective generalization across the full spin
parameter manifold.

In addition to these point estimates, Figure~\ref{fig:denoised_waveforms} (bottom panel)
reports a population-level uncertainty band around each denoised waveform
and its residual. We develop a procedure to get uncertainty estimate and we illustrate this using signals at fixed $\mathrm{SNR} =20$. For each spin category we construct these bands from an
independent evaluation ensemble of $N=1000$ signals held at fixed
$\mathrm{SNR}=20$: at each time sample $t$, the residual
$r_i(t) = \hat{h}_i(t) - h_i(t)$ is computed for every ensemble member $i$,
and the population $1\sigma$ band is defined as
\begin{equation}
\sigma(t) = \frac{p_{84}(t) - p_{16}(t)}{2},
\label{eq:sigma_band}
\end{equation}
where $p_{16}(t)$ and $p_{84}(t)$ are the 16th and 84th percentiles of the
residual distribution across the ensemble at that time sample; the
$2\sigma$ band is constructed analogously from the 2.5th/97.5th
percentiles. These bands are deterministic and population-calibrated ---
they characterize the typical residual magnitude for a given spin category
at fixed SNR --- rather than a per-event, posterior-style uncertainty of the kind returned
by Bayesian inference frameworks such as BayesWave. Whether these bands
are statistically honest, i.e. whether a stated $n\sigma$ band truly
encompasses the denoised residual an $n\sigma$ fraction of the time, is
assessed quantitatively in Section~\ref{sec:calibration}.

\subsection{\label{sec:calibration}Calibration of Population-Level Uncertainty Bands}

To assess whether the population-level uncertainty bands introduced in
Section~\ref{sec:qualitative} is statistically honest, we construct a
probability-probability (P-P) calibration diagnostic following standard
practice in Bayesian parameter-estimation validation
\citep{Cook:2006zir}, adapted here to a per-time-sample,
population-calibrated setting rather than a per-event posterior.

For each spin category, we hold out an evaluation ensemble of $N=100$
signals, entirely distinct from the calibration ensemble used to
construct the bands in Section~\ref{sec:qualitative}. At each time sample
$t$ within the $[-0.9,+0.1]$\,s evaluation window, we rank each held-out
signal's residual $r_i(t)$ against the empirical distribution of
calibration-set residuals at that same time sample, yielding a
probability-integral-transform (PIT) value $u_i(t) \in [0,1]$. Under
perfect calibration, the $u_i(t)$ are uniformly distributed on $[0,1]$; the
empirical coverage $Y(p) = \mathrm{fraction}(u \leq p)$, evaluated across
all (signal, time-sample) pairs --- $N \times T_{\rm eval} \approx
4.1\times10^{4}$ points per category --- should therefore track the
diagonal $Y(p) = p$.

% Figure~\ref{fig:pp_ensemble} shows the resulting P-P curves for the three
% trained-distribution spin categories. The pooled empirical coverage curve
% (red) tracks the diagonal closely in all three cases, with a
% Kolmogorov--Smirnov deviation of $0.019$ (non-spinning), $0.059$
% (aligned-spin), and $0.038$ (precessing-spin), and $96.8\%$/$99.7\%$/$100.0\%$
% of pooled points falling within the nominal $1\sigma$/$2\sigma$/$3\sigma$
% coverage regions for all three categories. Individual-signal P-P curves
% (green) show substantially more scatter, with a median per-signal KS
% deviation of $0.08$--$0.11$ and a maximum of up to $0.20$; this is the
% expected consequence of the small number of effectively independent trials
% contributed by a single waveform, whose time samples are strongly
% correlated through the smoothly varying chirp morphology rather than
% constituting independent draws. The close agreement between the pooled coverage curve
% and the diagonal, well within the sampling band expected for $n=100$
% trials (one trial per signal), indicates that the population-level uncertainty bands of
% Section~\ref{sec:qualitative} is statistically well calibrated: the
% reported $n\sigma$ band around the denoised waveform encompasses the true
% residual an $n\sigma$ fraction of the time.

\begin{figure}[htbp]
\centering
\includegraphics[width=0.7\columnwidth]{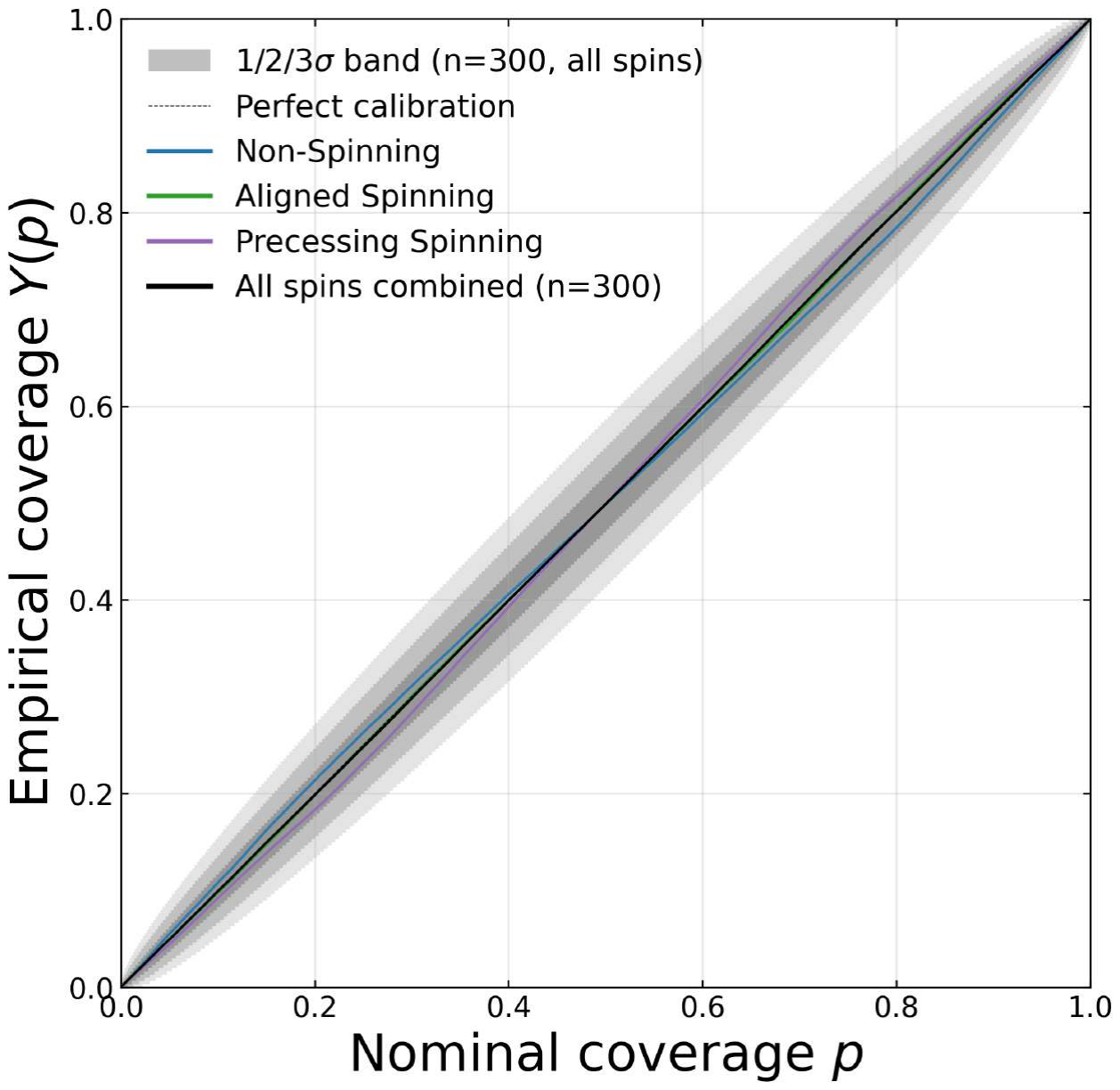}
\caption{Probability-probability (P-P) calibration diagnostic for the
population-level uncertainty bands of Figure~\ref{fig:denoised_waveforms}.
For each spin category, an evaluation ensemble of $N=100$ signals ---
held out from, and distinct from, the calibration ensemble used to
construct the bands in Section~\ref{sec:qualitative} --- is denoised,
and at each time sample within the $[-0.9,+0.1]$\,s merger window the
residual is converted to an empirical probability-integral-transform
(PIT) value against the calibration-set residual distribution at that
time sample. The pooled empirical coverage $Y(p)$ is shown for each spin
category individually (blue: non-spinning; green: aligned-spin; purple:
precessing-spin) and for all three categories combined (black, $n=300$),
plotted against the nominal coverage $p$; perfect calibration corresponds
to the diagonal (black dashed). The grey shaded region shows the exact
$1/2/3\sigma$ sampling-fluctuation band expected under perfect
calibration, computed from the binomial distribution with $n=300$
independent trials (one per signal, pooled across all three spin
categories, since time samples within a single waveform are highly
correlated and cannot be treated as independent draws). All pooled
curves lie close to the diagonal and within the shown sampling band,
indicating that the population uncertainty bands are, in aggregate,
statistically well calibrated across all three spin categories.}
\label{fig:pp_ensemble}
\end{figure}

Figure~\ref{fig:pp_ensemble} shows the resulting P-P curves for the three
trained-distribution spin categories, together with a grand-pooled curve
merging all $N=300$ signals across categories. The per-category pooled
coverage curves track the diagonal closely, with a Kolmogorov--Smirnov
deviation of $0.016$ (non-spinning), $0.005$ (aligned-spin), and $0.020$
(precessing-spin), and $99.8\%$/$100.0\%$/$100.0\%$ of pooled points
falling within the nominal $1\sigma$/$2\sigma$/$3\sigma$ coverage regions
for all three categories. The all-spins-combined curve achieves an even
tighter Kolmogorov--Smirnov deviation of $0.003$, with all points
($100.0\%$) falling within the $1\sigma$/$2\sigma$/$3\sigma$ regions.
Notably, all four curves --- including the three individual-category
curves --- lie within the single sampling band shown, which is sized to
$n=300$ (a stricter test than the $n=100$ per-category band each curve
would nominally be compared against), underscoring that the calibration
holds tightly even under this more conservative benchmark. This
indicates that the population-level uncertainty bands of
Section~\ref{sec:qualitative} are statistically well calibrated: the
reported $n\sigma$ band around the denoised waveform encompasses the
true residual an $n\sigma$ fraction of the time.

% ===========================================================
\subsection{\label{sec:population}Population-Level Performance Statistics}
% ===========================================================

To characterize the Multi-Frequency architecture's performance
across the full parameter space, we evaluate it on $N = 100$ test
injections from each of the three trained-distribution spin categories
(300 total), with luminosity distances drawn from the cosmologically
weighted distribution described in \cref{sec:distance_scaling}, yielding a
realistic SNR distribution spanning $\mathrm{SNR} \in [5, 80]$. To probe
generalization beyond the training distribution, we additionally evaluate
the same architecture on two out-of-distribution validation sets of
$N=100$ injections each (500 total across all five categories): an
extreme mass-ratio set extending the mass ratio to $q \in (6, 10]$,
beyond the trained $q \leq 6$ sample, and a
Gaussian aligned-spin set drawing component spins independently
from $\chi_{1,2} \sim \mathcal{N}(0.67, 0.1)$, motivated by the spin distribution expected for
hierarchical merger remnants \citep{Fishbach:2017dwv, Berti:2008af}, a distribution shape not represented in training even though individual spin magnitudes remain within the trained bounds. Both OOS and OOD sets otherwise follow the same mass range, distance sampling, and preprocessing as the trained-distribution categories. Metrics are computed using
Equations~\ref{eq:mse_norm} and~\ref{eq:mismatch} over the same
$[-0.9, +0.1]$\,s evaluation window.

\begin{figure*}[htbp]
\centering
\includegraphics[width=\columnwidth]{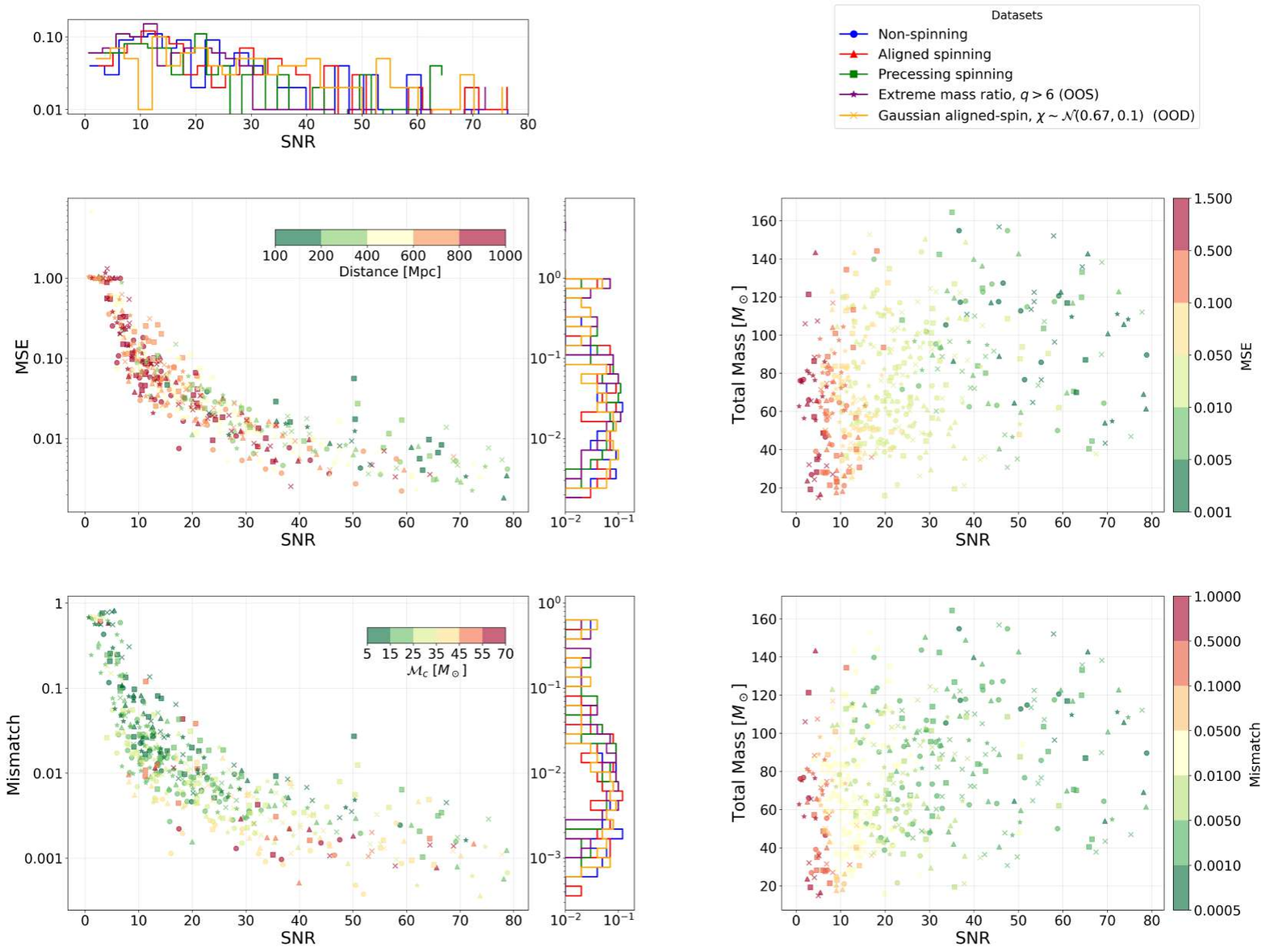}
\caption{
Population-level performance of the Multi-Frequency
architecture on $N = 100$ test injections per category (500 total across
five categories). Categories are distinguished by marker shape and color:
non-spinning (circles, blue), aligned-spin (triangles, red),
precessing-spin (squares, green), extreme mass-ratio $q>6$ (stars, purple, OOS), and Gaussian aligned-spin $\chi\sim\mathcal{N}(0.67,0.1)$ (crosses, orange, OOD).
Top panel: SNR distribution for each category, shown as
normalized step histograms on a logarithmic vertical axis. The bulk of
injections fall in $5 \lesssim \mathrm{SNR} \lesssim 70$ for all five
categories, consistent with the shared cosmologically weighted distance
sampling.
Middle left: $\mathrm{MSE}_{\rm norm}$ versus optimal SNR,
colored by luminosity distance $d_L$. The side histogram shows the
normalized $\mathrm{MSE}_{\rm norm}$ distribution per category.
Middle right: Total mass $M_{\rm tot}$ versus SNR, colored by
$\mathrm{MSE}_{\rm norm}$. At fixed SNR, higher-mass systems achieve lower
MSE across all five categories.
Bottom left: \textit{Mismatch} versus SNR, colored by
chirp mass $\mathcal{M}_c$. The side histogram shows the normalized
mismatch distribution per category.
Bottom right: Total mass versus SNR, colored by \textit{Mismatch}.
The two out-of-distribution categories broadly track the trained-distribution
trend of decreasing error with SNR and total mass, with a modest systematic
offset toward higher $\mathrm{MSE}_{\rm norm}$ and \textit{Mismatch} at
fixed SNR, quantified in Tables~\ref{tab:population} and
\ref{tab:population_ood}.}

\label{fig:population}
\end{figure*}

Figure~\ref{fig:population} summarizes the results for all five
categories. Table~\ref{tab:population} provides aggregate statistics
broken down by spin category and SNR bin for the three trained-distribution
categories; Table~\ref{tab:population_ood} provides the analogous
statistics for the two out-of-distribution sets.

\begin{table*}[htbp]
\centering
\caption{Population-level performance statistics for the Multi-Frequency architecture ($N = 100$ per spin category). Metrics are
computed over the $[-0.9, +0.1]$\,s evaluation window. SNR bins are
defined as low ($5 \leq \mathrm{SNR} < 15$), mid ($15 \leq \mathrm{SNR}
< 30$), and high ($30 \leq \mathrm{SNR} \leq 80$). Percentages give the
fraction of events in each category satisfying the quoted mismatch threshold.
}
\label{tab:population}
\begin{ruledtabular}
\begin{tabular}{lcccccc}
& \multicolumn{2}{c}{\textbf{Non-spinning}} &
\multicolumn{2}{c}{\textbf{Aligned-spin}} &
\multicolumn{2}{c}{\textbf{Precessing-spin}} \\
& $\mathrm{MSE}_{\rm norm}$ & \textit{Mismatch}
& $\mathrm{MSE}_{\rm norm}$ & \textit{Mismatch}
& $\mathrm{MSE}_{\rm norm}$ & \textit{Mismatch} \\
\hline
Mean   & 0.070 & 0.02851 & 0.112 & 0.05009 & 0.141 & 0.05024 \\
Median & 0.020 & 0.00603 & 0.024 & 0.00677 & 0.025 & 0.00857 \\
Std    & 0.169 & 0.09974 & 0.236 & 0.14007 & 0.425 & 0.14484 \\
\hline
$\textit{Mismatch} < 0.01$ & -- & 64\% & -- & 62\% & -- & 55\% \\
$\textit{Mismatch} < 0.03$ & -- & 90\% & -- & 80\% & -- & 81\% \\
\hline
\multicolumn{7}{l}{\textit{Median metrics by SNR bin:}} \\
Low ($N_{\rm ns}=34$, $N_{\rm al}=29$, $N_{\rm pr}=21$)
& 0.070 & 0.01551 & 0.103 & 0.02861 & 0.092 & 0.02580 \\
Mid ($N_{\rm ns}=33$, $N_{\rm al}=34$, $N_{\rm pr}=36$)
& 0.019 & 0.00582 & 0.024 & 0.00691 & 0.031 & 0.00815 \\
High ($N_{\rm ns}=31$, $N_{\rm al}=32$, $N_{\rm pr}=37$)
& 0.007 & 0.00193 & 0.007 & 0.00180 & 0.007 & 0.00241 \\
\end{tabular}
\end{ruledtabular}
\end{table*}

\begin{table*}[htbp]
\centering
\caption{Population-level performance statistics for the two
out-of-distribution validation sets ($N = 100$ each), evaluated over the
$[-0.9,+0.1]$\,s window using the same metrics and SNR binning as
Table~\ref{tab:population}. Extreme mass ratio extrapolates beyond
the trained mass-ratio range ($q>6$, out-of-support, ``OOS''); Gaussian
aligned-spin draws component spins from $\chi \sim \mathcal{N}(0.67,0.1)$
(hierarchical-merger-remnant-like), a distribution shape not seen in
training even though individual spin magnitudes remain within the trained
bound (``OOD'').}
\label{tab:population_ood}
\begin{ruledtabular}
\begin{tabular}{lcccc}
& \multicolumn{2}{c}{\textbf{Extreme mass ratio, $q>6$ (OOS)}} &
\multicolumn{2}{c}{\textbf{Gaussian aligned-spin (OOD)}} \\
& $\mathrm{MSE}_{\rm norm}$ & \textit{Mismatch}
& $\mathrm{MSE}_{\rm norm}$ & \textit{Mismatch} \\
\hline
Mean   & 0.246 & 0.0608 & 0.126 & 0.0527 \\
Median & 0.049 & 0.0152 & 0.023 & 0.0083 \\
Std    & 0.923 & 0.1481 & 0.245 & 0.1200 \\
\hline
$\textit{Mismatch} < 0.01$ & -- & 36\% & -- & 52\% \\
$\textit{Mismatch} < 0.03$ & -- & 72\% & -- & 76\% \\
\hline
\multicolumn{5}{l}{\textit{Median metrics by SNR bin:}} \\
Low ($N_{\rm oos}=41$, $N_{\rm ood}=29$)
& 0.111 & 0.02424 & 0.118 & 0.04651 \\
Mid ($N_{\rm oos}=29$, $N_{\rm ood}=29$)
& 0.032 & 0.01311 & 0.023 & 0.00880 \\
High ($N_{\rm oos}=21$, $N_{\rm ood}=37$)
& 0.011 & 0.00306 & 0.006 & 0.00195 \\
\end{tabular}
\end{ruledtabular}
\end{table*}

Several systematic trends are apparent. The most significant performance
driver is SNR: the median mismatch decreases monotonically from
$\textit{Mismatch} \approx 0.016$--$0.029$ in the low-SNR bin to
$\textit{Mismatch} \approx 0.002$--$0.003$ in the high-SNR bin across all
spin categories, while $\mathrm{MSE}_{\rm norm}$ decreases by more than
an order of magnitude over the same range. At the population level,
$55$--$64\%$ of all test events achieve mismatch below $0.01$ and
$80$--$90\%$ fall below $0.03$, confirming robust denoising performance
across the astrophysically motivated parameter space.

\begin{figure}[htbp]
\centering
\includegraphics[width=0.7\columnwidth]{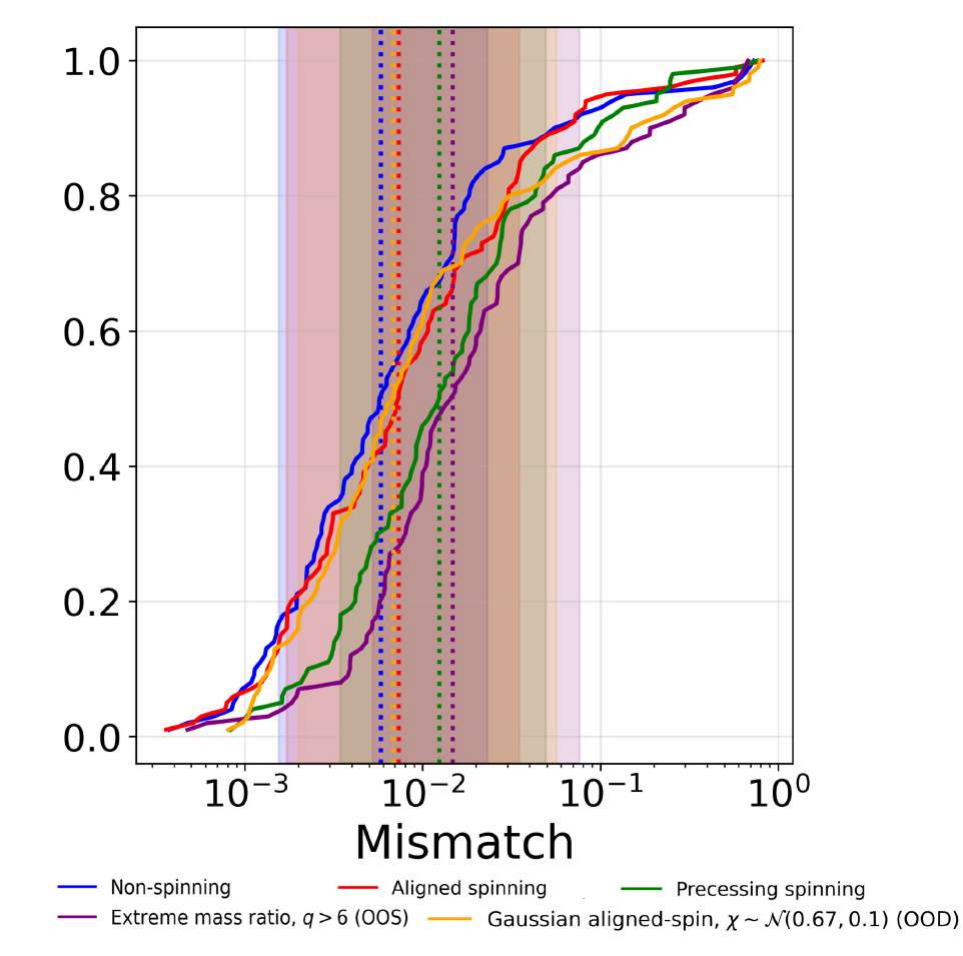}
\caption{
Cumulative mismatch distribution for the Multi-Frequency
architecture across all five categories ($N=100$ each): non-spinning
(blue), aligned-spin (red), precessing-spin (green), extreme mass-ratio
$q>6$ (purple, OOS), and Gaussian aligned-spin $\chi\sim\mathcal{N}(0.67,0.1)$
(orange, OOD). The x-axis shows mismatch $1-\mathcal{O}$ on a logarithmic
scale. Dotted vertical lines mark the median mismatch for each category;
shaded bands span the 16th--84th percentile range. Non-spinning
(median $1-\mathcal{O}=0.0054$) and aligned-spin (median $0.0055$)
distributions are nearly indistinguishable, precessing-spin (median
$0.0119$) shows the modest rightward shift already seen in the
three-category comparison, and the two OOS and OOD categories are
shifted further still: extreme mass-ratio (OOS, median $0.0152$) is the
most degraded of all five, while Gaussian aligned-spin (OOD, median
$0.0083$) falls between the aligned-spin and precessing-spin
distributions.}
\label{fig:mismatch_cum}
\end{figure}

Figure~\ref{fig:mismatch_cum} shows the cumulative mismatch distribution
for all five categories, making these thresholds visually explicit. The
non-spinning (median mismatch $0.0054$) and aligned-spin (median $0.0055$)
distributions remain nearly coincident, as in the three-category
comparison above, consistent with the harder reconstruction problem posed
by amplitude and phase modulations from orbital-plane precession in the
precessing-spin case (median $0.0119$). Both OOS and OOD categories are shifted to modestly higher mismatch relative to their
nearest trained-distribution counterpart: the extreme mass-ratio (OOS) set
(median $0.0152$) is shifted beyond even the precessing-spin distribution,
the most challenging of the three trained categories, while the Gaussian
aligned-spin (OOD) set (median $0.0083$) sits between the aligned-spin and
precessing-spin distributions despite testing a spin distribution shape
absent from training. The cumulative distributions retain the same overall shape
as the trained categories, with degradation concentrated in the low-SNR
tail (Table~\ref{tab:population_ood}).

A secondary trend is visible in the total-mass panels of
Figure~\ref{fig:population}: at fixed SNR, higher-mass systems
systematically achieve lower mismatch and lower MSE. This reflects
the dominant contribution of the merger and ringdown to the PSD-weighted
inner product for massive systems. The merger frequency scales approximately
as $f_{\rm merger} \propto M_{\rm tot}^{-1}$, placing it in the
$100$--$300$\,Hz band for $M_{\rm tot} \gtrsim 50\,M_\odot$ where the
aLIGO sensitivity is optimal, while the short ringdown duration
concentrates energy in a spectrally compact region that the network can
reconstruct with high fidelity. Lower-mass systems have longer,
frequency-swept inspiral phases that accumulate a larger fraction of
their SNR at frequencies below $100$\,Hz where signal-to-noise per
sample is lower, making reconstruction harder.

The ordering of spin categories by mismatch is non-spinning $\approx$
aligned-spin $<$ precessing-spin. Aligned-spin systems achieve nearly
identical median mismatch and MSE to non-spinning systems across all SNR
bins, despite the additional morphological complexity of spin-orbit
phasing modulations. Precessing systems show modestly degraded
performance, with median mismatch approximately $0.002$--$0.006$ higher
than the other categories at fixed SNR, and $\mathrm{MSE}_{\rm norm}$
approximately $20$--$50\%$ higher. The residual performance gap reflects
the genuine additional difficulty of the precessing denoising problem:
the time-varying orbital plane orientation produces amplitude and phase
modulations that are quasi-periodic rather than monotonically evolving,
requiring the network to track a higher-dimensional signal manifold.
Nonetheless, the large mean-median discrepancy in $\mathrm{MSE}_{\rm norm}$
(mean $\approx 4$--$6\times$ median for all categories) indicates that
the distributions are dominated by a small number of difficult low-SNR,
low-mass cases where reconstruction partially fails, rather than by
systematic degradation across the population.

Table~\ref{tab:population_ood} presents the analogous statistics for the
OOS and OOD sets. The extreme mass ratio (OOS) dataset degrades
performance relative to all three trained categories: $36\%$ of extreme
mass-ratio injections achieve mismatch below 0.01 (versus $55$--$64\%$
for the trained categories) and $72\%$ fall below $0.03$ (versus
$80$--$90\%$), with the degradation concentrated at low and mid SNR
(median mismatch $0.024$ and $0.013$, respectively, compared to
$0.016$--$0.029$ and $0.006$--$0.008$ for the trained categories), while
high-SNR performance approaches that of the trained categories (median
mismatch $0.0031$, modestly above the $0.0018$--$0.0024$ range of the
trained categories). This is consistent with extreme mass-ratio systems presenting
a genuinely harder reconstruction problem --- longer, weaker inspirals
dominated by the more massive component --- compounded by the network
never having observed $q>6$ during training. The Gaussian aligned-spin
(OOD) dataset, by contrast, shows performance intermediate between the
aligned-spin and precessing-spin trained categories ($52\%$ below
mismatch $0.01$, $76\%$ below $0.03$), despite drawing spins from a distribution shape never seen in training.

\section{\label{sec:real_data}Results:  LVK data}

We now turn from the simulated test population to genuine LIGO--Virgo--KAGRA
data. We first apply the trained Multi-Frequency model to three confirmed
binary black hole events spanning three observing runs
(Section~\ref{sec:real_events}), and then characterize its behavior on an
extended stretch of real detector noise containing no known
gravitational-wave signal (Section~\ref{sec:pure_noise_real}).

% ===========================================================
\subsection{\label{sec:real_events}Denoising of Real Gravitational Wave
Events}
% ===========================================================
\begin{figure*}%[htbp]
\centering
\subfloat[GW150914 (O1), GPS 1126259462.4\label{fig:real_gw150914}]{%
  \includegraphics[width=0.5\columnwidth]{%
    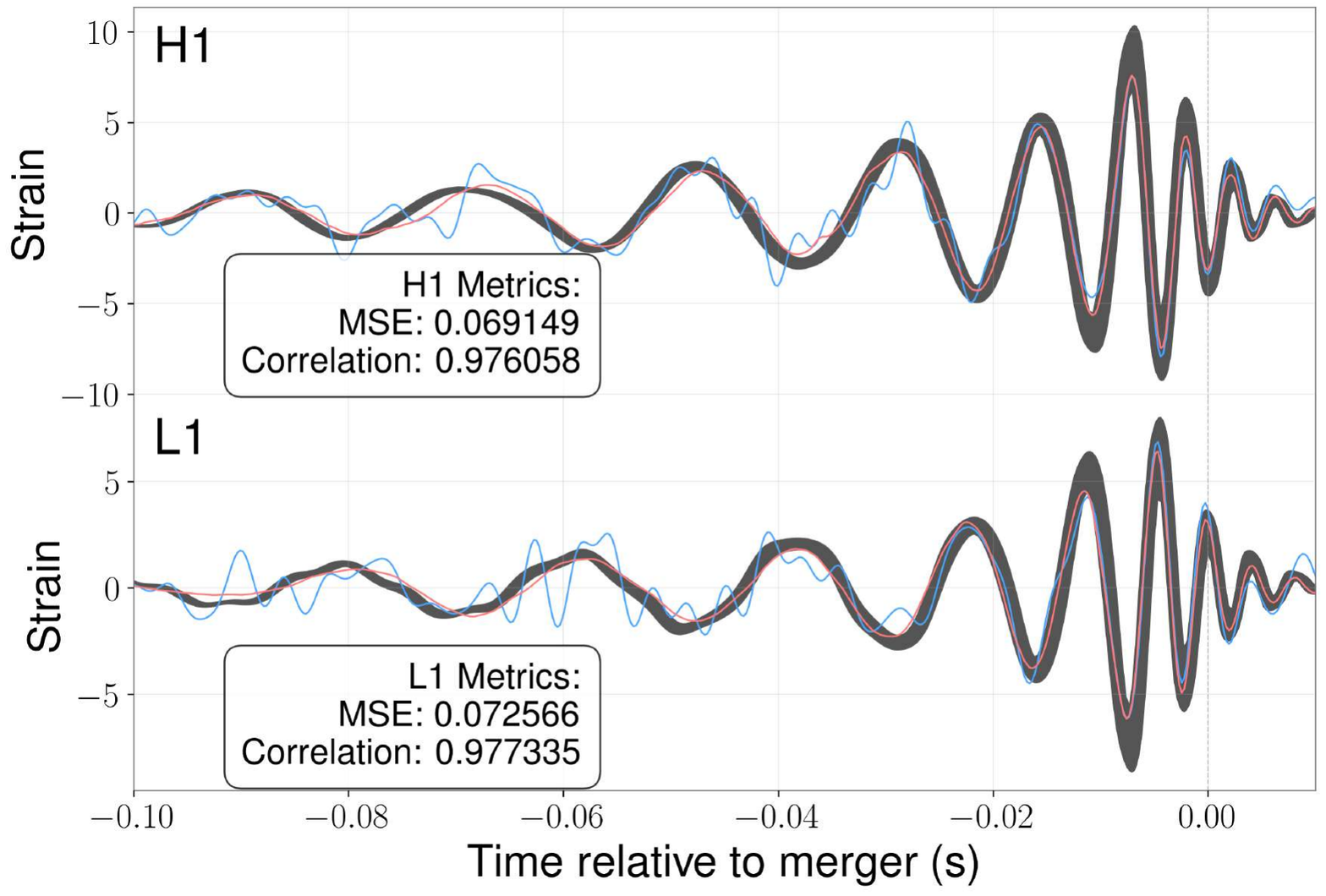}%
}%\hfill
\subfloat[GW200129 (O3b), GPS 1264316116.4\label{fig:real_gw200129}]{%
  \includegraphics[width=0.5\columnwidth]{%
    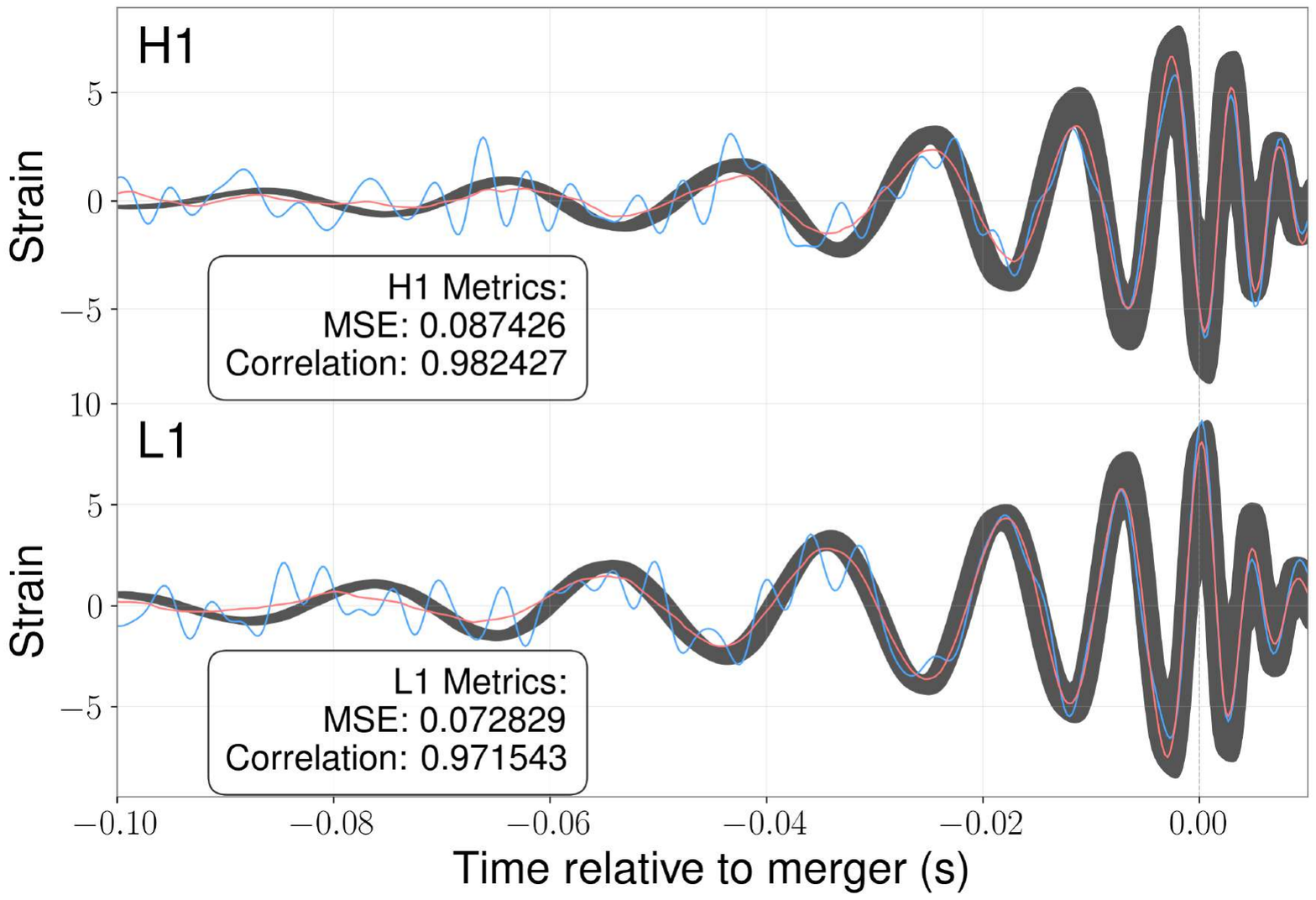}%
}\\%\hfill
\subfloat[GW231226 (O4), GPS 1387620938\label{fig:real_gw231226}]{%
  \includegraphics[width=0.5\columnwidth]{%
    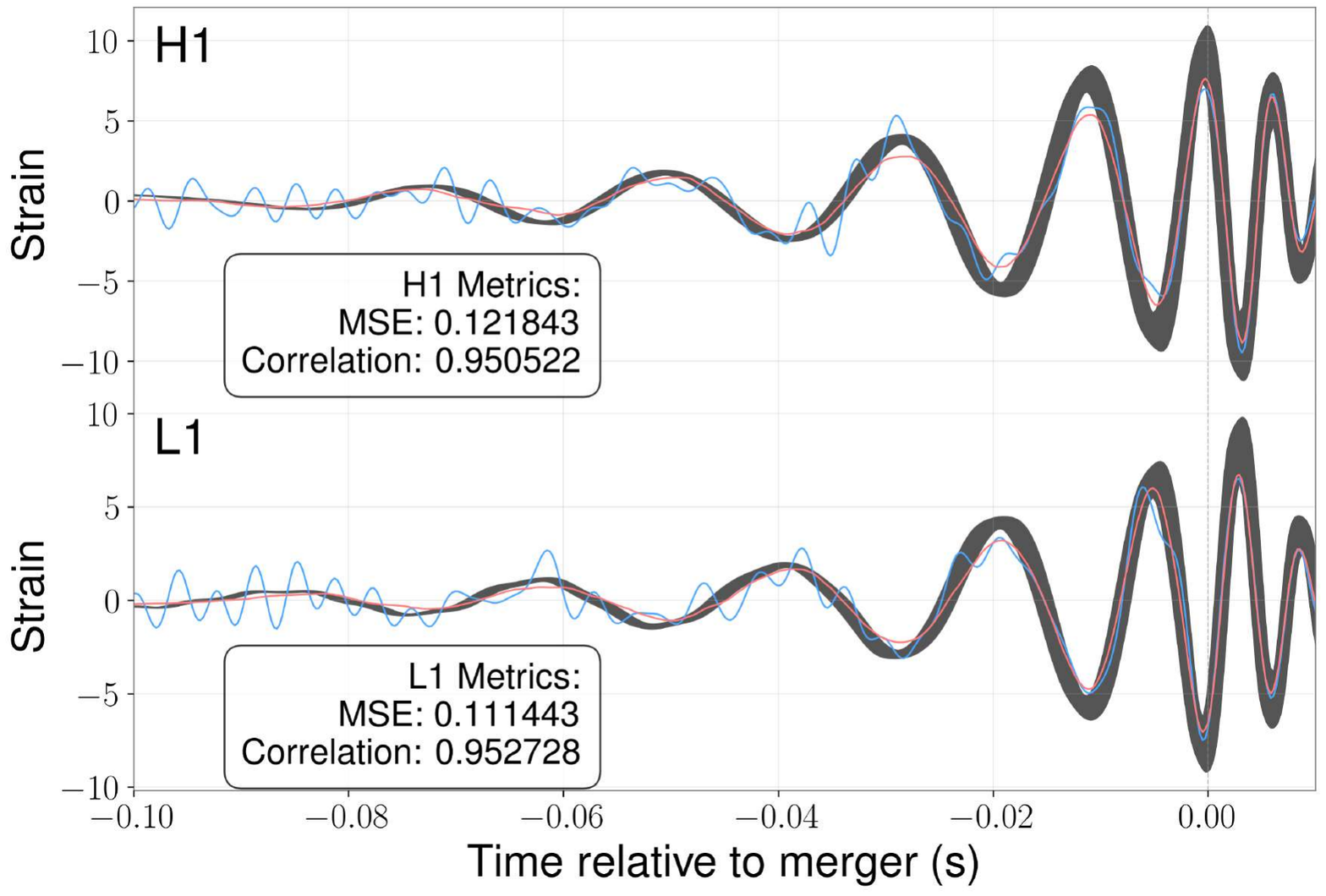}%
}%
\subfloat{%
  \includegraphics[width=0.5\columnwidth]{%
    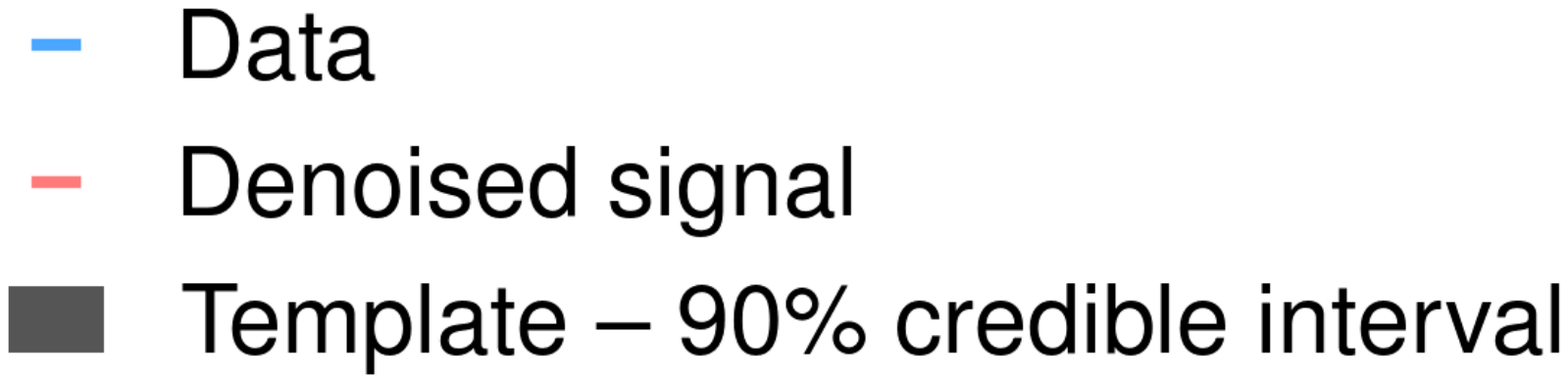}%
}
\label{fig:legend}
\caption{Denoised outputs from the Multi-Frequency architecture
applied to real LIGO strain data for three binary black hole events spanning
three observing runs. Each subfigure shows H1 (top) and L1 (bottom) data.
In each panel: LIGO open data after whitening and bandpass filtering
(blue), a posterior-ensemble 90\% credible-interval
band centered on the maximum-likelihood waveform template (grey
shading; cf.\ Fig.~6 of \citep{LIGOScientific:2016vlm} for the
analogous construction applied to GW150914 itself), and the model
denoised output (red). The vertical dashed line
marks the merger at $t = 0$. The displayed window ($[-0.1, +0.01]$\,s)
captures the high-signal-power region near merger; the full 2-second
preprocessed segment was used as model input. Metrics in each panel are
computed over the $[-0.1, +0.01]$\,s window and compared against the
maximum-likelihood template.}

\label{fig:real_events}
\end{figure*}

\begin{table*}[htbp]
\centering
\caption{Source parameters and denoising metrics for the three real
gravitational wave events. Source parameters are maximum-likelihood
estimates from the PE data release posteriors, used for template
generation. $\mathcal{M}_c$ is the chirp mass, $q = m_2/m_1 \leq 1$ is
the mass ratio, $\rho_{\rm net}$ is the network SNR, and $f_{\rm merger}$
is the approximate merger frequency estimated as
$f_{\rm merger} \approx 1.5 \times 220\,(100\,M_\odot/M_{\rm tot})$\,Hz.
Pearson correlation $r$ and energy-normalized MSE against the maximum likelihood template are evaluated over the
$[-0.1, +0.01]$\,s window around merger. The waveform column gives the
approximant used for maximum-likelihood template generation.}
\label{tab:real_events}
\begin{ruledtabular}
\resizebox{\textwidth}{!}{%
\begin{tabular}{lcccccccccccc}
\textbf{Event} & \textbf{Run} & $m_1$ & $m_2$ & $\mathcal{M}_c$ &
$\chi_{\rm eff}$ & $d_L$ & $\rho_{\rm net}$ & $f_{\rm merger}$ &
\textbf{Waveform} & $r_{\rm H1}$ & $r_{\rm L1}$ &
$\mathrm{MSE}_{\rm H1}$ / $\mathrm{MSE}_{\rm L1}$ \\
& & $(M_\odot)$ & $(M_\odot)$ & $(M_\odot)$ & & (Mpc) & & (Hz) & & & & \\
\hline
GW150914  & O1  & 38.22 & 32.21 & 30.42 & $-0.056$ &  456 & 26.0
& 469 & \texttt{IMRPhenomPv2}  & 0.976 & 0.977 & 0.069 / 0.073 \\
GW200129$^{a}$  & O3b & 43.83 & 31.61 & 31.90 & $+0.108$ &  952 & 26.8
& 437 & \texttt{IMRPhenomXPHM} & 0.982 & 0.972 & 0.087 / 0.073 \\
GW231226  & O4  & 48.24 & 42.46 & 39.31 & $-0.122$ & 1136 & 34.7
& 364 & \texttt{IMRPhenomXPHM} & 0.951 & 0.953 & 0.122 / 0.111 \\
\end{tabular}%
}
\end{ruledtabular}
{\small $^{a}$L1 strain incorporates \texttt{gwsubtract} glitch
mitigation~\citep{Davis:2022ird}; residual subtraction uncertainty
of $\pm 0.022$ in whitened strain units is concentrated in the
$20$--$50$\,Hz band~\citep{Payne:2022spz}.}
\end{table*}

We apply the Multi-Frequency architecture to strain data from
three confirmed binary black hole detections spanning three LVK
observing runs: GW150914~\citep{LIGOScientific:2016aoc} from O1,
GW200129~\citep{LIGOScientific:2021sio} from O3b, and GW231226~\citep{LIGOScientific:2025snk} from O4. These
events were selected to probe model generalization across distinct noise
epochs, as each observing run operated with a different detector
configuration and noise floor from the simulated aLIGO design sensitivity
PSD used in training.

\paragraph{Data preparation.}
For each event, strain data are retrieved from the Gravitational Wave Open 
Science Center (GWOSC)~\citep{LIGOScientific:2025snk} for both LIGO Hanford (H1) 
and LIGO Livingston (L1). The noise PSDs and maximum-likelihood waveform 
templates are extracted from the corresponding parameter estimation data 
release files using \texttt{pesummary}~\citep{Hoy:2020vys}, which provides 
direct access to the posterior samples which can be used to reconstruct waveforms. For each event we additionally draw a random ensemble of posterior samples
from the PE release, generate the corresponding time-domain waveform for
every sampled point using the same projection procedure as the
maximum-likelihood template, and propagate each realization through the
identical whitening, bandpass, and normalization pipeline applied to the
data. At each time sample, the 5th and 95th percentiles of the resulting ensemble of whitened waveforms define a 90\% credible-interval band about the template. This band reflects the parameter-estimation posterior's uncertainty on the
reconstructed waveform itself, and serves as a reference for judging the
denoised output; it is not a measure of the denoising network's own
uncertainty. A data segment centered on
the merger is extracted, whitened using the event-specific noise PSD from
the corresponding parameter estimation data release file, and bandpass
filtered to $40$--$300$\,Hz. This narrower bandpass, relative 
to the $20$--$2000$\,Hz training band, is chosen to exclude 
poorly conditioned low-frequency noise in the publicly 
distributed open data segments and to concentrate the 
comparison on the frequency range where the signal power is 
highest. Crucially, whitening uses each event's actual PE-release PSD rather than the simulated aLIGO design sensitivity PSD
employed during training---a deliberate choice that ensures the whitened
data faithfully reflect the true noise environment of each observing run
rather than an idealized approximation. The denoised output is compared
against the maximum-likelihood waveform template projected onto each
detector, generated from the PE posterior samples using the waveform
approximant listed in Table~\ref{tab:real_events}. The metrics are
evaluated over a $110$\,ms window spanning $100$\,ms before to $10$\,ms
after the merger, which encompasses the dominant signal power for the
high-mass systems considered here.

For GW200129, the standard GWOSC-distributed L1 strain incorporates the
LVK glitch mitigation performed with \texttt{gwsubtract}~\citep{Davis:2022ird},
which subtracted a transient noise artifact from the electro-optic modulator
system that temporally overlapped the signal. Payne et al.~\citep{Payne:2022spz}
demonstrated that this glitch is localized in the $20$--$50$\,Hz band and
that the \texttt{gwsubtract} subtraction carries a statistical uncertainty
of $\pm 0.022$ in whitened strain units. 

\paragraph{Results.}
Figure~\ref{fig:real_events} presents the denoised outputs for all three
events using Multi-Frequency architecture , and Table~\ref{tab:real_events} summarizes the source parameters
and per-detector metrics.

The model successfully recovers the merger morphology across all three
events and both detectors, demonstrating generalization from simulated
Gaussian noise to real detector data spanning a decade of observing
operations. The denoised output achieves high correlation with the maximum-likelihood template(Table~\ref{tab:real_events}) throughout the late inspiral and merger in all cases. The posterior-ensemble uncertainty band is narrow during the quiet early inspiral and widens over the high-amplitude late-inspiral and merger
cycles, tracking the local signal amplitude; the denoised output remains
within this band throughout the analysis
window for all three events, staying visibly within the band than
the raw whitened data does. 
% The band width itself varies across events,
% narrowest for GW150914 and widest for GW231226, with GW200129
% intermediate --- consistent with differences in the precision of each
% event's parameter-estimation posterior.

Across all three events and both detectors, the model achieves a minimum
Pearson correlation of $r > 0.95$ and a maximum
$\mathrm{MSE}_{\rm norm} < 0.13$ against the maximum-likelihood template,
despite having been trained exclusively on simulated Gaussian noise and
never exposed to real detector data. Each observing run has its own
characteristic noise environment---shaped by detector commissioning state,
instrumental noise sources, and non-stationary, non-Gaussian
transients---that differs from the purely stationary Gaussian noise
generated from the aLIGO Zero-Detuned High Power PSD used in training.
The consistency of denoising quality across events from O1, O3b, and O4
therefore demonstrates genuine generalization across noise epochs without
any fine-tuning or retraining on observed data.

Despite the model being trained exclusively on H1 strain data, it
achieves comparable reconstruction quality in L1 across all three events,
with correlations consistent to within a few percent. This
confirms that the learned denoising representations are not tied to
H1-specific instrumental features but capture generic signal morphology
that transfers to L1's distinct noise environment and arm orientation
without any detector-specific retraining.

Despite these domain gaps, all correlations exceed $r > 0.95$
across three observing runs, two different PE data releases, and a range
of source configurations spanning near-equal-mass to moderate mass-ratio
systems. This consistency confirms that the Multi-Frequency
architecture has learned signal features that transfer robustly to
real detector environments without any fine-tuning or retraining on
observed data.

\subsection{\label{sec:pure_noise_real}Behavior on Real Detector Noise: A Pure-Noise Background Study}

Section~\ref{sec:ablation} selects the pure-noise augmentation
fraction used during training ($30\%$) based on its effect on validation
loss, but does not itself characterize the trained model's output on
pure-noise input. Here we directly test that behavior, on an independent
sample of real H1 detector noise substantially larger than, and
disjoint from, the training and validation sets, verified to contain no
catalogued gravitational-wave event. We analyzed 24 hours of real H1
strain from the O4a observing run through our network, aggregated from
six $\sim$4-hour windows spread across the O4a run (GPS
$1368195220$--$1389456018$), each restricted to \texttt{H1\_CBC\_CAT3}
data-quality-passing time and excluded within $\pm120$\,s of every
catalogued event. The strain was chopped into non-overlapping 2\,s
segments, whitened following the procedure of Section~\ref{sec:whitening},
and passed through the trained Multi-Frequency model. After excluding
segments from chunks shorter than 300\,s or lacking sufficient
whitening-padding context, $43{,}083$ of $43{,}200$ raw segments remained,
corresponding to $11.97$ hours of windowed background evaluated over the
same $1$\,s analysis window used in Section~\ref{sec:qualitative}.

We quantify the network's output on each segment using an output-norm
statistic
\begin{equation}
\rho_{\rm out} = \left[4\int_{f_{\rm low}}^{f_{\rm Nyq}}
|\tilde{h}_{\rm out}(f)|^2\, df\right]^{1/2},
\label{eq:rho_out}
\end{equation}
with $f_{\rm low}=10$\,Hz, of the same functional form as the matched-filter
inner product of Equation~(\ref{eq:inner_product}) --- and indeed the same
definition as the injection optimal SNR of Equation~(\ref{eq:snr}) --- but
evaluated on the network's already-whitened output rather than the injected
signal, against a flat (unit) spectral density rather than $S_n(f)$ (the two
are equivalent once whitening has already been applied), and over the
$1$\,s analysis window of Section~\ref{sec:qualitative} rather than the
full $2$\,s segment. A value of $\rho_{\rm out}=0$ corresponds to perfect
suppression, the ideal target for a pure-noise input.

\begin{figure}[htbp]
\centering
\includegraphics[width=0.7\columnwidth]{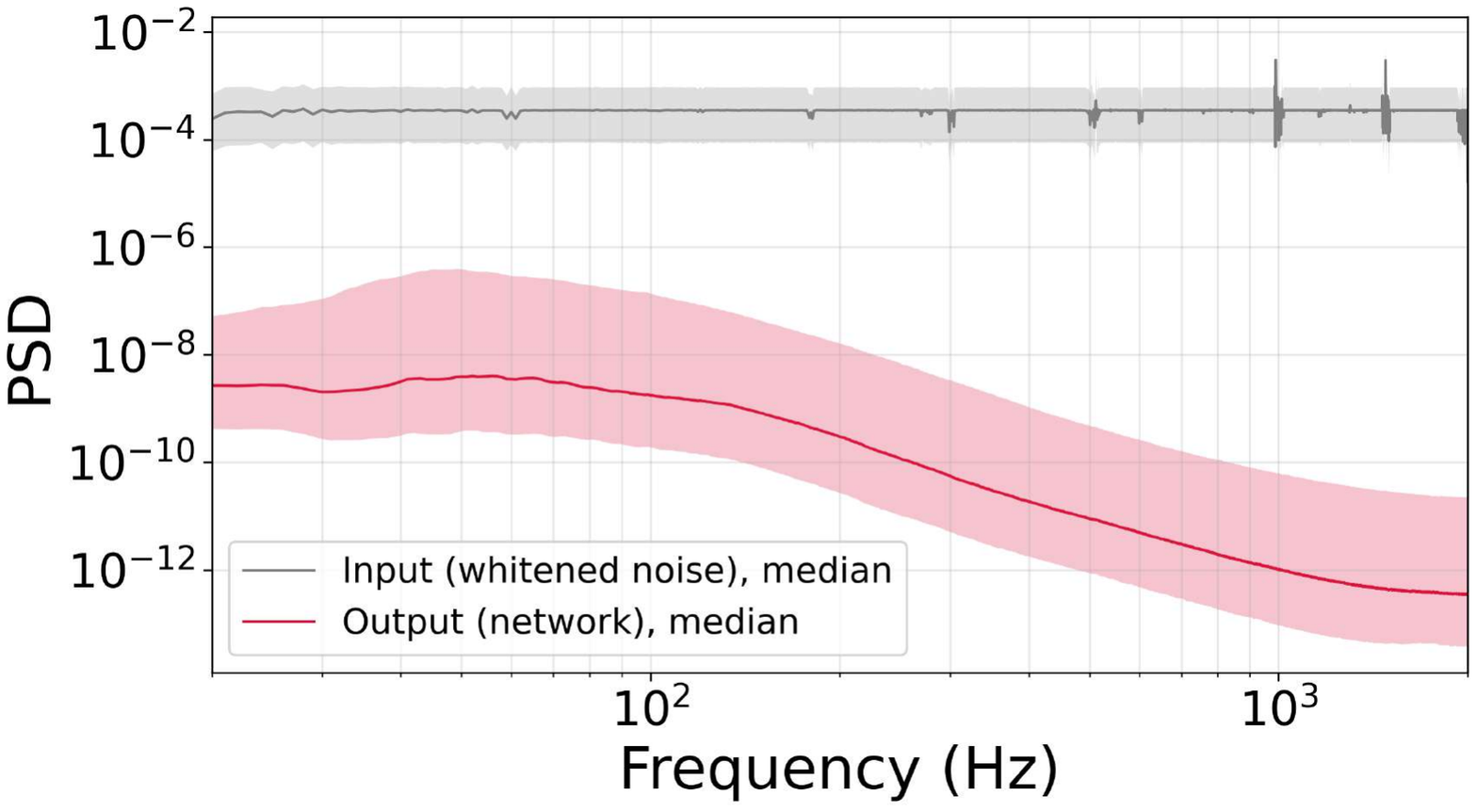}
\caption{Median power spectral density of the whitened input
(gray) and network output (crimson), computed over the
$[-1.0,0.0]$\,s analysis window across all $43{,}083$ real H1 noise-only
background segments. Shaded bands span the 16th--84th percentile across
segments. The output PSD lies orders of magnitude below the input across
the entire analysis band, with a maximum output/input ratio of $0.007$ at
$1.0$\,Hz and no frequency bin exceeding a factor of 3, confirming
near-total suppression on real, stationary detector noise.}
\label{fig:psd_comparison_pure}
\end{figure}

We find that the network reliably denoises data stretches containing
nearly Gaussian noise cleanly, suppressing $\rho_{\rm out}$ to
$0.0461$ (median) and $1.2058 \pm 78.0914$ (mean $\pm$ std) across the full background sample, consistent with near-total power suppression (Figure~\ref{fig:psd_comparison_pure}: median output PSD suppressed relative to input at every frequency in the analysis band, with a maximum output/input ratio of $0.007$ at $1.0$\,Hz).

\begin{figure}[htbp]
\centering
\includegraphics[width=\columnwidth]{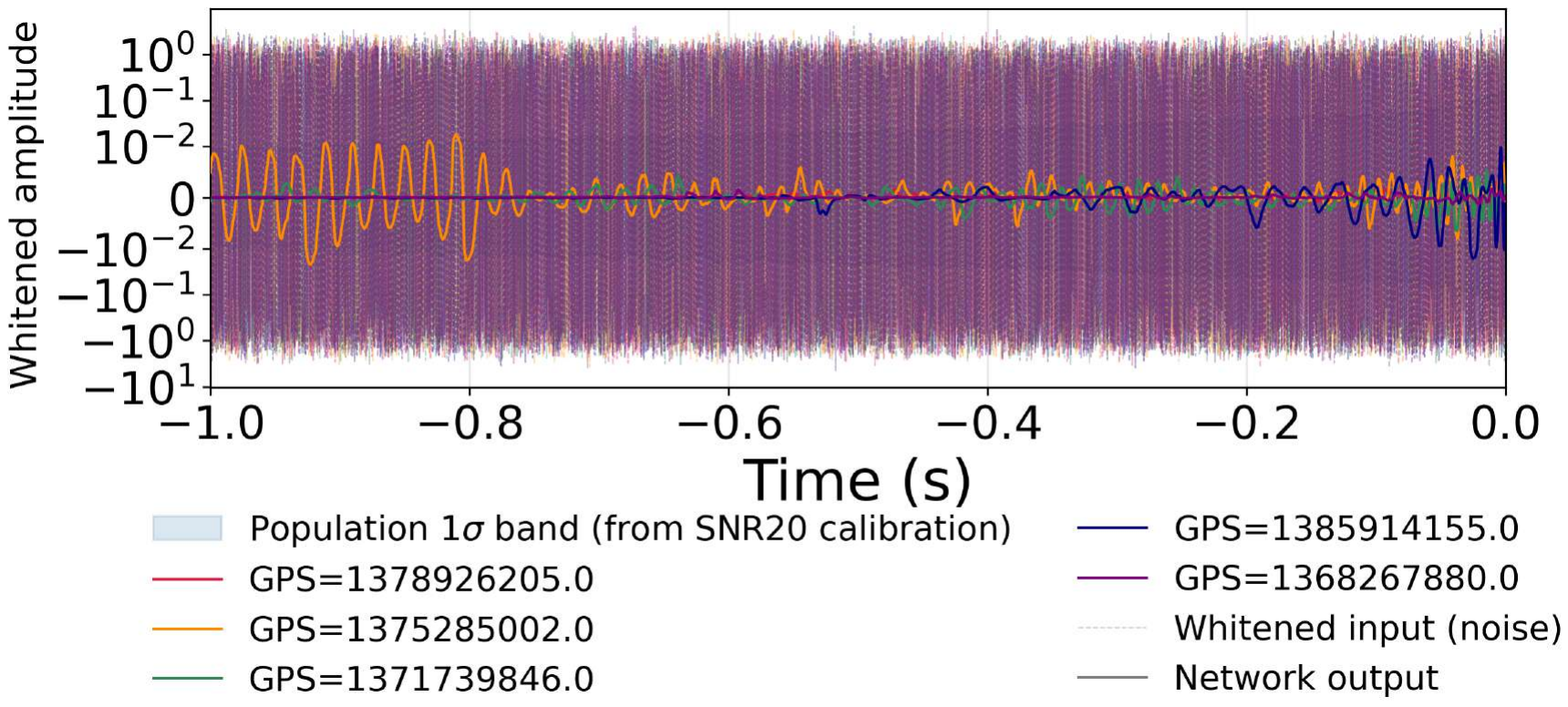}
\caption{Whitened network output (solid) for five randomly
selected real H1 noise-only background segments (dashed traces show the
corresponding whitened input), overlaid on the population $1\sigma$ band
from the SNR$=20$ calibration (Section~\ref{sec:qualitative}). All five
outputs remain suppressed within or below the calibration band throughout
the $[-1.0,0.0]$\,s analysis window, representative of the network's
behavior on the bulk of the real noise-only background.}
\label{fig:example_segment}
\end{figure}

Figure~\ref{fig:example_segment} illustrates this typical
behavior for five randomly selected background segments. For segments containing no signal, the denoised output does not resemble a chirp-type signature, unlike those containing a signal.

A rare, extreme high-$\rho_{\rm out}$ tail exists, however: the 95th and 99th percentiles are $2.17$ and $4.18$ respectively, with a maximum of $13{,}502.6$ across the full sample --- a tail heavy enough that the mean ($1.21$) and standard deviation ($78.09$) are dominated by a handful of extreme segments rather than reflecting the bulk of the distribution, which sits close to the median of $0.046$. We find that this tail is populated almost exclusively by segments containing real, non-Gaussian noise transients (glitches). Because the network was trained only on stationary simulated Gaussian noise (augmented with a $30\%$ pure-noise fraction, Section~\ref{sec:ablation}), it has no learned basis for suppressing genuinely non-Gaussian structure; instead, as illustrated by the two glitch examples in Figure~\ref{fig:pure_noise_examples}
(Appendix~\ref{sec:pure_noise_appendix}), the network removes the
surrounding Gaussian background while leaving the glitch's transient
structure largely intact --- effectively unveiling the glitch rather than inventing structure from nothing. This is consistent with the output peak coinciding with the input peak to within a few milliseconds ($2.2$--$5.1$\,ms for the two examples shown) and $56$--$61\%$ of input-window power retained in these segments, in contrast to the two representative near-median segments in the same figure, where under $10^{-6}$ of input power is retained and any residual output structure is uncorrelated in
time with the input.

\begin{figure}[htbp]
\centering
\includegraphics[width=\columnwidth]{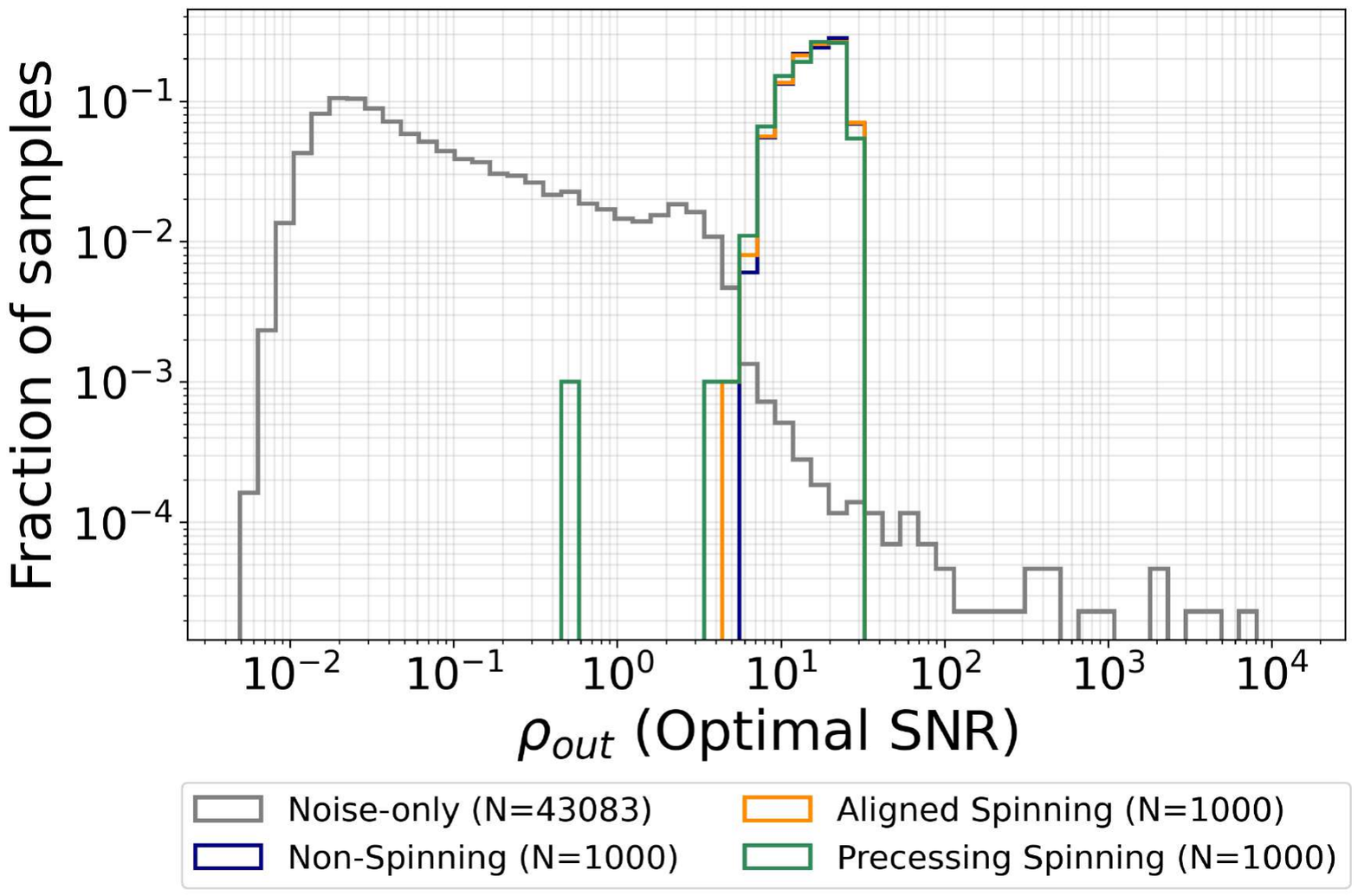}
\caption{Distribution of $\rho_{\rm out}$ for the real H1
noise-only background (gray, $N=43{,}083$) and for the stratified
SNR$=10$--$30$ injected signal population (colored, $N=1000$ per spin
category), both evaluated over the $[-1.0,0.0]$\,s analysis window. The
noise-only distribution is sharply peaked near zero with a long
non-Gaussian tail populated by glitches; the signal population clusters
around $\rho_{\rm out}\sim16.7$--$17.2$, overlapping the background only
in that glitch-driven tail. This is a single-detector (H1) characterization
with no coincidence testing against a second interferometer to suppress
glitches, unlike an operational search.}
\label{fig:rho_histogram}
\end{figure}

Figure~\ref{fig:rho_histogram} compares the resulting $\rho_{\rm out}$
distribution on this real noise-only background against $\rho_{\rm out}$
computed on an independent population of $N=1000$ injected signals per spin
category, drawn from a stratified SNR $\in [10,30]$ population (distinct
from the SNR$=20$ population used for case studies and calibration
elsewhere in this work). This background characterization is single-detector
(H1 only) and applies no multi-detector coincidence requirement of the kind
used in operational search pipelines to reject instrumental glitches; the
glitch-driven tail seen here therefore reflects the single-interferometer
noise environment directly.

\begin{figure*}[htbp]
\centering
\vspace{-3cm}
\includegraphics[width=\textwidth]{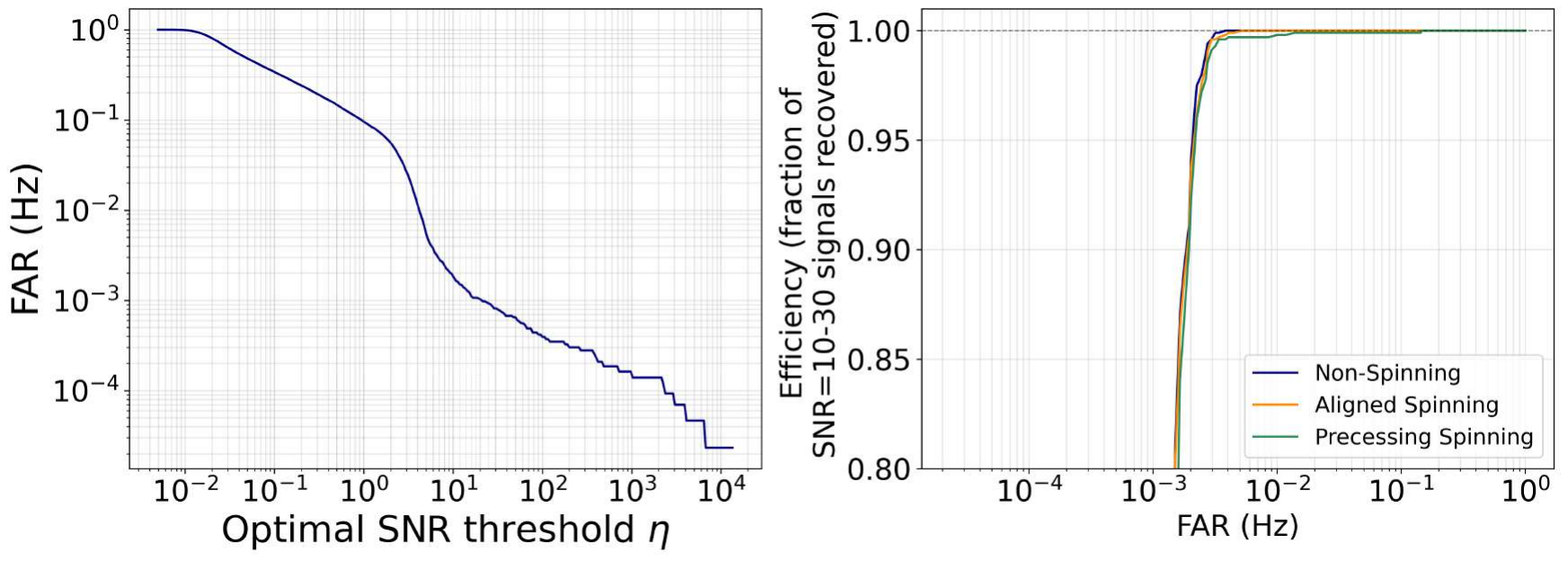}
\vspace{-3cm}
\caption{Left: false-alarm rate as a function of threshold $\eta$
on $\rho_{\rm out}$, computed from $11.97$\,hr of real H1 noise-only
background. Right: recovery efficiency (fraction of injected signals with
$\rho_{\rm out}\geq\eta$) as a function of the corresponding FAR, for the
stratified SNR$=10$--$30$ injected population, per spin category.
Efficiency remains near $100\%$ for FAR $\gtrsim 3\times10^{-3}$\,Hz and
falls off as the threshold is pushed further into the glitch-populated
tail of the background distribution.}
\label{fig:far_curve}
\end{figure*}

The noise-only distribution is sharply peaked near
zero with a long non-Gaussian tail, while the signal population clusters
around $\rho_{\rm out} \sim 16.7$--$17.2$ (mean, all three categories),
with the two distributions overlapping only in the glitch-populated tail of
the background. Table~\ref{tab:far_efficiency} and
Figure~\ref{fig:far_curve} quantify this separation directly: treating
$\rho_{\rm out}$ as a threshold statistic, the false-alarm rate computed
from the real noise-only background falls rapidly with threshold $\eta$,
while recovery efficiency on the injected SNR$=10$--$30$ population remains
at or near $100\%$ up to $\eta \sim 6$, declines steadily through the
crossover region, drops below $50\%$ between $\eta = 16$ and $18$ ---
close to the signal population's own median $\rho_{\rm out}$ --- and
reaches $0\%$ by $\eta \sim 30$, tracking the transition visible in
Figure~\ref{fig:rho_histogram}.

\begin{table}[htbp]
\centering
\caption{False-alarm rate (from $11.97$\,hr of real H1 noise-only
background) and per-category recovery efficiency (from the stratified
SNR$=10$--$30$ injected population, $N=1000$ per category) as a function of
threshold $\eta$ on $\rho_{\rm out}$.}
\label{tab:far_efficiency}
\footnotesize
\begin{ruledtabular}
\begin{tabular}{lccccc}
$\eta$ & FAR (Hz) & FAR (/day) & Eff.\ nons. & Eff.\ align. & Eff.\ prec. \\
\hline
0.1  & $3.40\times10^{-1}$ & $2.94\times10^{4}$ & 100.0\% & 100.0\% & 100.0\% \\
1    & $9.58\times10^{-2}$ & $8.28\times10^{3}$ & 100.0\% & 100.0\% &  99.9\% \\
4    & $1.16\times10^{-2}$ & $1.00\times10^{3}$ & 100.0\% & 100.0\% &  99.8\% \\
6    & $3.78\times10^{-3}$ & $3.27\times10^{2}$ & 100.0\% &  99.8\% &  99.6\% \\
8    & $2.44\times10^{-3}$ & $2.11\times10^{2}$ &  97.9\% &  97.6\% &  97.2\% \\
10   & $1.93\times10^{-3}$ & $1.66\times10^{2}$ &  90.3\% &  89.7\% &  88.5\% \\
12   & $1.51\times10^{-3}$ & $1.30\times10^{2}$ &  80.1\% &  79.9\% &  76.2\% \\
14   & $1.28\times10^{-3}$ & $1.10\times10^{2}$ &  67.8\% &  67.2\% &  65.5\% \\
16   & $1.11\times10^{-3}$ & $9.63\times10^{1}$ &  54.6\% &  56.0\% &  53.9\% \\
18   & $1.07\times10^{-3}$ & $9.23\times10^{1}$ &  44.0\% &  42.7\% &  41.8\% \\
20   & $1.04\times10^{-3}$ & $9.02\times10^{1}$ &  33.5\% &  31.3\% &  29.2\% \\
25   & $9.28\times10^{-4}$ & $8.02\times10^{1}$ &   8.2\% &   8.1\% &   6.5\% \\
30   & $8.12\times10^{-4}$ & $7.02\times10^{1}$ &   0.0\% &   0.0\% &   0.0\% \\
\end{tabular}
\end{ruledtabular}
\end{table}

As the scope of this work was to develop and investigate denoising
architectures for GW data, targeting specifically data where one already
knows of the existence of GW signal(s) in it, this analysis is offered as a
motivating observation rather than a validated search: it indicates the
network's output statistic can, in principle, distinguish real detector
noise from the presence of a signal, a capability we defer to a subsequent,
dedicated study of detection performance. The main limitation exposed here
is glitches, which we have deliberately not removed or mitigated in this
analysis; a glitch-aware treatment, potentially using this same
architecture for glitch classification, is deferred to forthcoming work.

\section{\label{sec:conclusion}Conclusions}

We have developed a deep-learning framework for gravitational-wave denoising that addresses key limitations of prior work by (i) incorporating physically motivated architectural inductive biases, (ii) performing controlled, like-for-like comparisons across model families, and (iii) validating performance on real detector data. The main findings are as follows.

We introduced a Multi-Scale Frequency-Aware (\emph{Multi-Frequency}) architecture that routes signal content through parallel branches specialized to the inspiral-, merger-, and ringdown-dominated frequency bands. This decomposition mitigates the time--frequency resolution trade-off inherent to fixed-kernel designs, while the cross-frequency integration stage enables information sharing across bands. Across all models tested, Multi-Frequency attains the best reconstruction quality with substantially fewer parameters than the largest baselines, illustrating that domain-aligned inductive bias can outperform brute-force scaling.

We performed a controlled comparison of five distinct paradigms---recurrent, convolutional--recurrent, transformer-based, multi-scale skip-connection, and frequency-decomposed---using identical datasets, training procedures, and evaluation metrics. Performance broadly tracks architectural capability: convolutional structure yields the largest improvement over purely recurrent baselines, consistent with the chirp's hierarchical time--frequency structure, while self-attention provides additional gains by capturing long-range temporal dependencies (albeit with slower convergence for long sequences). Time-resolved analyses further show that Multi-Frequency is most advantageous in the early inspiral, where its dedicated low-frequency branch can extract weak, slowly evolving signal content that competing architectures often miss.

An ablation study over the fraction of pure-noise training samples shows that $30\%$ yields the best trade-off between noise rejection and waveform fidelity. With no pure-noise exposure ($0\%$), models tend to hallucinate signal-like structure in noise-only inputs; with too much ($50\%$), the effective waveform diversity is reduced, leading to earlier convergence and poorer minima.

By training on a BBH parameter space spanning non-spinning, aligned-spin, and generically precessing systems, we extend existing denoising studies beyond restricted mass ranges and non-spinning configurations. Population-level tests show robust performance across an astrophysically motivated SNR distribution.

We further introduced population-level uncertainty bands around each denoised waveform, constructed from the residual distribution of an independent evaluation ensemble, and validated their statistical calibration using population-wide P-P diagnostics: reported $n\sigma$ bands were found to encompass the true residual an $n\sigma$ fraction of the time to within a few percent. Testing the architecture on two evaluation sets extending beyond the training sample and distribution --- extrapolated mass ratios and a shifted spin distribution shape --- showed graceful performance degradation rather than failure.

Finally, applying the models to three confirmed events spanning multiple observing runs (GW150914, GW200129, and GW231226) demonstrates successful transfer from simulated Gaussian noise to real detector data without fine-tuning. Whitening with each event's parameter-estimation PSD (rather than the design-sensitivity PSD used in training) yields an appropriate noise model for each epoch. In all cases, Pearson correlations exceed $r>0.95$ against the maximum-likelihood template, and performance transfers from H1-trained models to L1 data with comparable fidelity, indicating that the learned representations capture detector-agnostic signal morphology.

Several limitations of the current framework motivate future work.
Training was performed exclusively on H1 strain data with simulated
Gaussian noise; future extensions should incorporate multi-detector
training and real noise realizations to further close the domain gap
with operational deployment. A background study on 24 hours of real H1 detector noise
(Section~\ref{sec:pure_noise_real}) indicates that the network's output
statistic separates real noise-only data from the presence of an injected
signal with high fidelity, a capability that motivates a dedicated,
rigorously validated detection study as future work. The dominant
limitation exposed by this study is real noise transients (glitches):
because the network was trained only on stationary simulated noise, it has
no learned basis for suppressing genuinely non-Gaussian structure, and
essentially all of the rare high-amplitude outputs on real background were
traced to glitches rather than the network hallucinating structure from
Gaussian noise. Extending training to include glitch classes, potentially
using this same architecture, is deferred to forthcoming work, building on 
recent glitch-aware waveform reconstruction frameworks 
\citep{Chatterjee:2024obg}. Because H1 and L1 data are denoised
independently rather than through a coherent multi-detector likelihood as
in cWB or BayesWave, nothing in the present architecture guarantees that
separate single-detector reconstructions of the same event correspond to
consistent intrinsic source parameters, underscoring the model-dependent
character of this framework relative to coherent, model-independent
reconstruction methods. A similar contrast between model-agnostic (cWB, BayesWave) and modelled reconstructions was recently highlighted for the short-duration, intermediate-mass-ratio signal GW231123 \citep{Chatterjee:2025avc}. The parameter space was restricted to
quasi-circular BBH systems; extending to binary neutron star and
neutron star--black hole systems, eccentric orbits, and higher-mass
intermediate mass ratio inspirals would substantially broaden
applicability. The training SNR distribution was cosmologically weighted
to peak at $\mathrm{SNR} \sim 10$--$30$; dedicated training for
sub-threshold and marginal-SNR events could improve performance for
third-generation detectors such as Einstein Telescope and Cosmic
Explorer, where the event rate at low SNR will dominate. Finally, a broader bandpass and lower-frequency training would be required for lighter systems
and for next-generation detectors sensitive below $10$\,Hz. Transfer
learning from our pre-trained Multi-Frequency model provides a natural
starting point for these extensions, enabling adaptation to new source
classes and noise environments without retraining from scratch.

\ack{
R.R. thanks ICTS for logistical support. The models in this work were trained and analyzed using the high-performance computing resources provided by the Sonic Cluster at ICTS.
}

\funding{
R.R. was supported by the Prime Minister's Research Fellowship (PMRF),
Government of India. P.K. was supported by the Department of Atomic Energy,
Government of India, under project nos. RTI4019 and RTI4013, and by the Ashok
and Gita Vaish Early Career Faculty Fellowship at the International Centre for
Theoretical Sciences. This material is based upon work supported by NSF's LIGO
Laboratory, which is a major facility fully funded by the National Science
Foundation.
}
% R.R. acknowledges the Prime Minister’s Research Fellowship (PMRF) scheme for providing fellowship. R.R. also acknowledges ICTS for logistical support. P.K. acknowledges support of the Department of Atomic Energy, Government of India, under project nos. RTI4019 and RTI4013; and by the Ashok and Gita Vaish Early Career Faculty Fellowship at the International Centre for Theoretical Sciences. The models are trained and analyzed using the high-performance computing resources provided by Sonic Cluster at ICTS. This material is based upon work supported by NSF's LIGO Laboratory which is a major facility fully funded by the National Science Foundation.

\data{
The trained model weights and architecture definitions underlying 
this article are publicly available on Zenodo at 
\url{https://doi.org/10.5281/zenodo.20475968}~\cite{Raha:2025zenodo}. 
The release includes \texttt{.keras} model files for all five 
architectures described in ~\cref{sec:architectures} --- 
the LSTM Encoder-Decoder (trained with $0\%$, $30\%$, and $50\%$ 
pure noise augmentation), CNN-LSTM Hybrid, ACRED-Net, U-Net with 
Attention Gates, and Multi-Frequency architecture 
(each trained with $30\%$ pure noise augmentation). 
The gravitational wave strain data used for real-event validation 
(Section~\ref{sec:real_events}) are publicly available through 
GWOSC~\cite{LIGOScientific:2025snk} at 
\url{https://gwosc.org}. Any additional data underlying this 
article will be shared on reasonable request to the corresponding 
author.
}

\roles{
\noindent Rohan Raha\ \orcidlink{0009-0002-2354-2884}\
\href{https://orcid.org/0009-0002-2354-2884}{0009-0002-2354-2884}\\
Conceptualization
(equal), Data curation (lead), Formal analysis (lead), Investigation (lead),
Methodology (equal), Software (lead), Validation (lead), Visualization
(lead), Writing -- original draft (lead), Writing -- review \& editing
(equal).
\\
\noindent Prayush Kumar\ \orcidlink{0000-0001-5523-4603}\
\href{https://orcid.org/0000-0001-5523-4603}{0000-0001-5523-4603}\\
Conceptualization
(equal), Formal analysis (supporting), Funding acquisition (lead),
Investigation (supporting), Methodology (equal), Project administration
(lead), Resources (lead), Supervision (lead), Validation (supporting),
Writing -- original draft (supporting), Writing -- review \& editing
(equal).
}

\appendix

\section{\label{sec:methods}Data Preparation}
% \subsection{\label{sec:data}Training and Testing Dataset}

\subsection{Dataset Composition}

\begin{figure}[!htbp]
\centering
\includegraphics[width=0.7\columnwidth]{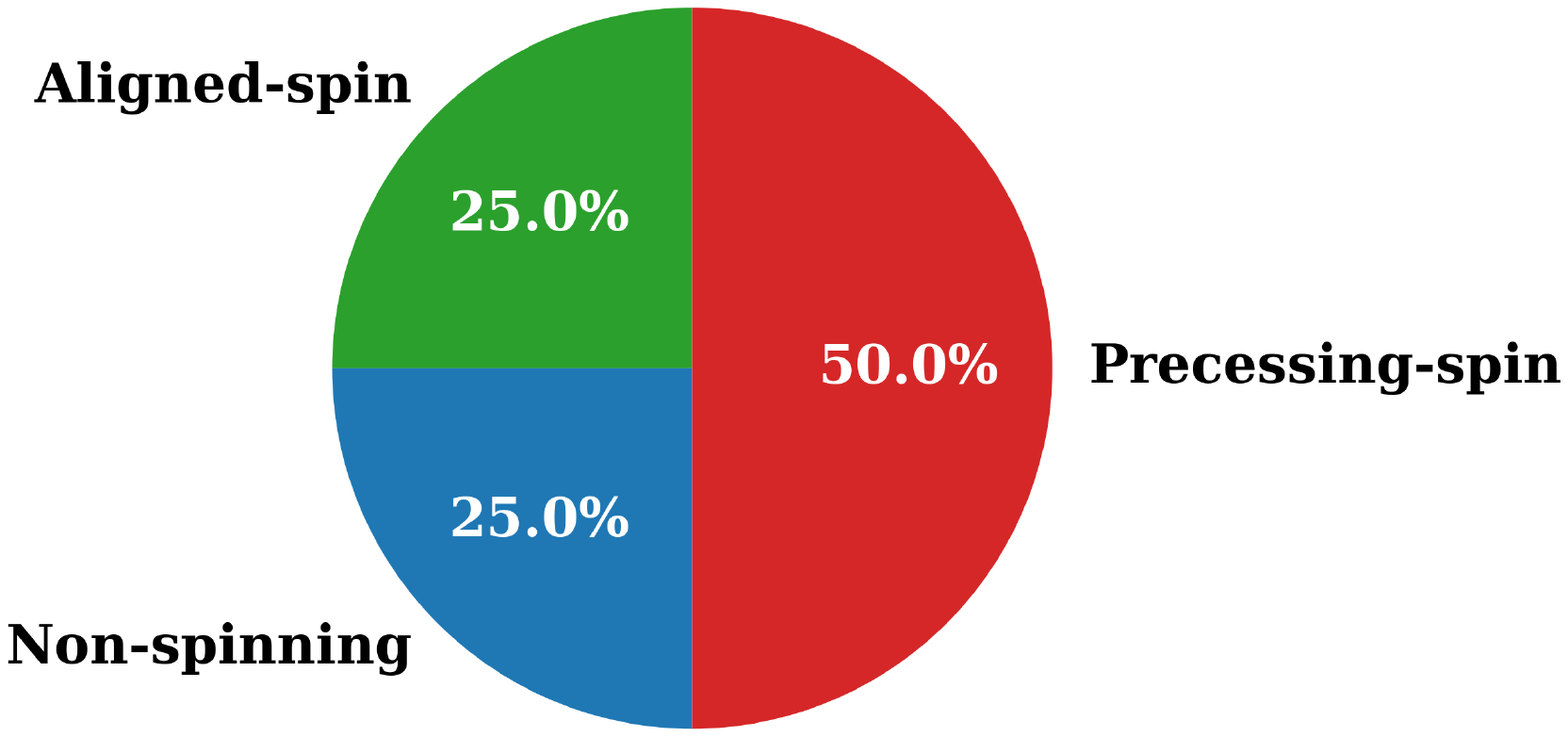}%
\caption{Compositional breakdown of both training and testing datasets ($N=20{,}000$ waveforms each). The three waveform categories are: Precessing-spin systems (red, 50\%) with generic 3D spin orientations and $\chi_p > 0$; Aligned-spin systems (green, 25\%) with spins parallel or anti-parallel to the orbital angular momentum ($s_{1x}=s_{1y}=s_{2x}=s_{2y}=0$); and Non-spinning systems (blue, 25\%) with $|\vec{s}_1|=|\vec{s}_2|=0$.}
\label{fig:dataset_composition}
\end{figure}

As summarized in \cref{sec:summary_data}, the training and testing
datasets each comprise 20{,}000 gravitational wave signals
distributed across three waveform categories, shown in
Figure~\ref{fig:dataset_composition}:
\begin{itemize}
    \item \textbf{Precessing-spin systems} (50\%, red): Full 3D spin dynamics with precessing spin parameter, $\chi_p > 0$, representing the generic astrophysical case where black hole spins are misaligned with the orbital angular momentum
    \item \textbf{Aligned-spin systems} (25\%, green): Spins parallel or anti-parallel to orbital angular momentum ($s_{1x} = s_{1y} = s_{2x} = s_{2y} = 0$, but $s_{1z}, s_{2z} \neq 0$), producing no orbital precession
    \item \textbf{Non-spinning systems} (25\%, blue): Zero spin magnitudes ($|\vec{s}_1| = |\vec{s}_2| = 0$), the simplest case with no spin-induced effects
\end{itemize}
Both datasets follow identical compositional distributions and parameter ranges.

\subsection{\label{sec:parameter}Waveform Model and Parameter Space}

The foundation of our denoising framework relies on a comprehensive dataset of gravitational wave signals spanning the binary black hole (BBH) parameter space. We employ the Python-based gravitational wave astronomy toolkit PyCBC~\cite{Usman:2015kfa}\footnote{PyCBC software is available at \url{https://github.com/gwastro/pycbc}.} to generate synthetic waveforms that accurately model the inspiral, merger, and ringdown phases of BBH coalescences.

\paragraph{Waveform Approximant Selection}

For waveform generation, we employ two variants of the effective-one-body (EOB) waveform family: \texttt{SEOBNRv4\_opt}~\cite{Bohe:2016gbl, Husa:2015iqa} for non-spinning and aligned-spin systems, and \texttt{SEOBNRv4P}~\cite{Bohe:2016gbl, Husa:2015iqa} for precessing-spin systems. These approximants are particularly well-suited for our analysis as they provide accurate modeling of the complete coalescence (including inspiral, merger, and ringdown phases), capture spin-precession dynamics crucial for systems with non-aligned spins, and have been extensively validated against numerical relativity simulations~\cite{Bohe:2016gbl, Pan:2013rra}. The waveforms are generated at a sampling rate of 4,096~Hz and subsequently processed to extract the final 2 seconds of evolution, with the merger occurring 0.1 seconds before the end of the time series. This temporal window ensures adequate coverage of the high-frequency merger and ringdown phases along with the low-frequency inspiral.

\paragraph{ Mass Parameter Space}

We sample the BBH mass parameter space ensuring uniform coverage in the symmetric mass ratio $\eta$, which is directly related to the gravitational wave luminosity~\cite{Cutler:1994ys}. The symmetric mass ratio is defined as:
\begin{equation}
\eta = \frac{m_1 m_2}{(m_1 + m_2)^2}
\end{equation}
where $m_1 \geq m_2$ are the component masses of the binary system. The individual component masses are constrained to the range $m_{\{1,2\}} \in [5, 100] M_{\odot}$, consistent with observed stellar-mass black hole populations from LVK observations~\cite{KAGRA:2021vkt}. 
The mass ratio $q = m_2/m_1 \leq 1$ is constrained to $q \geq 1/6$, 
corresponding to a minimum symmetric mass ratio of:
$\eta_{\text{min}} = \dfrac{q_{\text{min}}}{(1 + q_{\text{min}})^2} 
\approx 0.122$,
while the maximum $\eta = 0.25$ occurs for equal-mass binaries ($q=1$).

\begin{figure}[htbp]
\centering
\includegraphics[width=\columnwidth]{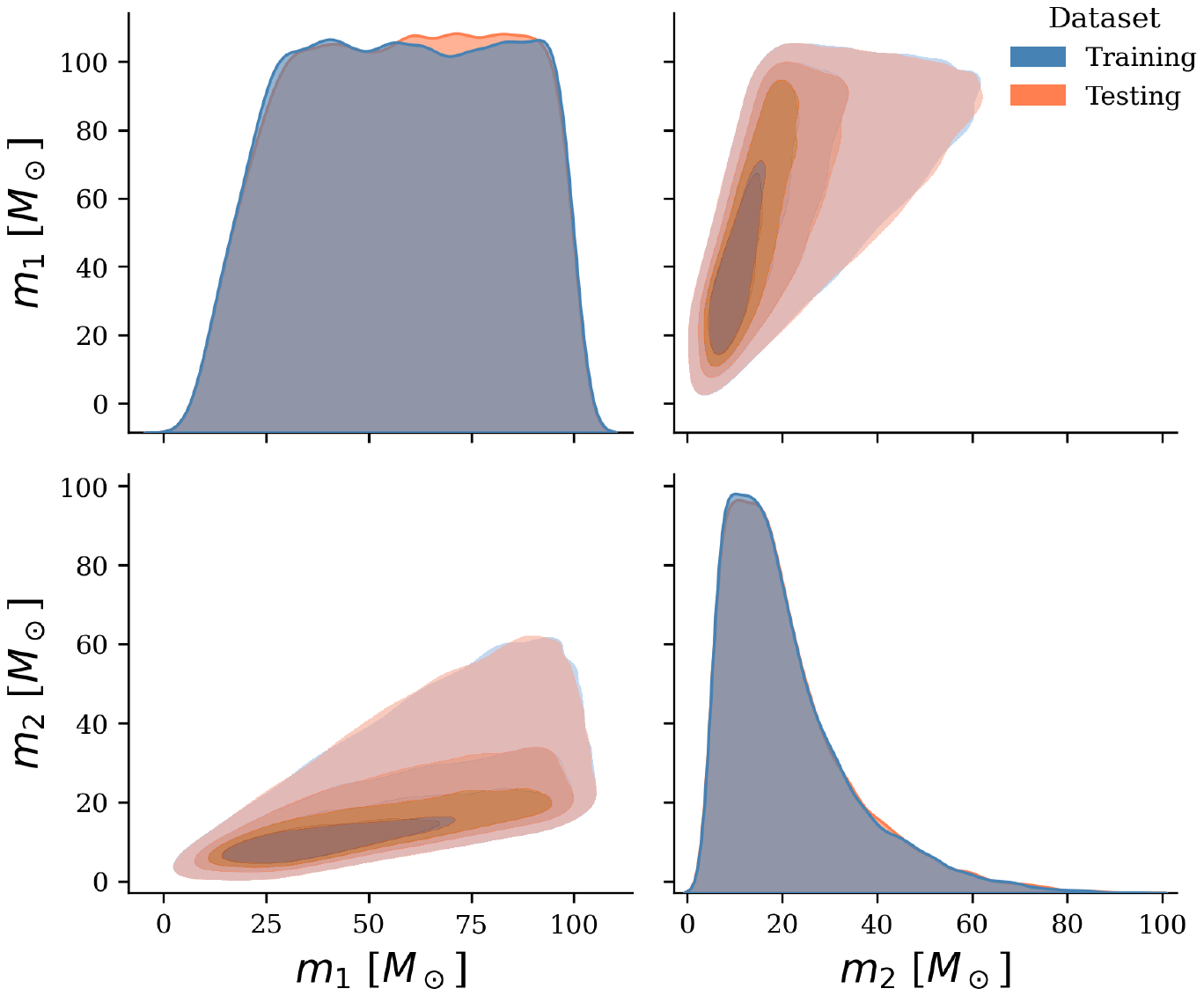}%
\caption{Mass parameter distributions for training (blue) and testing (orange) datasets. Component mass distribution in the $(m_1, m_2)$ plane, with $m_{1,2} \in [5, 100]\,M_{\odot}$. The color scale reflects the number of waveforms per bin; the higher density near equal-mass systems arises from uniform sampling in $\eta$ rather than $(m_1, m_2)$ space. Overlapping contours in all panels confirm identical and unbiased parameter sampling across datasets. }
\label{fig:mass_distributions}
\end{figure}

Figure~\ref{fig:mass_distributions} shows the resulting mass distributions 
for both training and testing datasets. The uniform coverage in $\eta$ --- 
rather than in $(m_1, m_2)$ space ensures that asymmetric mass-ratio systems receive equal statistical 
weight to near-equal-mass systems, preventing the training dataset from 
being biased toward the more densely sampled equal-mass regime.

In order to achieve uniform distribution in $\eta$, we employ a direct sampling approach:
\begin{enumerate}
    \item Sample $\eta$ uniformly from $[\eta_{\text{min}}, \eta_{\text{max}}]$.
    \item Compute the mass ratio using the inverse relationship: 
    $q = \left(1 - 2\eta - \sqrt{1 - 4\eta}\right)/(2\eta)$,
    which yields $q \leq 1$ by construction for 
    $\eta \in [\eta_{\rm min}, 0.25]$.
    \item Determine the valid total mass range $M = m_1 + m_2$:
    \begin{equation}
    M \in \left[\frac{5(1+q)}{q},\; 100(1+q)\right].
    \end{equation}
    \item Sample $M$ uniformly from this range.
    \item Calculate individual masses: $m_1 = M/(1+q)$ and 
    $m_2 = q\,m_1 = Mq/(1+q)$.  
\end{enumerate}

The extended tail toward higher total masses arises from low-$q$ 
systems (extreme mass ratios) where one component approaches 
$100\,M_{\odot}$ while the other is substantially lighter.

\paragraph{Spin Parameter Space}

The spin angular momenta of the two black holes are characterized by dimensionless spin vectors $\vec{s}_{1,2} = \vec{S}_{1,2}/m_{1,2}^2$, where each component has magnitude $|\vec{s}_{1,2}| \leq 0.99$, consistent with the Kerr bound~\cite{Kerr:1963ud}. We employ a method to generate uniformly distributed spin vectors:
\begin{enumerate}
    \item Sample polar angle $\theta$ using inverse transform:
    \[
        \theta = \arccos(1 - 2U), \quad U \sim \text{Uniform}(0, 1)
    \]
    
    \item Sample azimuthal angle $\phi$ uniformly from $[0, 2\pi]$.
    
    \item Convert to unit Cartesian coordinates:
    \[
    \begin{aligned}
        x &= \sin\theta \cos\phi, \\
        y &= \sin\theta \sin\phi, \\
        z &= \cos\theta
    \end{aligned}
    \]
    
    \item Sample magnitude $|\vec{s}|$ uniformly from $[10^{-6}, 0.99]$.
    
    \item Scale:
    \[
        \vec{s} = |\vec{s}| \cdot (x, y, z)
    \]
\end{enumerate}

This procedure ensures isotropic spin direction sampling while maintaining the Kerr bound.
From the component spins, we compute two physically meaningful quantities:
\begin{itemize}
    \item \textit{Effective aligned spin} $\chi_{\text{eff}}$, which characterizes the net spin component along the orbital angular momentum:
    \begin{equation}
    \chi_{\text{eff}} = \frac{m_1 s_{1z} + m_2 s_{2z}}{m_1 + m_2}
    \end{equation}
    where $s_{1z}, s_{2z}$ are the spin components along the orbital axis. This parameter directly affects the inspiral rate and merger time~\cite{Ajith:2009bn}.
    
    \item \textit{Effective precessing spin} $\chi_p$, which quantifies the spin-induced orbital precession~\cite{Schmidt:2014iyl}:
    \begin{equation}
    \chi_p = \max\left(|\vec{s}_{1,\perp}|, \, \frac{3+4q}{4+3q} \,q \, |\vec{s}_{2,\perp}|\right)
    \end{equation}
    where $|\vec{s}_{i,\perp}| = \sqrt{s_{ix}^2 + s_{iy}^2}$ are the in-plane spin magnitudes.
\end{itemize}

\begin{figure*}[htbp]
\centering
  \includegraphics[width=\textwidth]{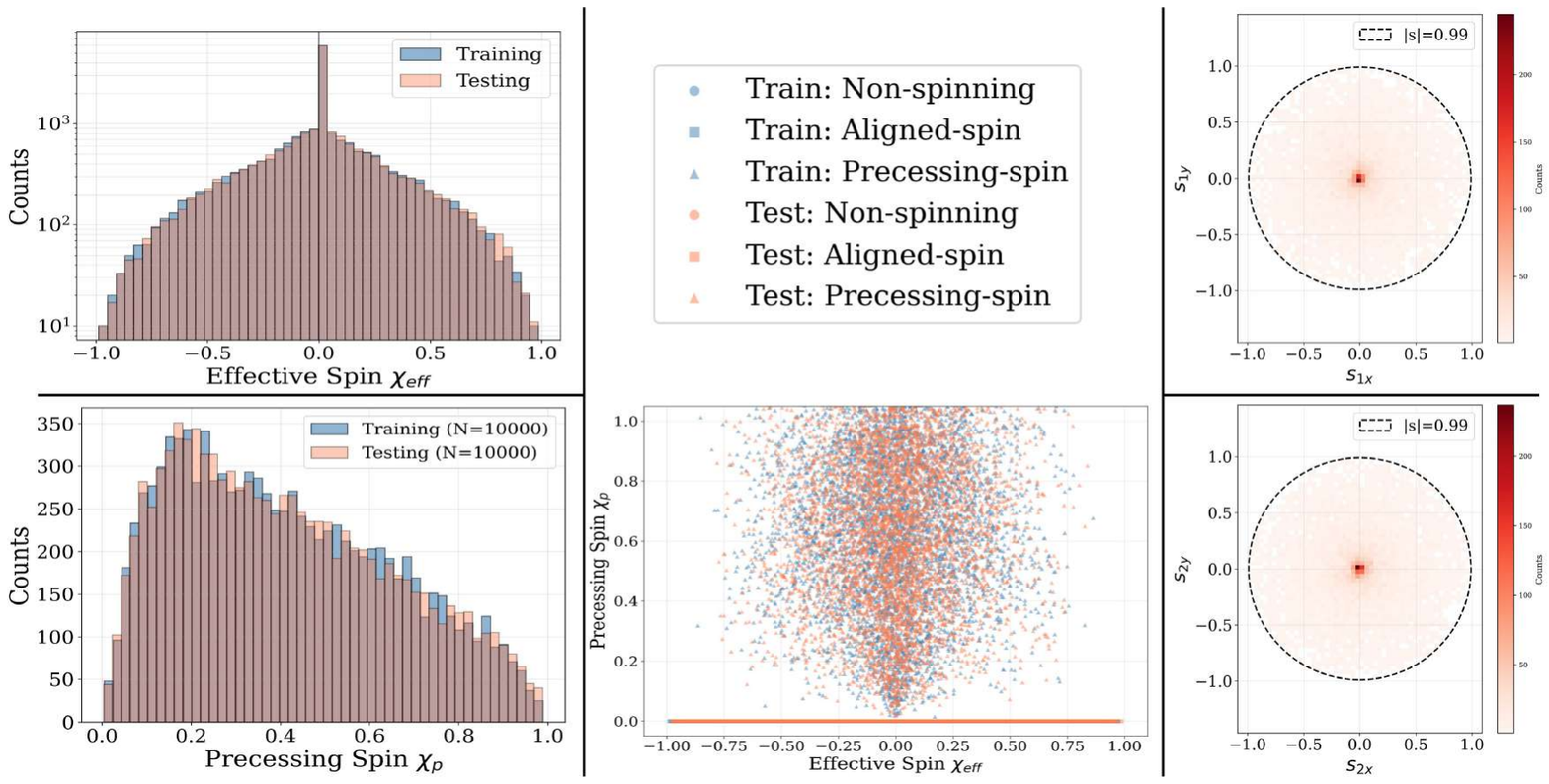}%
\caption{Spin parameter distributions for training and testing datasets.
Left: Effective spin $\chi_{\rm eff}$ (top, logarithmic scale)
spanning $[-1,1]$ for all 20000 samples, peaking near zero due to presence of non-spinning samples. Precessing spin $\chi_p$ (bottom) for only 10000 precessing samples,
peaking at $\chi_p \sim 0.3$--$0.7$ with values bounded below 
unity, since the suppression factor $q(4+3q)/(3+4q)$ 
(reaching ${\sim}0.2$ at $q=1/6$) reduces the secondary spin 
contribution at extreme mass ratios. Middle: The 2D $(\chi_{\rm eff}, \chi_p)$ panel shows non-spinning (blue, $N=5000$) systems
clustered at the origin, aligned-spin (green, $N=5000$) along $\chi_p = 0$,
and precessing (red, $N=10{,}000$) spanning the full plane including
$\chi_p > 1$ at high $q$.
 Right: In-plane spin components $(s_{1x}, s_{1y})$ (top) and
$(s_{2x}, s_{2y})$ (bottom) for only 10000 precessing systems, confirming isotropic
azimuthal coverage within $|\vec{s}| \leq 0.99$ (black dashed circle). The
mild central concentration reflects uniform magnitude sampling combined with
isotropic direction sampling.
Together, these confirm comprehensive and unbiased coverage of the BBH spin
parameter space.}
\label{fig:spin_distributions}
\end{figure*}

Figure~\ref{fig:spin_distributions} confirms comprehensive coverage of the 
BBH spin parameter manifold. The central concentration of $\chi_{\rm eff}$ 
near zero is a consequence of including non-spinning samples in the distribution. The scatter in 
$\chi_p$ at fixed $\chi_{\rm eff}$ reflects the degeneracy between aligned 
and in-plane spin components, which influence the waveform through distinct 
mechanisms --- inspiral phasing for $\chi_{\rm eff}$ and amplitude/phase 
modulations for $\chi_p$. The isotropic azimuthal distribution of in-plane 
spin components confirmed in Figure~\ref{fig:spin_distributions} (right panel) validates that 
our sampling algorithm introduces no directional bias.

\paragraph{Extrinsic Parameters}

The extrinsic parameters describe the orientation and location of the binary system relative to the detector and do not affect the intrinsic waveform morphology in the source frame. These include the orbital inclination angle $\iota$, coalescence phase angle $\phi_c$, right ascension $\alpha$ and declination $\delta$ for the source location on the sky, and the polarization angle $\psi$ relating the source frame to the radiation frame.

The inclination $\iota \in [0, \pi]$ and coalescence phase 
$\phi_c \in [0, 2\pi]$ are both sampled uniformly. The inclination controls the 
relative amplitudes of the two polarizations: face-on systems 
($\iota \sim 0, \pi$) exhibit dominant plus-polarized emission, while 
edge-on systems ($\iota \sim \pi/2$) produce comparable contributions 
from both polarizations. The coalescence phase sets the orbital 
configuration at merger and becomes degenerate with the polarization 
angle in the absence of spin precession, and with the mean periastron 
anomaly in the absence of eccentricity.

To ensure isotropic sky coverage, right ascension $\alpha \in [0, 2\pi]$ 
is drawn uniformly, while declination $\delta$ is sampled via the 
transformation $\cos\theta \sim \text{Uniform}(-1, 1)$ followed by 
$\delta = \pi/2 - \theta$. This equal-area sampling avoids the 
pole-concentration that would arise from naive uniform sampling in 
$(\alpha, \delta)$ coordinates~\cite{Schutz:1985jx}, and the resulting 
sky distributions are uniform in a 2-sphere to ensure comprehensive coverage of all possible observational configurations in the sky. The polarization angle $\psi \in [0, 2\pi]$ is sampled 
uniformly; for non-precessing systems it is degenerate with other extrinsic parameters and primarily governs the projection of $h_+$ and $h_\times$ onto 
differently oriented detectors~\cite{Schutz:1985jx}.

We restrict our analysis to quasi-circular orbits with eccentricity $e = 0$ at the reference frequency. This assumption is well-justified for the majority of stellar-mass BBH systems, as gravitational wave emission circularizes the orbit efficiently during the inspiral phase~\cite{Peters:1963ux}. Eccentric BBH mergers, while theoretically possible in dense stellar environments~\cite{Samsing:2017xmd}, constitute a relatively small fraction of the detected population, and are reserved for future extensions of this work.

\subsection{\label{sec:strain}Detector Strain}

Gravitational waves from BBH coalescences are characterized by two independent polarization states: the plus polarization $h_+(t)$ and the cross polarization $h_{\times}(t)$, which correspond to orthogonal quadrupolar deformation patterns of spacetime. These polarizations are modelled in the source frame with the $z$-axis aligned with the orbital angular momentum vector.
The relative amplitude and phase 
evolution between $h_+$ and $h_\times$ encodes the spin-precession 
dynamics, distinguishing precessing waveforms from non-precessing and 
aligned-spin systems.

The strain observed by a detector $d$ is given by the projection of these polarizations onto the detector response tensor, which can be written as a linear combination~\cite{Schutz:1985jx, Allen:2005fk}:
\begin{equation}
h_d(t) = F_+^d(\alpha, \delta, \psi, t) \, h_+(t - \Delta t_d) + F_{\times}^d(\alpha, \delta, \psi, t) \, h_{\times}(t - \Delta t_d)
\label{eq:detector_strain}
\end{equation}
where $F_{+,\times}^d$ are the detector's antenna pattern functions, which depend on the source's right ascension $\alpha$, declination $\delta$, polarization angle $\psi$, and time $t$ (accounting for Earth's rotation). The time delay $\Delta t_d$ represents the light travel time from the geocenter to the detector, computed as:
\begin{equation}
\Delta t_d = \frac{\vec{r}_d \cdot \hat{n}}{c}
\end{equation}
where $\vec{r}_d$ is the detector's position vector relative to Earth's center, $\hat{n}$ is the unit vector pointing toward the source, and $c$ is the speed of light.
The antenna pattern functions $F_{+,\times}$ range from $-1$ to $+1$ and 
determine the linear combination of $h_+$ and $h_\times$ measured by the 
detector for a given sky position and polarization angle. Their 
time-dependent variation due to Earth's rotation occurs on $\sim$24-hour 
timescales and is negligible for the 2-second waveform durations in our 
dataset; all waveforms are evaluated at a fixed GPS end time 
$t_{\rm end} = 1592529720$ (approximately June 24, 2030, 01:21:42 UTC), 
consistent with the anticipated aLIGO design sensitivity operational 
period.

The detector strain projection is implemented using PyCBC's \texttt{Detector} class~\cite{Usman:2015kfa}, which provides antenna pattern functions and time delays for each observatory. For each waveform in our dataset, we:
\begin{enumerate}
    \item Generate intrinsic waveforms $(h_+, h_{\times})$ at a fiducial distance of 1 Mpc using the appropriate approximant (\texttt{SEOBNRv4\_opt} or \texttt{SEOBNRv4P})
    \item Randomly sample extrinsic parameters $(\alpha, \delta, \psi)$ following the distributions described in \cref{sec:parameter}.
    \item Compute detector-specific antenna pattern functions $F_{+,\times}^d(\alpha, \delta, \psi, t_{\text{end}})$ at the GPS end time $t_{\text{end}} = 1592529720$
    \item Calculate geometric time delays $\Delta t_d$ from Earth's geocenter to each detector
    \item Project waveforms onto detector frames using Equation~(\ref{eq:detector_strain})
    \item Store the resulting strain time series $\{h_{\text{H1}}, h_{\text{L1}}, h_{\text{V1}}\}$ along with all source parameters
\end{enumerate}

Although our training uses H1 
strain exclusively, the projection formalism of 
Equation~(\ref{eq:detector_strain}) applies identically to any detector.

\subsection{Signal Morphology and Characteristics}

\begin{figure*}[htbp]
\centering
\includegraphics[width=\columnwidth]{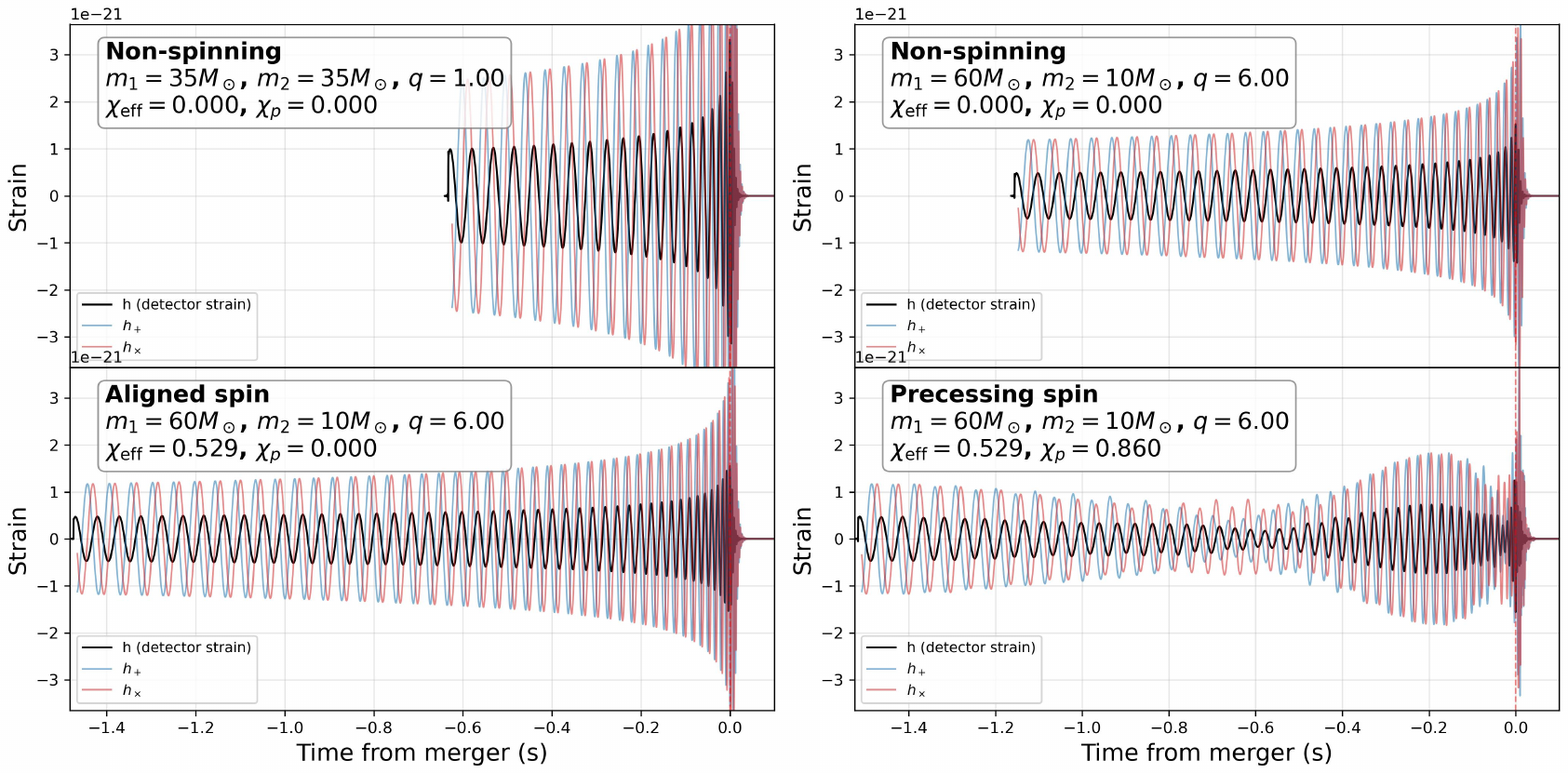}
\caption{Representative waveforms demonstrating morphological diversity across spin configurations
and mass ratios for binary black hole systems with total mass $M_{\text{tot}} = 70\,M_{\odot}$,
generated with $f_{\text{lower}} = 20$ Hz for visualization purposes. Each panel shows the
detector strain $h(t)$ at H1 (solid black), along with the plus polarization $h_+$ (blue) and
cross polarization $h_\times$ (red) overlaid as fainter lines. Top row: Non-spinning systems
($\chi_{\text{eff}} = 0.000$, $\chi_p = 0.000$) for equal-mass ($q = 1$, $m_1 = m_2 = 35\,M_{\odot}$,
left) and high mass-ratio ($q = 1/6$, $m_1 = 60\,M_{\odot}$, 
$m_2 = 10\,M_{\odot}$, right)
configurations, generated with \texttt{SEOBNRv4\_opt}. Bottom row: High mass-ratio systems ($q = 1/6$)
with aligned-spin configuration (left; $\chi_{\text{eff}} = 0.529$, $\chi_p = 0.000$,
\texttt{SEOBNRv4\_opt}) and precessing-spin configuration (right; $\chi_{\text{eff}} = 0.529$,
$\chi_p = 0.860$, \texttt{SEOBNRv4P}). The vertical red dashed line marks merger at $t = 0$\,s.
High mass-ratio systems produce shorter observable waveforms compared to equal-mass systems due to
faster inspiral evolution. In the precessing system (bottom right), amplitude modulations from
orbital plane precession are imprinted distinctly on each polarization, with $h_+$ and $h_\times$
dephasing visibly relative to one another. These illustrative waveforms highlight the morphological
diversity — across mass ratio, spin magnitude, and spin orientation — that motivates comprehensive
parameter space coverage in our training dataset.}
\label{fig:waveform_comparison}
\end{figure*}

To illustrate the diversity of BBH waveform morphologies in our dataset, we examine the influence of mass ratio and spin configurations on the observed signals through both visual inspection and quantitative analysis. Figure~\ref{fig:waveform_comparison} presents representative waveforms across different spin configurations (non-spinning, aligned-spin, and precessing-spin) and two mass ratios ($q=1$ and $q=6$) for systems with total mass $M_{\text{tot}} = 70\,M_{\odot}$, generated with $f_{\text{lower}} = 20$ Hz to demonstrate morphological diversity within a 2-second observation window. Table~\ref{tab:waveform_parameters} provides comprehensive temporal and amplitude measurements, revealing systematic dependencies crucial for understanding the full range of signal characteristics encountered during training.

\paragraph{Mass Ratio and Chirp Mass Effects:}
The chirp mass $\mathcal{M}_c = (m_1 m_2)^{3/5}/(m_1+m_2)^{1/5}$ controls 
the orbital decay rate and hence the inspiral duration. For quasi-circular 
inspirals, the time to coalescence from a given gravitational wave 
frequency $f$ scales as~\cite{Cutler:1994ys}:
\begin{equation}
\tau_{\text{coal}}(f) = \frac{5}{256} \frac{c^5}{G} 
\frac{1}{\mathcal{M}_c^{5/3} (\pi f)^{8/3}},
\end{equation}
where $G$ is Newton's constant and $c$ is the speed of light. The 
$\tau \propto \mathcal{M}_c^{-5/3}$ scaling directly explains the 
compressed inspiral phase visible for high mass-ratio systems in 
Figure~\ref{fig:waveform_comparison}, and is confirmed quantitatively 
by the cycle counts and durations in Table~\ref{tab:waveform_parameters}. 
Additionally, the reduced symmetric mass ratio $\eta$ for asymmetric 
binaries (low-q systems) decreases the gravitational wave luminosity via 
$h \propto \eta M / d_L$~\citep{Blanchet:2006zz}, resulting in systematically lower merger 
amplitudes at fixed total mass and distance, as reflected in 
Table~\ref{tab:waveform_parameters}.

\paragraph{Spin-Alignment Effects:}
The effective aligned spin $\chi_{\rm eff}$ modulates the inspiral through 
spin-orbit coupling terms in the post-Newtonian expansion~\cite{Ajith:2009bn}, 
quantified by spin-dependent corrections to the binding energy:
\begin{equation}
E_{\text{bind}} \propto \eta \left[ -\frac{1}{2v^2} + \chi_{\text{eff}} 
v^3 + \mathcal{O}(v^4) \right],
\end{equation}
where $v = (G M \Omega / c^3)^{1/3}$ is the characteristic velocity and 
$\Omega$ is the orbital frequency~\cite{Buonanno:1998gg}. Positive 
$\chi_{\rm eff}$ increases the binding energy, extending the inspiral 
duration and adding pre-merger cycles, while anti-aligned spins extract 
angular momentum and accelerate coalescence~\cite{Damour:2001tu}.

\begin{table*}[htbp]
\centering
\caption{Quantitative characteristics of representative BBH waveforms shown in Figure~\ref{fig:waveform_comparison}. All waveforms have total mass $M_\text{tot} = 70\,M_\odot$ and are generated with $f_\text{lower} = 20$~Hz. Amplitudes are computed at a fiducial distance of 1~Mpc. Cycle counts are determined by zero-crossing analysis: pre-merger cycles counted from $t=0$ to $t=1.9$~s, post-merger cycles from $t=1.9$~s to $t=2.0$~s. The pre-merger duration $\tau_\text{pre}$ is measured from $f_\text{lower}$ to merger.}
\label{tab:waveform_parameters}
\begin{tabular}{@{}lcccccccccc@{}}
\toprule
\textbf{Configuration} & $m_1$ & $m_2$ & $\mathcal{M}$ & $M$ & $q$ & $\chi_{\text{eff}}$ & $\chi_p$ & \textbf{Cycles} & \textbf{Max Amp.} & $\tau_{\text{pre}}$ \\
 & $(M_{\odot})$ & $(M_{\odot})$ & $(M_{\odot})$ & $(M_{\odot})$ & & & & \textbf{(pre / post)} & $(10^{-21})$ & (s) \\
\midrule
Non-spinning, $q=1$ & 35 & 35 & 30.47 & 70 & 1.0 & $0.00$ & $0.00$ & $46.0$ / $27.0$ & $3.34$ & $2.62$ \\
Non-spinning, $q=6$ & 60 & 10 & 19.86 & 70 & 6.0 & $0.00$ & $0.00$ & $102.5$ / $24.5$ & $1.50$ & $6.96$ \\
\midrule
Aligned spin, $q=6$  & 60 & 10 & 19.86 & 70 & 6.0 & $0.53$ & $0.00$ & $116.5$ / $27.0$ & $1.56$ & $7.93$ \\
Precessing spin, $q=6$ & 60 & 10 & 19.86 & 70 & 6.0 & $0.53$ & $0.86$ & $119.0$ / $41.5$ & $1.44$ & $8.03$ \\
\bottomrule
\end{tabular}
\end{table*}

\paragraph{Spin-Precession Signatures:}
Spin-induced orbital precession produces time-dependent viewing angles 
that periodically modulate the projected strain amplitude, with the 
instantaneous amplitude varying by factors of $\sim$2--3 over precession 
timescales of $\sim$0.1--1 seconds~\cite{Schmidt:2014iyl}. This 
systematic amplitude suppression relative to aligned-spin counterparts, 
visible in Figure~\ref{fig:waveform_comparison} and quantified in 
Table~\ref{tab:waveform_parameters}, constitutes a key observational 
signature for discriminating between aligned and precessing 
systems~\cite{Vitale:2016avz}. The morphological complexity of precessing 
waveforms motivates both the use of \texttt{SEOBNRv4P} and the 50\% 
allocation to precessing systems in our training dataset.

\paragraph{Ringdown and Post-Merger Dynamics:}
Precessing systems exhibit enhanced post-merger cycle counts relative to 
aligned-spin counterparts (Table~\ref{tab:waveform_parameters}), arising 
from the excitation of higher-order quasi-normal modes (QNMs) due to the 
misaligned merger configuration~\cite{Berti:2005ys}. The in-plane spin 
components $\chi_p$ cause the binary to merge with non-optimal alignment, 
populating subdominant QNM multipoles beyond the fundamental 
$(\ell, m, n) = (2, 2, 0)$ mode~\cite{London:2014cma}, producing extended 
ringdown structure that the denoising network must learn to distinguish 
from noise transients. The dominant $(\ell=2, m=2, n=0)$ 
QNM damping timescale scales linearly with the final black hole mass as 
$\tau_{220} \approx 11.2\,(M_{\rm final}/M_{\odot})\,\mu$s~\cite{Berti:2005ys}, 
yielding $\tau_{220} \sim 0.8$\,ms for the $M = 70\,M_{\odot}$ systems 
in Table~\ref{tab:waveform_parameters}.

\subsection{Simulated Detector Noise}
\label{sec:simulated_noise}

Realistic training and evaluation of denoising algorithms for gravitational wave detection requires synthetic detector noise that accurately represents the stochastic background observed in ground-based interferometers. We generate simulated noise realizations using the aLIGO Zero-Detuned High Power noise power spectral density (PSD), which characterizes the anticipated sensitivity of Advanced LIGO at design sensitivity~\cite{Aasi_2015}.

The aLIGO Zero-Detuned High Power PSD represents the expected noise characteristics of Advanced LIGO operating at its design sensitivity with zero detuning of the signal recycling cavity, optimized for broadband sensitivity across the detector's observable frequency band~\cite{Harry:2010zz}. The analytical form of this PSD is implemented through the \texttt{aLIGOZeroDetHighPower} method in PyCBC~\cite{Usman:2015kfa}, which provides a faithful representation of the anticipated detector noise characteristics.

We generate colored Gaussian noise realizations in the time domain following standard procedures for gravitational wave data analysis~\cite{Allen:2005fk}. Each noise realization is produced by inverse Fourier transforming frequency-domain Gaussian random variates scaled by the square root of the PSD, ensuring the correct statistical properties in both time and frequency domains. Specifically, for a target PSD $S_n(f)$, we generate noise time series $n(t)$ such that the expectation value of the noise spectrum satisfies:
\begin{equation}
\langle \tilde{n}(f) \tilde{n}^*(f') \rangle = \frac{1}{2} S_n(|f|) \delta(f - f'),
\end{equation}
where $\tilde{n}(f)$ denotes the Fourier transform of $n(t)$ and the factor of $1/2$ accounts for the one-sided PSD convention~\cite{Maggiore:2007ulw}.

\begin{figure}[htbp]
\centering
\includegraphics[width=\columnwidth]{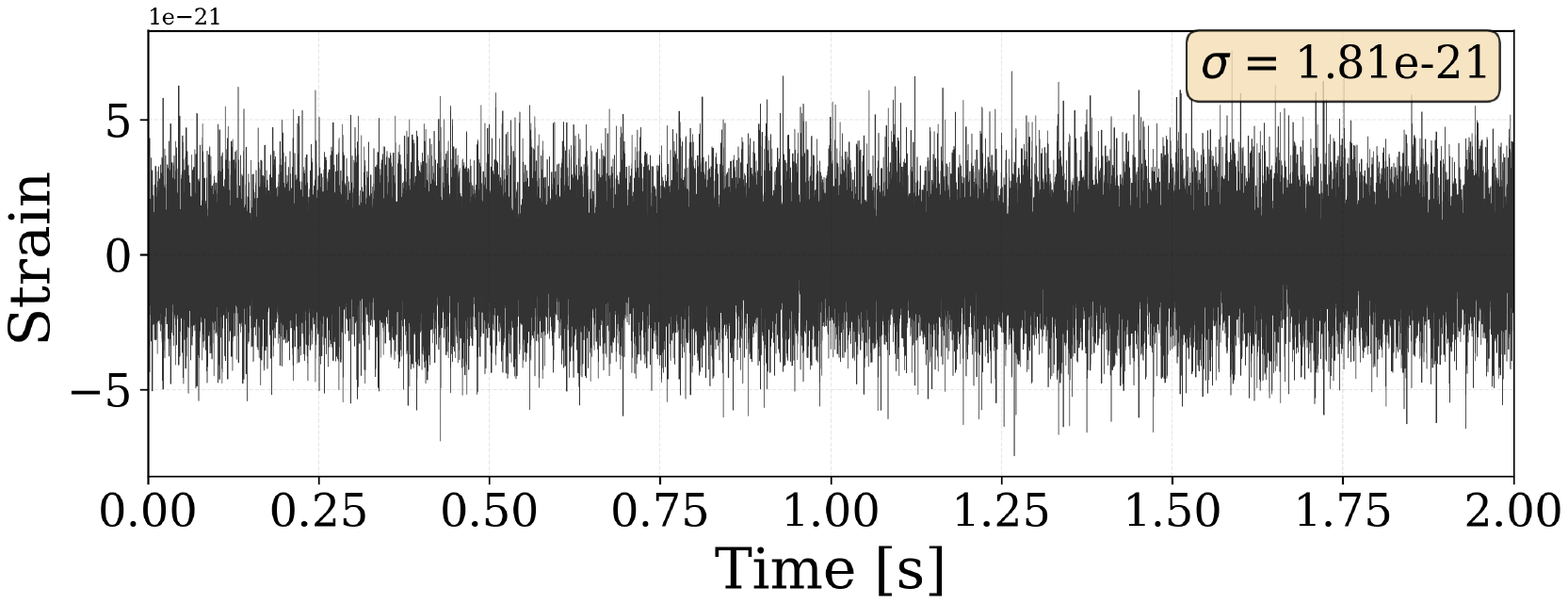}
\caption{Representative 2-second time series of simulated aLIGO detector noise. The stochastic fluctuations reflect the colored Gaussian noise characteristics prescribed by the aLIGO Zero-Detuned High Power PSD, with amplitude consistent with strain sensitivity at design configuration ($h \sim 10^{-23}$--$10^{-22}$).}
\label{fig:noise_sample}
\end{figure}

The noise generation is implemented using the \texttt{pycbc.noise.noise\_from\_psd} method. Figure~\ref{fig:noise_sample} displays a representative 16-second noise realization in the time domain, exhibiting the characteristic stochastic fluctuations expected from detector noise. The amplitude scale is consistent with the strain sensitivity of aLIGO at design configuration, with typical fluctuations on the order of $10^{-23}$--$10^{-22}$ in dimensionless strain units.
% , which employs the following procedure:
% \begin{enumerate}
%     \item Define frequency parameters: resolution $\Delta f = 1/T$ where 
%     $T = 16$~s, sampling rate $f_s = 16{,}384$~Hz, and 
%     $N_{\rm freq} = T f_s / 2 + 1$ frequency bins.
    
%     \item Generate the target PSD $S_n(f_k)$ using the aLIGO Zero 
%     Detuning High Power model, setting $S_n(f_k) = 0$ for 
%     $f_k < f_{\rm low} = 10$~Hz and at the DC and Nyquist components.
    
%     \item Draw independent complex Gaussian variates 
%     $Z_k = A_k + iB_k$, $A_k, B_k \sim \mathcal{N}(0,1)$, and scale 
%     as $\tilde{n}(f_k) = Z_k \cdot \sqrt{S_n(f_k)/(4\Delta f)}$, 
%     enforcing Hermitian symmetry $\tilde{n}(f_{N-k}) = \tilde{n}^*(f_k)$ 
%     for real-valued output.
    
%     \item Apply the inverse FFT to obtain the time-domain colored 
%     Gaussian noise realization $n(t_j) = \text{IFFT}[\tilde{n}(f_k)]$.
% \end{enumerate}
% 

We generate a large number of noise realizations (50,000) to ensure adequate statistical sampling for training deep neural networks, where diverse noise instantiations are critical for learning robust denoising representations. 

\begin{figure}[htbp]
\centering
\includegraphics[width=\columnwidth]{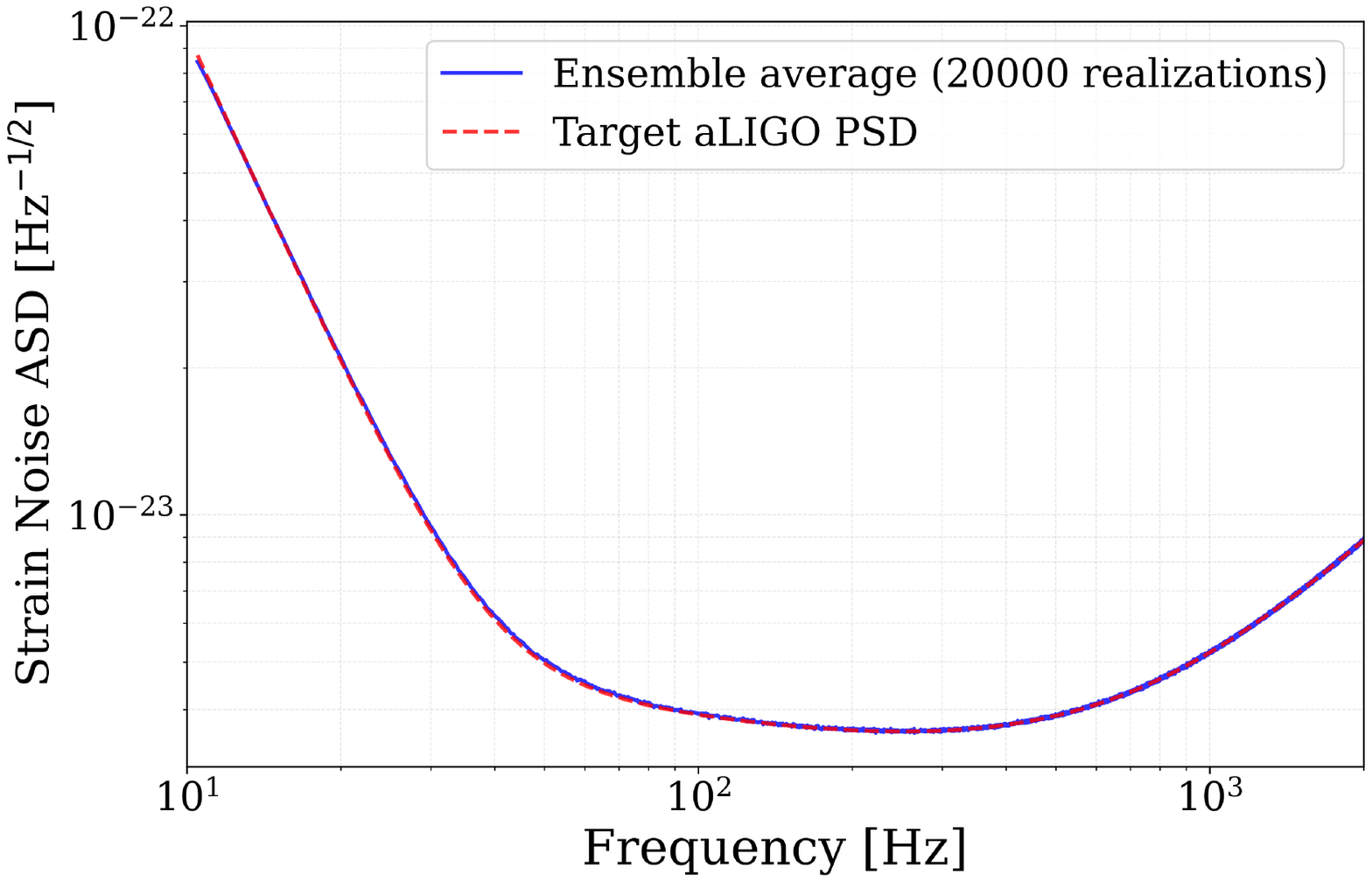}
\caption{Statistical validation of 50000 simulated noise realizations.  Ensemble-averaged power spectral density validation. The empirical PSD computed from the noise realizations (blue solid line) closely matches the target aLIGO Zero-Detuned High Power PSD (red dashed line), confirming accurate reproduction of the desired noise characteristics across the observable frequency band from 10~Hz to 2~kHz.}
\label{fig:noise_validation}
\end{figure}

Figure~\ref{fig:noise_validation} presents statistical validation of the noise dataset. It demonstrates that the ensemble-averaged power spectrum matches the target aLIGO PSD, confirming the accuracy of our noise generation procedure.

\subsection{Training and Validation Dataset Preparation}

For network training and evaluation, each of the 20,000 waveforms in the training and testing sets is combined with randomly selected noise realizations from our pool of independent noise instances. We select a random noise realization of $2$s to be combined with the $2$s strain signal. This injection procedure produces noisy data samples of the form:
\begin{equation}
d(t) = n(t) + h(t; \boldsymbol{\theta}),
\end{equation}
where $d(t)$ is the observed strain data, $n(t)$ is the detector noise realization, and $h(t; \boldsymbol{\theta})$ is the gravitational wave signal with parameters $\boldsymbol{\theta}$. The signal-to-noise ratio (SNR) of each injection is controlled through appropriate scaling of the waveform amplitude, with specific SNR distributions. The preprocessing and augmentation strategies applied to these noisy samples are detailed in \cref{sec:preprocessing}.

\section{\label{sec:preprocessing}Data Preprocessing}

This appendix details each stage of the preprocessing pipeline
summarized in \cref{sec:summary_preprocessing}, following the
progressive transformation of raw detector-frame waveforms through
distance scaling, whitening, bandpass filtering, and normalization.

\subsection{\label{sec:distance_scaling}Distance Scaling and 
Cosmological Realism}

Each waveform, initially generated at a fiducial distance of $d_0 = 1$~Mpc, 
is rescaled to a randomly sampled luminosity distance $d_L \in [1, 1000]$~Mpc, 
encompassing the detection horizon for stellar-mass BBH systems at aLIGO 
design sensitivity~\cite{KAGRA:2021vkt}. For a uniformly distributed 
population of sources in comoving volume, two competing effects shape the 
distance distribution: the quadratic increase in comoving volume 
($\propto d_L^2$) and the time-dilation suppression of the observed event 
rate ($\propto 1/(1+z)^3$), yielding~\cite{Schutz:1985jx}:
\begin{equation}
p(d_L) \propto \frac{d_L^2}{(1+z(d_L))^3},
\label{eq:distance_distribution}
\end{equation}
where $z$ is related to $d_L$ through the Planck18 cosmological 
parameters~\cite{Planck:2018vyg}, implemented via numerical interpolation 
of pre-computed $(d_L, z)$ pairs.

The waveform amplitude is rescaled as:
\begin{equation}
h(t; d_L) = \frac{h(t; d_0 = 1\,\text{Mpc})}{d_L},
\label{eq:distance_scaling}
\end{equation}
and noisy detector data are then constructed by adding an independent 
colored Gaussian noise realization:
\begin{equation}
d(t) = h(t; d_L) + n(t).
\label{eq:noisy_data}
\end{equation}

\subsection{\label{sec:whitening}Frequency-Domain Whitening}

\paragraph{Mathematical Framework}

Whitening transforms the colored detector noise into approximately
unit-variance white noise~\cite{Allen:2005fk}:
\begin{equation}
\tilde{s}_{\text{white}}(f) = \frac{\tilde{s}(f)}{\sqrt{S_n(f)}},
\label{eq:whitening}
\end{equation}
applied identically to both the clean signal $h(t)$ and noisy data $d(t)$.

\paragraph{Implementation Procedure}

Our whitening procedure follows standard gravitational wave data analysis 
practices~\cite{Allen:2005fk, Maggiore:2007ulw}:
\begin{enumerate}
    \item Transform time-domain signals to the frequency domain via FFT 
    with resolution $\Delta f = 1/T$, where $T = 2$~s.
    
    \item Apply Equation~(\ref{eq:whitening}) to both $h(t)$ and $d(t)$.
    
    \item Suppress low-frequency content below $f_{\rm low} = 10$~Hz by 
    inflating the PSD to $S_n(f < 10\,{\rm Hz}) = 10^{-40}$, several 
    orders of magnitude above typical in-band values 
    ($\sim 10^{-44}$--$10^{-47}$), effectively implementing a high-pass 
    filter via the whitening division.
\end{enumerate}

\begin{figure}[htbp]
\centering
\includegraphics[width=\columnwidth]{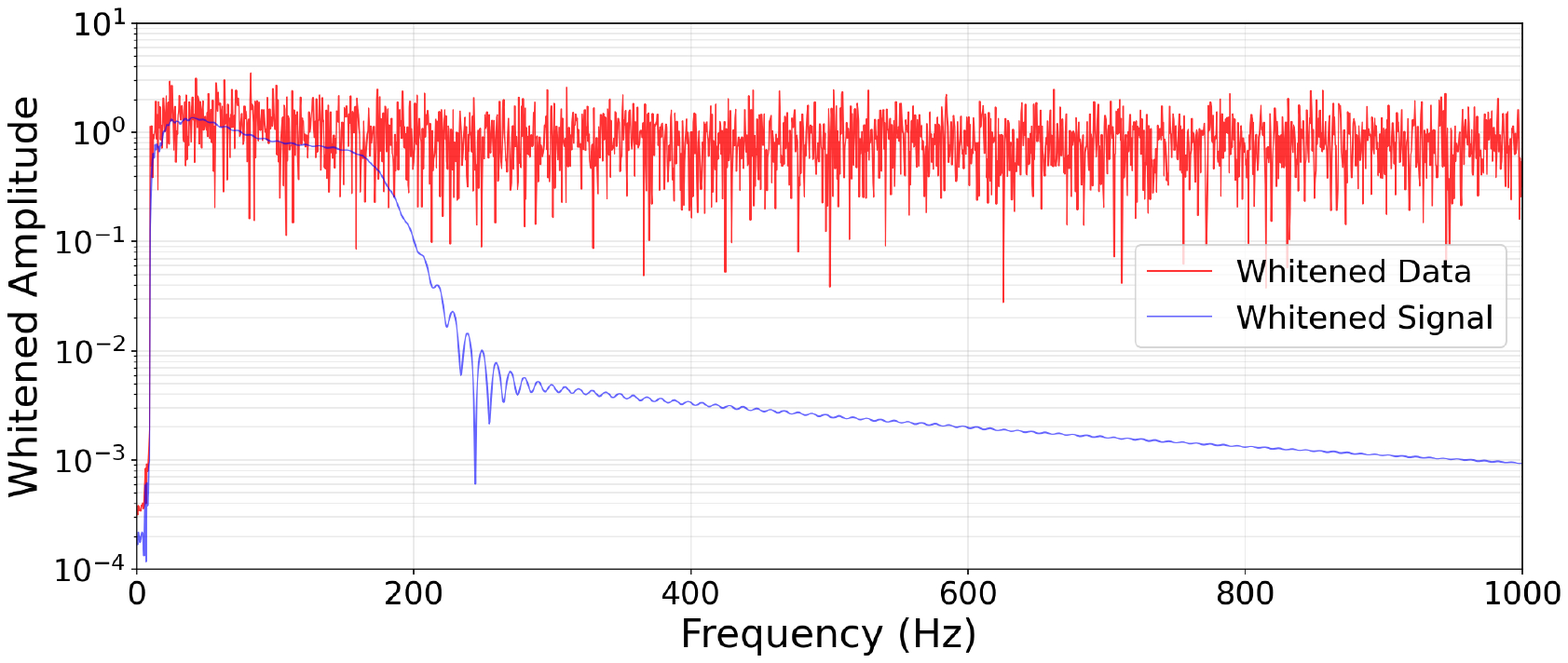}
\caption{Frequency-domain representation after PSD-based whitening according 
to Equation~(\ref{eq:whitening}). The colored noise spectrum (red) has been 
transformed into approximately unit-variance white noise across the observable 
band, while the clean signal spectrum (blue) exhibits enhanced power at 
frequencies where the detector has reduced sensitivity. Whitening equalizes the 
statistical properties of noise fluctuations, enabling the neural network to 
treat all frequencies on equal footing.}
\label{fig:whitening_frequency}
\end{figure}

The whitened signals are subsequently bandpass filtered and converted back 
to the time domain as described in 
~\crefrange{sec:bandpass}{sec:time_domain_conversion}. 
Figure~\ref{fig:whitening_frequency} shows the frequency-domain 
result after whitening, confirming the flattened noise spectrum and 
enhanced signal power at low-sensitivity frequencies.

\paragraph{Signal-to-Noise Ratio Calculation}

Following whitening, we compute the optimal SNR for each injection using 
the matched-filtering statistic~\cite{Allen:2005fk, Cutler:1994ys}:
\begin{equation}
\text{SNR}^2 = 4 \int_{f_{\text{low}}}^{f_{\text{high}}} 
\frac{|\tilde{h}(f)|^2}{S_n(f)} \, df,
\label{eq:snr}
\end{equation}
where $f_{\rm low} = 10$~Hz and $f_{\rm high} = 2048$~Hz, which 
characterizes the detectability of each injection and is used to 
stratify the training distribution in 
\cref{sec:dataset_statistics}.

\subsection{\label{sec:bandpass}Bandpass Filtering}

Following whitening, we apply a sharp frequency-domain bandpass filter 
restricting the signal content to $f \in [20, 2000]$~Hz. The low-frequency 
cutoff at 20~Hz excludes residual seismic noise and poorly calibrated 
spectral regions below the detector's reliable sensitivity 
band~\cite{KAGRA:2021vkt}, while the high-frequency cutoff at 2000~Hz 
removes shot-noise-dominated content that carries negligible astrophysical 
information. For our target mass range ($m_{\{1,2\}} \in [5, 100]\,M_\odot$), 
the dominant merger and ringdown frequencies scale as 
$f_{\rm ringdown} \sim 1.5 \times 220\,(100\,M_\odot/M_{\rm final})$~Hz~\cite{Berti:2005ys}, 
placing typical ringdown frequencies in the 100--500~Hz range and well 
within the retained band. 

\subsection{\label{sec:time_domain_conversion}Time-Domain Conversion 
and Normalization}

\begin{figure}[htbp]
\centering
\includegraphics[width=\columnwidth]{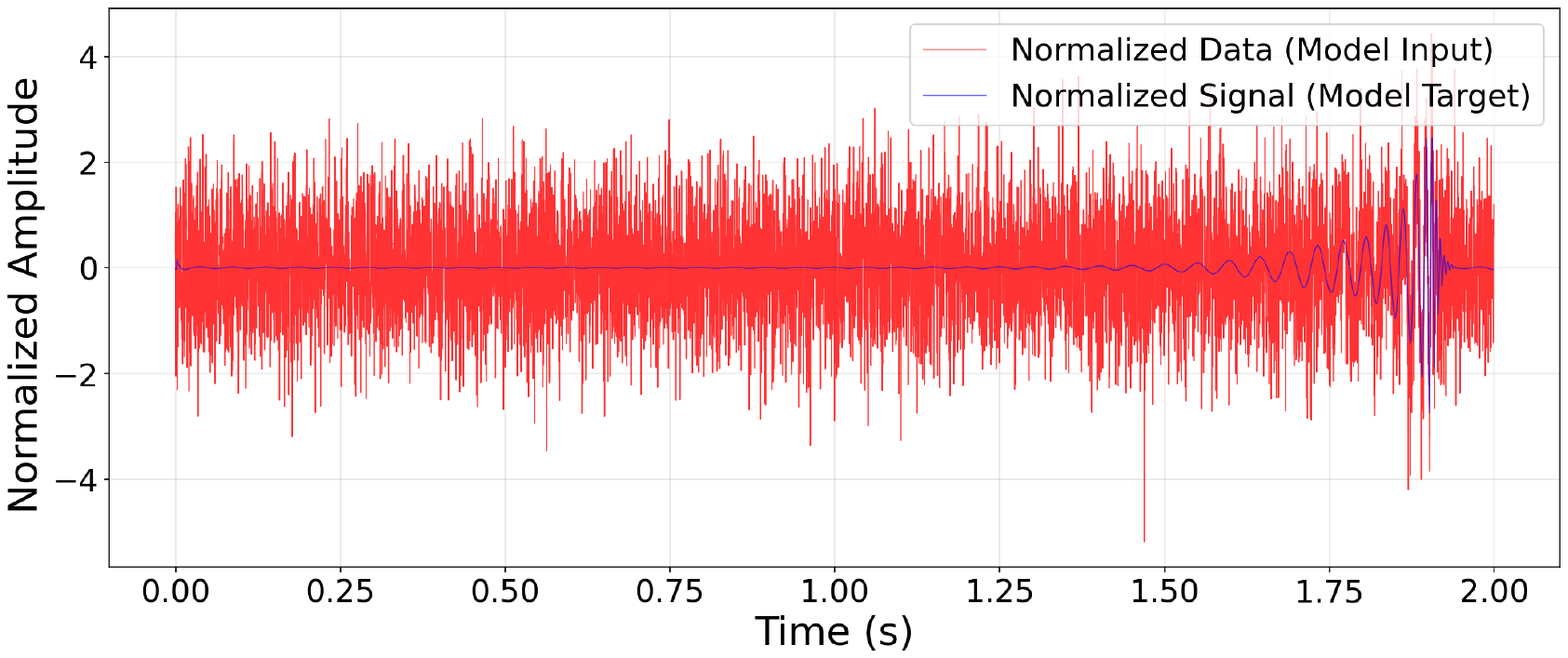}
\caption{Final preprocessed signals after standard deviation normalization 
according to Equation~(\ref{eq:normalization}). Both the noisy input (red) 
and clean target (blue) have been scaled to unit variance, with typical 
amplitude fluctuations now centered around $\pm 1$ in dimensionless 
normalized units. This normalized time series serves as the direct input 
to the denoising neural network, with the noisy signal provided as the 
model input and the clean signal as the target output for supervised 
learning.}
\label{fig:final_normalized}
\end{figure}

After frequency-domain whitening and bandpass filtering, the processed 
signals are converted back to the time domain via inverse FFT. The final 
preprocessing stage normalizes both the clean and noisy signals by the 
standard deviation of the noisy time series:
\begin{equation}
d_{\text{norm}}(t) = \frac{d_{\text{filtered}}(t)}{\sigma_{\text{noisy}}}, 
\quad h_{\text{norm}}(t) = \frac{h_{\text{filtered}}(t)}{\sigma_{\text{noisy}}}.
\label{eq:normalization}
\end{equation}
Both signals share the same normalization factor, preserving the relative signal-to-noise structure of each injection, ensuring that 
high-SNR and low-SNR samples retain their distinctive characteristics 
for the network. The final normalized signals are shown in 
Figure~\ref{fig:final_normalized}.

\subsection{\label{sec:dataset_statistics}Training and Testing Dataset 
Statistics}

\begin{figure}[htbp]
\centering
  \includegraphics[width=\columnwidth]{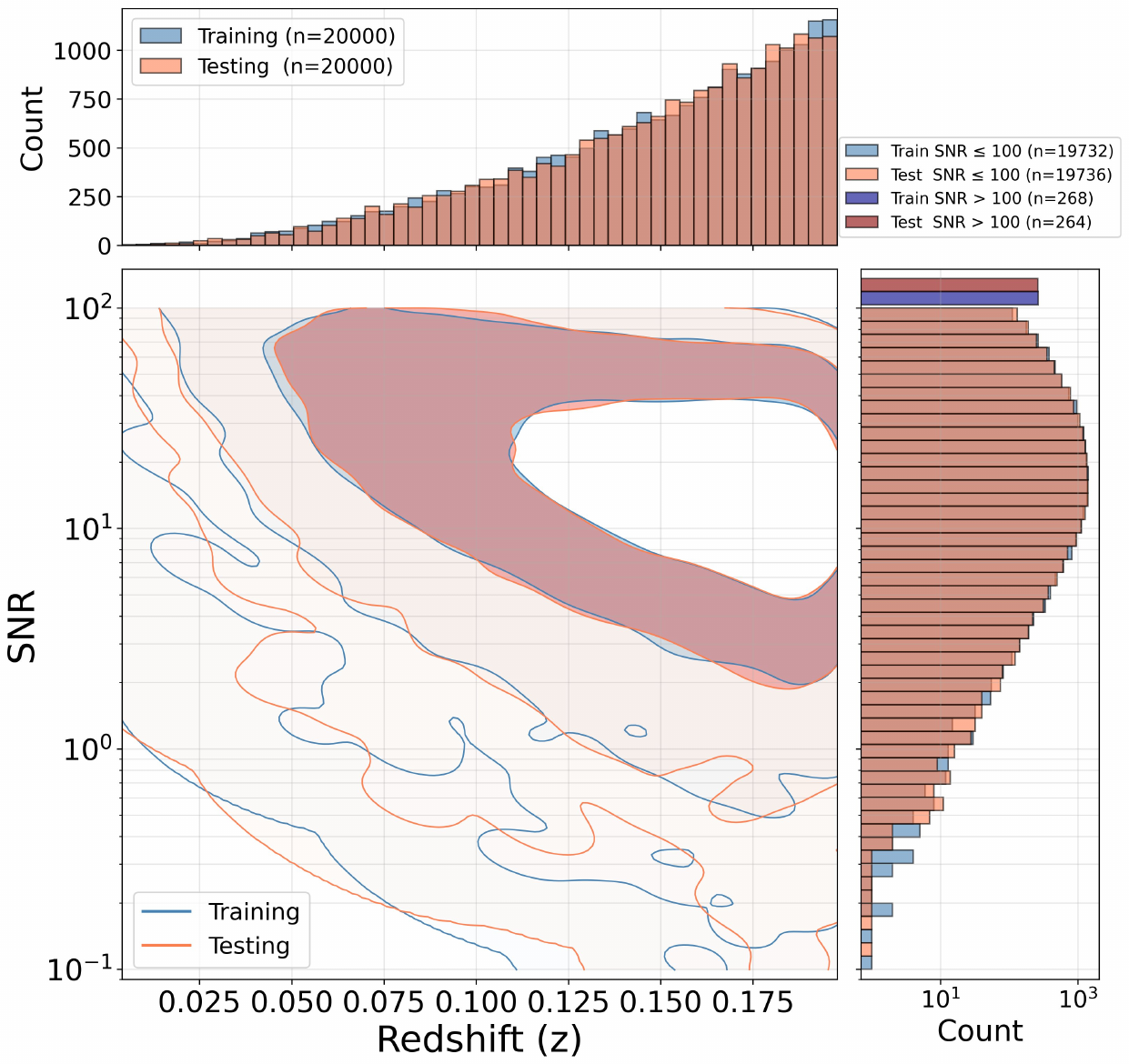}%
\caption{Corner plot showing the joint and marginal distributions of redshift
and optimal SNR for the training (blue) and testing (red) datasets.
\textit{Top left:} Marginal redshift distribution, spanning $z \sim 0$--$0.2$,
consistent with our luminosity distance upper limit of 1000\,Mpc under
Planck18 cosmology. \textit{Bottom left:} Joint KDE density in the
redshift--SNR plane; contours mark the 10th--90th percentiles of the
estimated density, computed in $(z,\,\log_{10}\,\mathrm{SNR})$ space to
avoid distortion on the logarithmic SNR axis. \textit{Bottom right:}
Marginal SNR distribution on a log--log scale. Dark-coloured overflow bars
indicate samples with $\mathrm{SNR} > 100$, representing a small fraction
of nearby, high-mass or favourably oriented systems. Training and testing
distributions are consistent across all panels, confirming unbiased dataset
splitting.}
\label{fig:dataset_distributions}
\end{figure}

Figure~\ref{fig:dataset_distributions} summarizes the key statistical 
properties of the training and testing datasets resulting from our 
cosmologically weighted distance sampling. The redshift distribution spans 
$z \in [0, 0.2]$ with peak density at $z \sim 0.18$, corresponding to the 
$d_L \sim 900$~Mpc peak where the competing effects of increasing comoving 
volume and cosmological time dilation balance. The approximately uniform 
comoving volume distribution confirms that our $d_L^2/(1+z)^3$ weighting 
correctly reflects the astrophysical source distribution. The resulting SNR 
distribution peaks in the range $\mathrm{SNR} \sim 10$--$30$, with 
$\sim$98.7\% of samples having $\mathrm{SNR} \leq 100$, representative of 
real gravitational wave observations where most detected events have modest 
SNR.

\begin{figure}[htbp]
\centering
\subfloat[SNR--distance distribution\label{fig:snr_distance_2d}]{%
  \includegraphics[width=0.5\columnwidth]{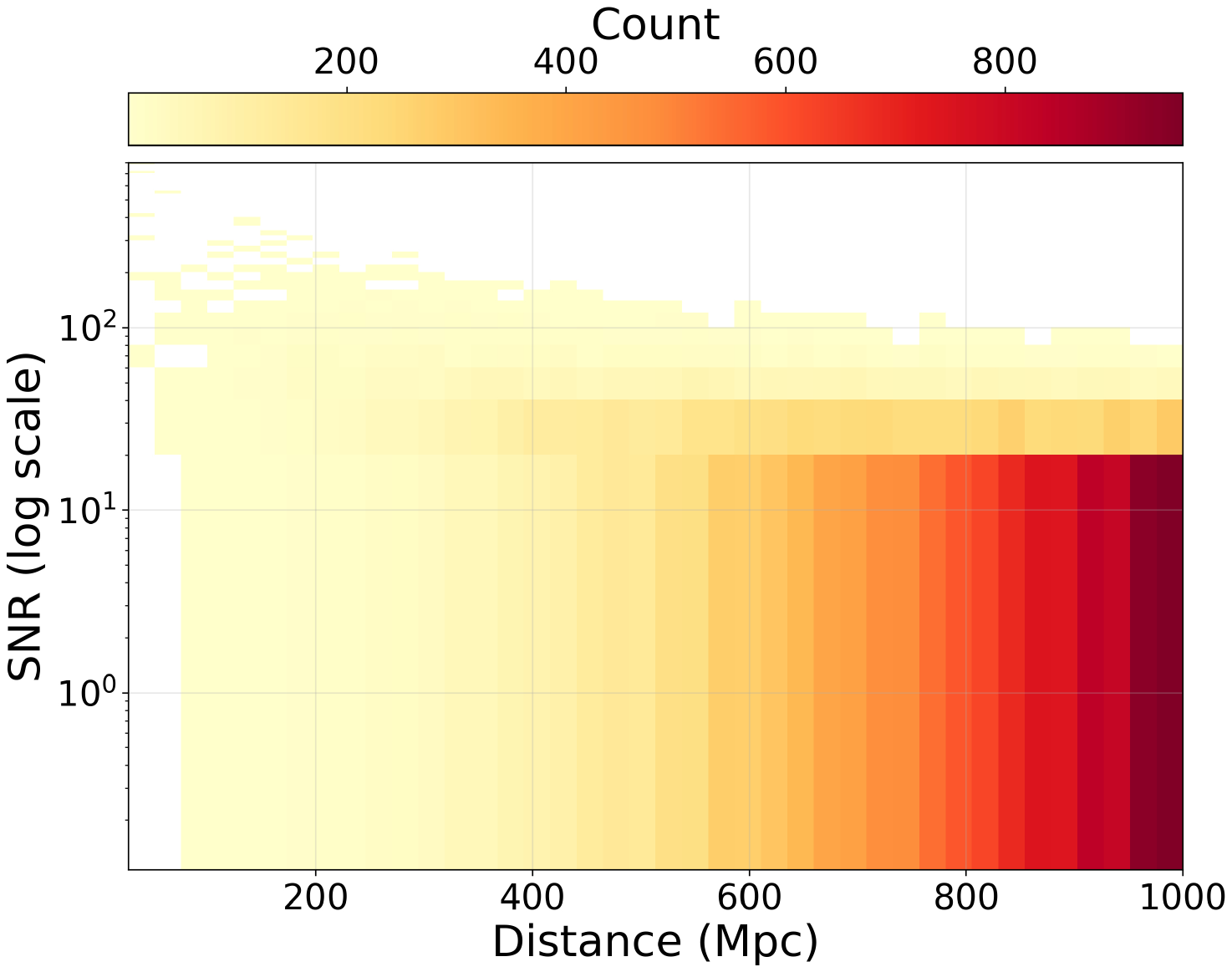}%
}\hfill
\subfloat[Chirp mass--distance distribution\label{fig:chirp_mass_distance_2d}]{%
  \includegraphics[width=0.5\columnwidth]{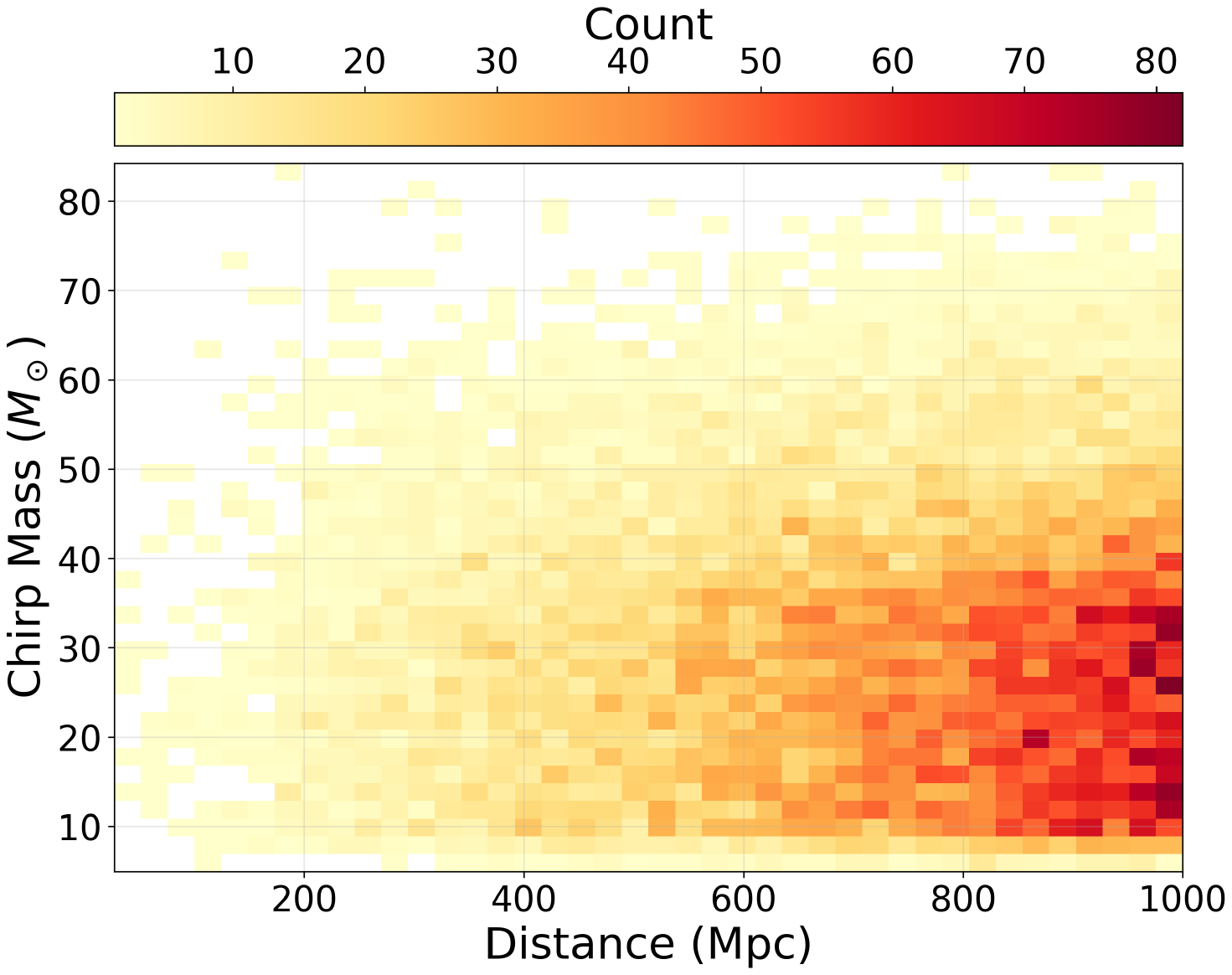}%
}
\caption{Two-dimensional histograms showing joint distributions for the 
training dataset, with color scales indicating event density in each bin. 
\textbf{(a)} Joint distribution of optimal SNR and luminosity distance 
$d_L$. The expected inverse SNR--distance relationship is clearly visible, 
with the vertical banding reflecting discrete sampling of the luminosity 
distance array. The logarithmic vertical scale for SNR spans three orders 
of magnitude from $\text{SNR} \sim 0.1$ to $\text{SNR} \sim 100$. 
\textbf{(b)} Joint distribution of chirp mass $\mathcal{M}_c$ and luminosity distance. Coverage is 
relatively uniform across $\mathcal{M} \sim 10$--$80\,M_{\odot}$ at all 
distances, with a mild concentration at $\mathcal{M} \sim 20$--$40\,M_{\odot}$ 
arising from the convolution of uniform-$\eta$ sampling with the component 
mass constraints. Consistent distributions between training and testing 
datasets confirm unbiased sampling across both mass and distance 
parameters.}
\label{fig:snr_chirpmass_distance}
\end{figure}

Figure~\ref{fig:snr_chirpmass_distance} shows the joint SNR--distance and 
chirp mass--distance distributions for the training dataset. The expected 
inverse SNR--distance correlation is clearly visible, with the scatter at 
fixed distance arising from variations in binary masses, spins, and orbital 
inclinations. The chirp mass coverage is approximately uniform across 
$\mathcal{M} \in [10, 80]\,M_\odot$ at all distances, with SNR $\propto 
\mathcal{M}^{5/6}$ at fixed distance, ensuring the network is trained 
across the full range of inspiral morphologies.

\section{\label{sec:architectures}Neural Network Architectures}

This appendix provides complete specifications for the five neural
network architectures compared in this work, which are summarized
together with their key design dimensions in
\cref{sec:summary_architectures} and
Table~\ref{tab:architecture_comparison} of the main text. The
subsections below detail the mathematical formulation, layer
configurations, and design rationale for each architecture in turn.

\subsection{\label{sec:arch_lstm}Baseline LSTM Encoder-Decoder}

The encoder-decoder LSTM architecture~\cite{Hochreiter:1997yld, 
sutskever2014, Cho:2014sdq} serves as our baseline, establishing the 
minimum performance achievable through pure recurrent processing without 
convolutions, attention, or skip connections. The encoder compresses the 
noisy input into a low-dimensional latent representation, while the 
decoder reconstructs the clean waveform, allowing us to isolate and 
quantify performance gains attributable to the more sophisticated 
components introduced in subsequent architectures.

\paragraph{Architecture Specification}

\begin{figure*}[htbp]
\centering
\includegraphics[width=0.9\textwidth]{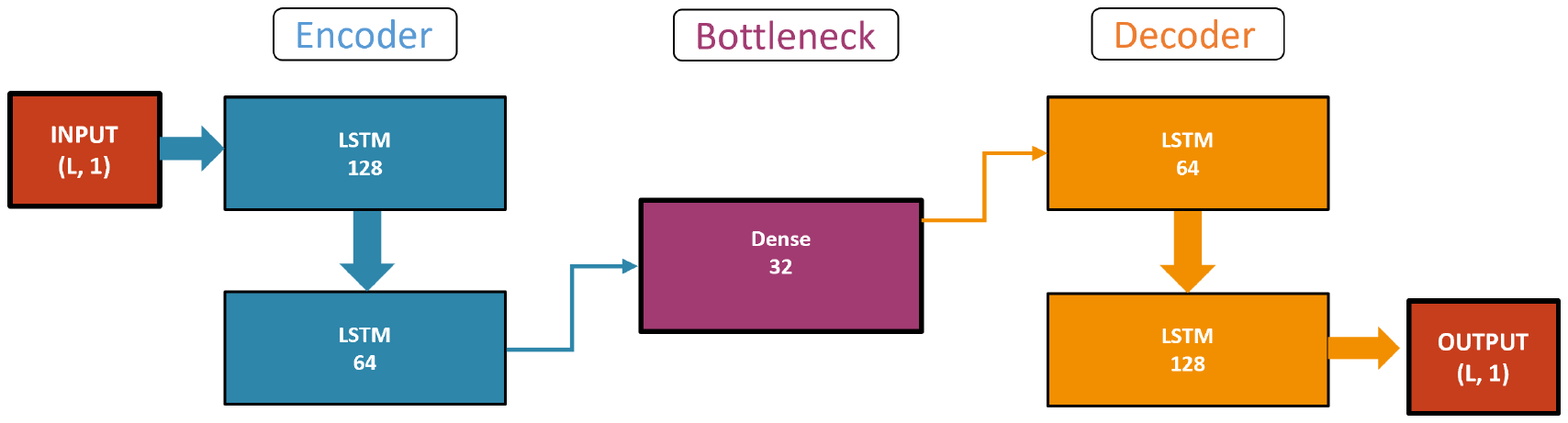}
\caption{Schematic diagram of the baseline LSTM encoder-decoder 
architecture. The encoder comprises two stacked LSTM layers (128 and 64 
units) that progressively compress the noisy input into a 32-dimensional 
latent representation via a dense bottleneck. The symmetric decoder mirrors 
this structure with two LSTM layers (64 and 128 units) that reconstruct the 
clean waveform. Layer normalization and dropout (rate 0.2) after each LSTM 
layer ensure training stability and prevent overfitting. The architecture 
contains approximately 250K trainable parameters.}
\label{fig:arch_lstm_diagram}
\end{figure*}

The architecture processes input sequences $\mathbf{x} \in \mathbb{R}^{T 
\times 1}$ ($T = 8192$ time steps) through symmetric encoder and decoder 
pathways, illustrated in Figure~\ref{fig:arch_lstm_diagram}.

\textit{Encoder:}
Two stacked LSTM layers with decreasing hidden dimensionalities compress 
the input:
\begingroup
\small
\begin{equation}
\begin{aligned}
\mathbf{h}_1^{(t)} &= \text{LSTM}_1(\mathbf{x}^{(t)}, 
\mathbf{h}_1^{(t-1)}; \mathbf{W}_1), \quad d_1 = 128 \\
\mathbf{h}_1^{(t)} &\leftarrow \text{LayerNorm}(\text{Dropout}
(\mathbf{h}_1^{(t)}, p=0.2)) \\
\mathbf{h}_2^{(t)} &= \text{LSTM}_2(\mathbf{h}_1^{(t)}, 
\mathbf{h}_2^{(t-1)}; \mathbf{W}_2), \quad d_2 = 64 \\
\mathbf{h}_2^{(t)} &\leftarrow \text{LayerNorm}(\text{Dropout}
(\mathbf{h}_2^{(t)}, p=0.2))
\end{aligned}
\label{eq:lstm_encoder}
\end{equation}
\endgroup

\textit{Bottleneck:}
A dense layer reduces the encoder output to a 32-dimensional latent 
representation:
\begingroup
\small
\begin{equation}
\mathbf{z}^{(t)} = \text{LayerNorm}(\text{ReLU}(\mathbf{W}_b 
\mathbf{h}_2^{(t)} + \mathbf{b}_b)), \quad \mathbf{z}^{(t)} \in 
\mathbb{R}^{32}
\label{eq:lstm_bottleneck}
\end{equation}
\endgroup

\textit{Decoder:}
The decoder mirrors the encoder with reversed dimensionalities:
\begingroup
\small
\begin{equation}
\begin{aligned}
\mathbf{h}_3^{(t)} &= \text{LSTM}_3(\mathbf{z}^{(t)}, 
\mathbf{h}_3^{(t-1)}; \mathbf{W}_3), \quad d_3 = 64 \\
\mathbf{h}_3^{(t)} &\leftarrow \text{LayerNorm}(\text{Dropout}
(\mathbf{h}_3^{(t)}, p=0.2)) \\
\mathbf{h}_4^{(t)} &= \text{LSTM}_4(\mathbf{h}_3^{(t)}, 
\mathbf{h}_4^{(t-1)}; \mathbf{W}_4), \quad d_4 = 128 \\
\mathbf{h}_4^{(t)} &\leftarrow \text{LayerNorm}(\text{Dropout}
(\mathbf{h}_4^{(t)}, p=0.2))
\end{aligned}
\label{eq:lstm_decoder}
\end{equation}
\endgroup

\textit{Output Projection:}
A time-distributed dense layer with linear activation produces the 
single-channel output:
\begin{equation}
\hat{\mathbf{y}}^{(t)} = \mathbf{W}_o \mathbf{h}_4^{(t)} + \mathbf{b}_o, 
\quad \hat{\mathbf{y}}^{(t)} \in \mathbb{R}^1
\label{eq:lstm_output}
\end{equation}

% \paragraph{Regularization and Implementation Details:}
Layer normalization~\citep{Ba:2016jcy} after each LSTM layer is preferred 
over batch normalization~\citep{Ioffe:2015ovl} as it operates independently 
on each sample, making it better suited for variable-length sequence 
processing. Dropout ($p = 0.2$)~\cite{Srivastava:2014kpo} prevents 
overfitting to noise patterns in the training data, and all weight matrices 
use Glorot uniform initialization~\cite{Glorot2010} to maintain stable 
activation magnitudes across layers. The architecture contains 
$\sim$243K trainable parameters, making it the most parameter-efficient 
model in our comparison.

\begin{figure*}[htbp]
\centering
\includegraphics[width=0.9\textwidth]{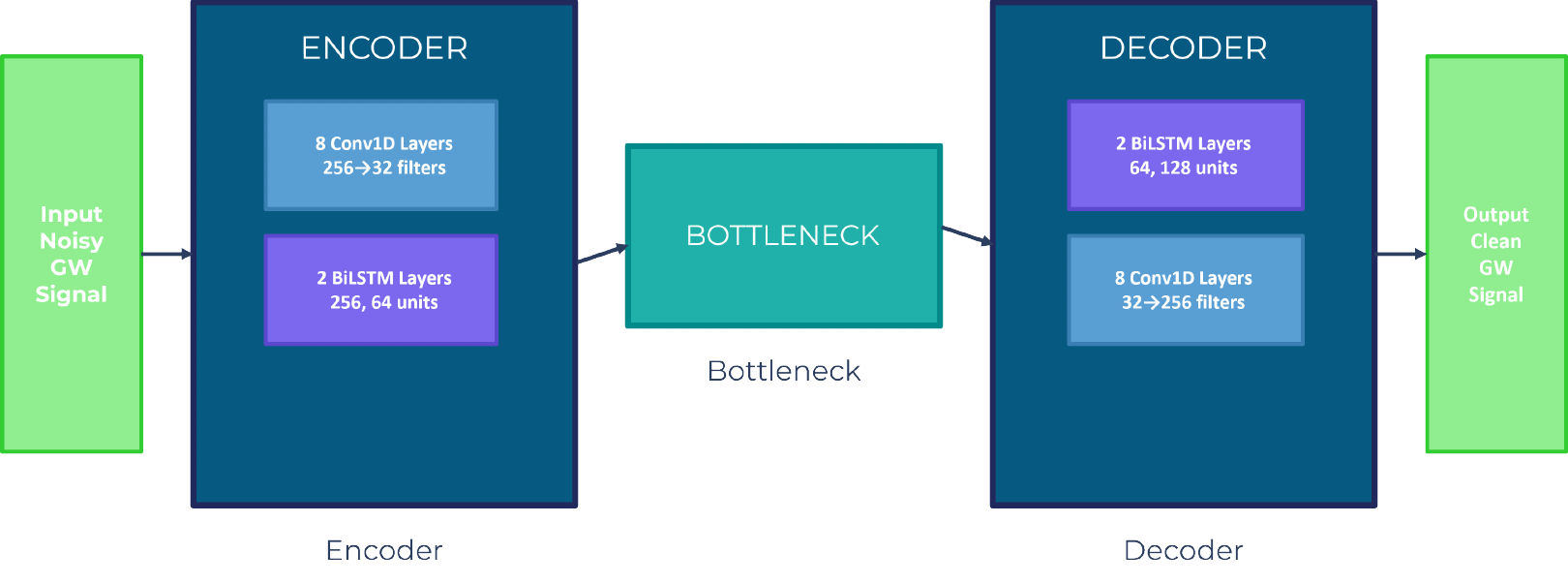}
\caption{Schematic diagram of the CNN-LSTM hybrid architecture. The encoder 
begins with 8 convolutional layers (kernel sizes 512$\to$48, filter counts 
128$\to$32) for multi-scale feature extraction, followed by 2 LSTM layers 
(128 and 64 units) for temporal integration. After a 32-dimensional dense 
bottleneck, the decoder applies 2 LSTM layers (64 and 128 units) followed 
by 8 convolutional layers (mirroring the encoder configuration) for 
hierarchical reconstruction. Batch normalization after each convolution and 
layer normalization after each LSTM ensure training stability. The 
architecture contains approximately 26.5M parameters.}
\label{fig:arch_cnn_lstm_diagram}
\end{figure*}

\subsection{\label{sec:arch_cnn_lstm}CNN-LSTM Hybrid Architecture}

% \paragraph{Motivation for Convolutional Preprocessing}
While LSTMs capture long-range temporal dependencies, they process inputs 
sequentially without explicit awareness of local temporal structure. 
Convolutional layers~\cite{Lecun:1998iof} complement this by extracting 
translation-invariant local features through parameter sharing, motivating hybrid CNN-LSTM architectures~\cite{Zheng2014, Sainath2015} \textbf{, including CNN-LSTM denoising autoencoders previously applied to 
BBH signal extraction \citep{Chatterjee:2021lit, Chatterjee:2024alf}}. For 
gravitational wave signals, this hierarchical processing aligns naturally 
with BBH waveform structure: large convolutional kernels capture the 
slowly evolving low-frequency inspiral, while small kernels resolve 
high-frequency merger transients and ringdown oscillations. Stacking 
convolutional layers with progressively decreasing kernel sizes therefore 
implements a physically motivated multi-scale feature pyramid.

\paragraph{Architecture Specification:}
The encoder combines convolutional feature extraction with recurrent 
temporal processing, while the decoder mirrors this design to reconstruct 
the clean waveform, as illustrated in 
Figure~\ref{fig:arch_cnn_lstm_diagram}.

\textit{Encoder Convolutions:} Eight convolutional layers arranged in a pyramid with decreasing kernel 
sizes and filter counts extract multi-scale features:
\begingroup
\small
\begin{equation}
\begin{aligned}
\mathbf{c}_i &= \text{ReLU}(\text{BatchNorm}(\text{Conv1D}_{k_i, 
f_i}(\mathbf{c}_{i-1}))), \quad i = 1, \ldots, 8 \\
(k, f) &= \{(512, 128), (384, 112), (256, 96), (192, 80), \\
&\quad\;\; (128, 64), (96, 48), (64, 40), (48, 32)\}
\end{aligned}
\label{eq:cnn_encoder_conv}
\end{equation}
\endgroup
where $\mathbf{c}_0 = \mathbf{x}$, $k_i$ is the kernel size, and $f_i$ 
the number of filters. Large kernels ($k = 512 \approx 125$~ms at 4096~Hz) 
capture low-frequency inspiral evolution while small kernels 
($k = 48 \approx 12$~ms) resolve high-frequency merger transients. Batch 
normalization~\cite{Ioffe:2015ovl} after each convolution and He normal 
initialization~\cite{He:2015dtg} for convolutional weights ensure training 
stability.

\textit{Recurrent Processing and Bottleneck:}
Following convolutional feature extraction, the encoder applies two stacked 
LSTM layers (128 and 64 units) with layer normalization and dropout 
(Equation~\ref{eq:lstm_encoder}), followed by the 32-dimensional dense 
bottleneck (Equation~\ref{eq:lstm_bottleneck}). The decoder mirrors this 
with two LSTM layers (64 and 128 units, Equation~\ref{eq:lstm_decoder}), 
whose output feeds into the decoder convolutional layers.

\textit{Decoder Convolutions:}
Eight convolutional layers with reversed kernel and filter ordering 
progressively reconstruct the waveform:
\begingroup
\small
\begin{equation}
\begin{aligned}
\mathbf{d}_i &= \text{ReLU}(\text{BatchNorm}(\text{Conv1D}_{k_i, 
f_i}(\mathbf{d}_{i-1}))), \quad i = 1, \ldots, 8 \\
(k, f) &= \{(48, 32), (64, 40), (96, 48), (128, 64), \\
&\quad\;\; (192, 80), (256, 96), (384, 112), (512, 128)\}
\end{aligned}
\label{eq:cnn_decoder_conv}
\end{equation}
\endgroup

\textit{Output Projection:}
A final $1 \times 1$ convolution projects to the single-channel output:
\begin{equation}
\hat{\mathbf{y}}^{(t)} = \text{Conv1D}_{1, 1}(\mathbf{d}_8^{(t)})
\label{eq:cnn_lstm_output}
\end{equation}

The architecture contains $\sim$26.5M trainable parameters, dominated by 
the 16 convolutional layers, with convolutional GPU parallelization 
resulting in training times comparable to the baseline despite the 
$\sim$109$\times$ increase in parameter count.

\subsection{\label{sec:arch_transformer}Transformer-Based Architecture: 
ACRED-Net}
Recurrent networks suffer from vanishing gradients over long sequences, 
limiting their ability to model dependencies between distant time 
steps~\cite{Bengio1994, pascanu2013}. Attention 
mechanisms~\cite{Bahdanau:2014ghw, Vaswani:2017lxt} circumvent this by 
allowing each output element to directly attend to all input elements with 
$\mathcal{O}(1)$ path length, regardless of temporal separation. Our 
transformer-based architecture, ACRED-Net (Attentive Convolutional 
Recurrent Encoder-Decoder Network), combines multi-head self-attention with 
convolutional feature extraction and bidirectional LSTMs. For gravitational 
wave denoising, this enables direct modeling of correlations between early 
inspiral and late ringdown features, data-dependent feature aggregation 
adapting to varying signal morphologies, and efficient parallelization 
across the temporal dimension.

\paragraph{Architecture Specification}

\begin{figure*}[htbp]
\centering
\includegraphics[width=\textwidth]{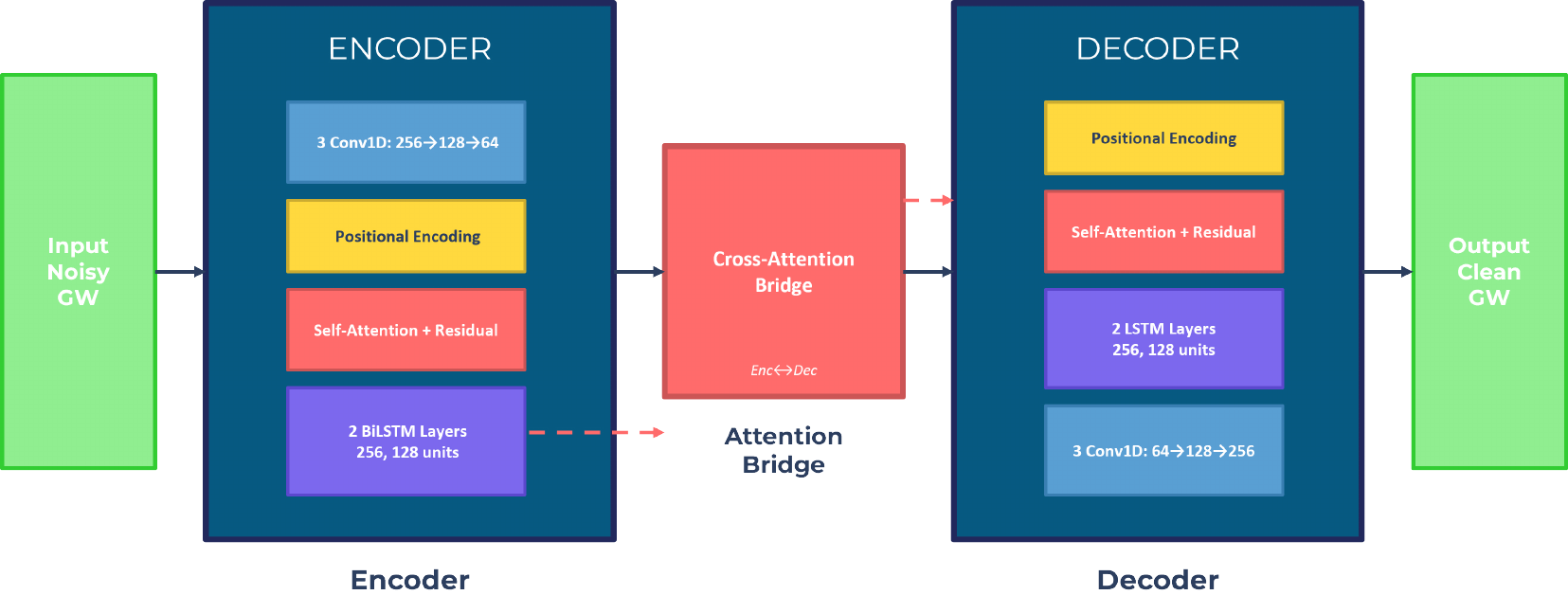}
\caption{Schematic diagram of ACRED-Net, the transformer-based architecture 
combining convolutional feature extraction, self-attention mechanisms, and 
recurrent processing. The encoder applies 3 convolutional blocks (filters 
256$\to$64, kernels 512$\to$128) followed by sinusoidal positional 
encoding, multi-head self-attention (1 head, key dimension 4), and 2 
bidirectional LSTM layers (256 and 128 units). The decoder begins with 
positional encoding and self-attention, processes with 1 LSTM layer (256 
units), applies cross-attention to encoder outputs, processes with another 
LSTM layer (128 units), and refines with 3 convolutional blocks (filters 
64$\to$256, kernels 512$\to$128). Residual connections and layer 
normalization stabilize training. The architecture contains approximately 
22.0M parameters and operates in float64 precision.}
\label{fig:arch_transformer_diagram}
\end{figure*}

The complete ACRED-Net architecture is illustrated in 
Figure~\ref{fig:arch_transformer_diagram}, incorporating positional 
encoding, multi-head self-attention, cross-attention, and convolutional 
refinement within an encoder-decoder framework.

\textit{Positional Encoding:}
Since attention mechanisms lack inherent sequence ordering, we inject 
positional information using sinusoidal encodings~\cite{Vaswani:2017lxt}:
\begin{equation}
\begin{aligned}
PE_{(t, 2i)} &= \sin(t / 10000^{2i/d}) \\
PE_{(t, 2i+1)} &= \cos(t / 10000^{2i/d})
\end{aligned}
\label{eq:positional_encoding}
\end{equation}
where $t \in [0, T-1]$ is the time step index, $i \in [0, d/2-1]$ is the 
dimension index, and $d$ is the embedding dimension. The varying wavelengths 
across dimensions enable attention to relative positions at multiple 
scales~\cite{Vaswani:2017lxt}.

\textit{Multi-Head Self-Attention:}
The encoder applies multi-head self-attention to compute context-dependent 
representations:
\begingroup
\small
\begin{equation}
\begin{aligned}
\text{Attention}(\mathbf{Q}, \mathbf{K}, \mathbf{V}) &= 
\text{softmax}\!\left(\frac{\mathbf{Q}\mathbf{K}^T}{\sqrt{d_k}}\right)
\mathbf{V} \\
\text{MultiHead}(\mathbf{X}) &= \text{Concat}(\text{head}_1, \ldots, 
\text{head}_h)\mathbf{W}^O
\end{aligned}
\label{eq:multihead_attention}
\end{equation}
\endgroup
where $h = 1$ attention heads each with key dimension $d_k = 4$, and 
$\mathbf{W}^O$ projects the concatenated outputs back to the original 
dimensionality. The self-attention output is combined with the input via a 
residual connection~\cite{He:2015wrn} and normalized:
\begingroup
\small
\begin{equation}
\mathbf{X}' = \text{LayerNorm}(\mathbf{X} + \text{MultiHead}(\mathbf{X} 
+ PE))
\label{eq:encoder_residual}
\end{equation}
\endgroup

\textit{Encoder Convolutions and LSTMs:}
The attention-enhanced features pass through 3 convolutional blocks with 
batch normalization:
\begingroup
\small
\begin{equation}
\mathbf{X}^{(l)} = \text{BatchNorm}(\text{ReLU}(\mathbf{W}^{(l)} * 
\mathbf{X}^{(l-1)} + \mathbf{b}^{(l)})), \quad l = 1, 2, 3
\label{eq:acred_encoder_conv}
\end{equation}
\endgroup
where $\mathbf{W}^{(1)} \in \mathbb{R}^{512 \times 1 \times 256}$, 
$\mathbf{W}^{(2)} \in \mathbb{R}^{256 \times 256 \times 128}$, and 
$\mathbf{W}^{(3)} \in \mathbb{R}^{128 \times 128 \times 64}$, followed by 
2 bidirectional LSTM layers with layer normalization and dropout ($p=0.2$):
\begingroup
\small
\begin{equation}
\mathbf{H}^{(l)} = \text{Dropout}(\text{LayerNorm}(\text{BiLSTM}_{n_l}
(\mathbf{H}^{(l-1)}))), \quad l = 1, 2
\label{eq:acred_encoder_lstm}
\end{equation}
\endgroup
where $(n_1, n_2) = (256, 128)$ units per direction, yielding 
$\mathbf{H}_{\rm enc} \in \mathbb{R}^{T \times 256}$.

\textit{Decoder Self-Attention and Cross-Attention:}
The decoder applies positional encoding and self-attention 
(Equation~\ref{eq:multihead_attention}) to a $1\times1$ convolution of the 
encoder output, then incorporates cross-attention that queries encoder 
representations from decoder states:
\begingroup
\small
\begin{equation}
\begin{split}
\text{CrossAttention}(\mathbf{X}_{\text{dec}}, \mathbf{X}_{\text{enc}}) 
= \text{MultiHead}(&\mathbf{X}_{\text{dec}}\mathbf{W}^Q, \\
&\mathbf{X}_{\text{enc}}\mathbf{W}^K, \; 
\mathbf{X}_{\text{enc}}\mathbf{W}^V)
\end{split}
\label{eq:cross_attention}
\end{equation}
\endgroup

The cross-attention output is concatenated with the decoder LSTM output and 
fused via a dense layer:
\begingroup
\small
\begin{equation}
\mathbf{X}_{\text{combined}} = \text{Dense}([\mathbf{X}_{\text{LSTM}}, 
\mathbf{X}_{\text{cross}}])
\label{eq:decoder_combination}
\end{equation}
\endgroup
This provides a learned mechanism for the decoder to selectively extract 
relevant features from the encoder, improving upon the fixed bottleneck 
of simpler encoder-decoder architectures~\cite{Bahdanau:2014ghw}.

\textit{Decoder Convolutions and Output:}
Following the second decoder LSTM (128 units), three convolutional blocks 
with increasing filter counts (64$\to$256) and kernel sizes (128$\to$512) 
reconstruct the waveform, concluding with a $1\times1$ convolution 
producing the final single-channel output. The architecture contains 
$\sim$22.0M trainable parameters, with the $\mathcal{O}(T^2 d)$ attention 
complexity dominating the computational cost for $T = 8192$ time steps.

\subsection{\label{sec:arch_unet}U-Net with Attention Gates}

The U-Net architecture~\cite{Ronneberger:2015cwd} introduces two key 
innovations for dense prediction tasks: symmetric encoder-decoder structure 
with skip connections that preserve high-resolution features bypassed by 
the bottleneck compression, and progressive downsampling creating a 
multi-scale feature hierarchy. For gravitational wave denoising, skip 
connections preserve fine-grained merger transients while multi-scale 
processing simultaneously captures extended inspiral evolution. We augment 
the standard U-Net with attention gates~\cite{oktay2018, SCHLEMPER2019} 
that learn to weight skip connections based on their relevance to the 
current decoding stage, suppressing irrelevant encoder features while 
emphasizing signal-dominated regions such as those near merger.

\paragraph{Architecture Specification}

\begin{figure*}[htbp]
\centering
\includegraphics[width=\textwidth]{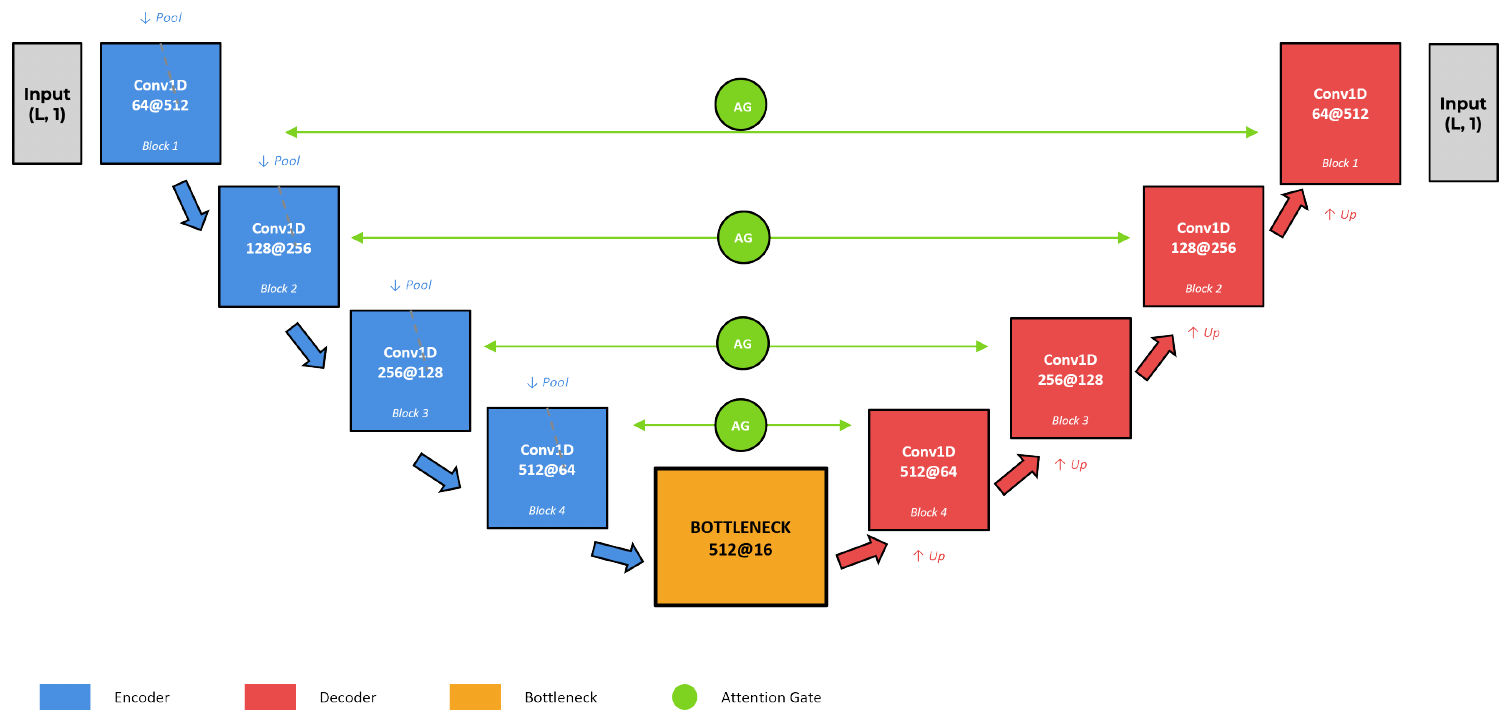}
\caption{Schematic diagram of the U-Net architecture with attention gates. 
The encoder path (left, green) comprises 4 blocks of double convolutions 
(filters 64$\to$512, kernels 512$\to$64) with batch normalization, ReLU 
activation, and dropout, followed by max-pooling (stride 2) for 
downsampling. The bottleneck (center, red) applies double convolutions with 
512 filters and kernel size 16. The decoder path (right, blue) performs 
upsampling (stride 2) followed by attention gates (magenta) that weight 
skip connections from corresponding encoder levels, concatenation, and 
double convolutions. The attention gates learn to selectively propagate 
relevant features while suppressing irrelevant encoder activations. The 
architecture achieves multi-scale feature extraction through progressive 
downsampling, reducing sequence length from 8192 (input) to 256 
(bottleneck) and back to 8192 (output). Total parameters: $\sim$172M.}
\label{fig:arch_unet_diagram}
\end{figure*}

The complete architecture is illustrated in 
Figure~\ref{fig:arch_unet_diagram}, comprising symmetric encoding and 
decoding pathways connected by gated skip connections at multiple 
resolutions.

\textit{Encoder Path (Contracting):}
Four convolutional blocks with progressively increasing filter counts and 
decreasing kernel sizes compress the input through a 5-level resolution 
hierarchy ($8192 \to 4096 \to 2048 \to 1024 \to 512$ samples):
\begin{equation}
\begin{alignedat}{2}
\mathbf{c}_i^{(1)} &=
  \text{ReLU}\!\left(
    \text{BatchNorm}\!\left(
      \text{Conv1D}_{k_i, f_i}(\mathbf{c}_{i-1})
    \right)
  \right), \\[4pt]
\mathbf{c}_i^{(2)} &=
  \text{Dropout}\!\Big(
    \text{ReLU}\!\Big(
      \text{BatchNorm}\!\big(\\[-2pt]
        &\hspace{1.5em}\text{Conv1D}_{k_i, f_i}(\mathbf{c}_i^{(1)})
      \big)
    \Big), 
   p_i
  \Big), \\[4pt]
\mathbf{p}_i &=
  \text{MaxPool}\!\left(
    \mathbf{c}_i^{(2)},\, \text{pool\_size}=2
  \right), \\[4pt]
(k, f, p) &\in
  \{(512, 64, 0.1),\,
    (256, 128, 0.1),\\[-2pt]
  &\quad (128, 256, 0.2),\,
    (64, 512, 0.2)\}.
\end{alignedat}
\label{eq:unet_encoder}
\end{equation}
Increasing dropout rates ($p = 0.1 \to 0.2$) provide stronger 
regularization at deeper, higher-capacity layers. The encoder outputs 
$\{\mathbf{c}_1^{(2)}, \mathbf{c}_2^{(2)}, \mathbf{c}_3^{(2)}, 
\mathbf{c}_4^{(2)}\}$ are stored for skip connections.

\textit{Bottleneck:}
The lowest-resolution representation is processed with two convolutions 
using 512 filters, kernel size 16, and elevated dropout ($p = 0.3$):
\begin{equation}
\begin{aligned}
\mathbf{b}^{(1)} &=
  \text{ReLU}\!\left(
    \text{BatchNorm}\!\left(
      \text{Conv1D}_{16, 512}(\mathbf{p}_4)
    \right)
  \right), \\[4pt]
\mathbf{b}^{(2)} &=
  \text{Dropout}\!\Big(
    \text{ReLU}\!\Big(
      \text{BatchNorm}\!\big(
        \text{Conv1D}_{16, 512}(\mathbf{b}^{(1)})
      \big)
    \Big),\, p = 0.3
  \Big)
\end{aligned}
\label{eq:unet_bottleneck}
\end{equation}

\textit{Attention Gates:}
At each skip connection, an attention gate~\cite{oktay2018} weights the 
encoder features using a gating signal from the decoder:
\begin{equation}
\begin{aligned}
\psi &= \text{sigmoid}(\mathbf{W}_{\psi} \text{ReLU}(\mathbf{W}_g 
\mathbf{g}_i + \mathbf{W}_x \mathbf{c}_i^{(2)})) \\
\mathbf{c}_i^{\text{att}} &= \psi \odot \mathbf{c}_i^{(2)}
\end{aligned}
\label{eq:attention_gate}
\end{equation}
where $\mathbf{W}_g$, $\mathbf{W}_x$, and $\mathbf{W}_{\psi}$ are learned 
$1\times1$ convolutions, $f_{\rm inter}$ is halved relative to the input 
channels, and $\odot$ denotes element-wise multiplication. The sigmoid 
activation produces soft attention coefficients in $[0, 1]$.

\textit{Decoder Path (Expanding):}
The decoder upsamples, applies attention-gated skip connections, and 
reconstructs through double convolutions mirroring the encoder:
\begingroup
\small
\begin{equation}
\begin{aligned}
\mathbf{u}_i &=
  \text{ReLU}\!\Big(
    \text{BatchNorm}\!\Big(
      \text{Conv1D}_{k_i, f_i/2}\!\big(
        \text{UpSample}(\\[-2pt]
  &\hspace{5.5em}\mathbf{d}_{i-1},\, 2)
      \big)
    \Big)
  \Big), \\[4pt]
\mathbf{c}_i^{\text{att}} &=
  \text{AttentionGate}(\mathbf{c}_i^{(2)}, \mathbf{u}_i), \\[4pt]
\mathbf{m}_i &=
  \text{Concatenate}\!\left([\mathbf{c}_i^{\text{att}}, \mathbf{u}_i]
  \right), \\[4pt]
\mathbf{d}_i^{(1)} &=
  \text{ReLU}\!\left(
    \text{BatchNorm}\!\left(
      \text{Conv1D}_{k_i, f_i}(\mathbf{m}_i)
    \right)
  \right), \\[4pt]
\mathbf{d}_i^{(2)} &=
  \text{Dropout}\!\Big(
    \text{ReLU}\!\Big(
      \text{BatchNorm}\!\big(
        \text{Conv1D}_{k_i, f_i}(\mathbf{d}_i^{(1)})
      \big)
    \Big),\, p_i
  \Big)
\end{aligned}
\label{eq:unet_decoder}
\end{equation}
\endgroup
where $i = 4, 3, 2, 1$ proceeds from low to high resolution.

\textit{Output Projection:}
A final $1\times1$ convolution projects to the single-channel output:
\begin{equation}
\hat{\mathbf{y}} = \text{Conv1D}_{1, 1}(\mathbf{d}_1^{(2)})
\label{eq:unet_output}
\end{equation}

The architecture contains $\sim$172M trainable parameters, making it the 
largest in our comparison. Despite this, progressive downsampling reduces 
computational cost at deeper layers: bottleneck operations (sequence length 
512) are $16\times$ cheaper than full-resolution operations, with training 
time approximately $2\times$ that of the baseline LSTM encoder-decoder.

\subsection{\label{sec:arch_multifreq}Multi-Scale Frequency-Aware (Multi-Frequency)
Architecture}

All previous architectures treat gravitational wave signals as generic time 
series without explicit consideration of their frequency-domain structure. 
BBH waveforms possess distinctive spectral characteristics: low-frequency 
components (20--100~Hz) encode the early inspiral spanning seconds, 
mid-frequency components (100--500~Hz) capture the late inspiral and 
merger, and high-frequency components (500--2000~Hz) contain the ringdown 
quasi-normal mode oscillations. Fixed-kernel convolutional architectures 
must compromise between temporal resolution and receptive field, while 
recurrent and attention-based architectures, although avoiding this 
trade-off, do not explicitly separate frequency bands for specialized 
processing.

We introduce the Multi-Frequency architecture, which addresses 
these limitations through parallel processing branches explicitly 
specialized for different frequency ranges, combining insights from 
multi-scale signal processing~\cite{Mallat1989}, parallel pathway 
networks~\cite{Szegedy:2014nrf}, and frequency-aware deep 
learning~\cite{hu2020}. Unlike classical wavelet 
decomposition~\cite{Mallat1989, Daubechies1992} which uses fixed bases, 
our architecture learns optimal frequency-selective filters end-to-end, 
while the cross-frequency integration layer 
(Equation~\ref{eq:multifreq_low_integration}) explicitly models spectral 
energy transfer between adjacent bands that independent wavelet coefficient 
processing would ignore~\cite{Donoho1995}.

\paragraph{Architecture Specification}

\begin{figure*}[htbp]
\centering
\includegraphics[width=\textwidth]{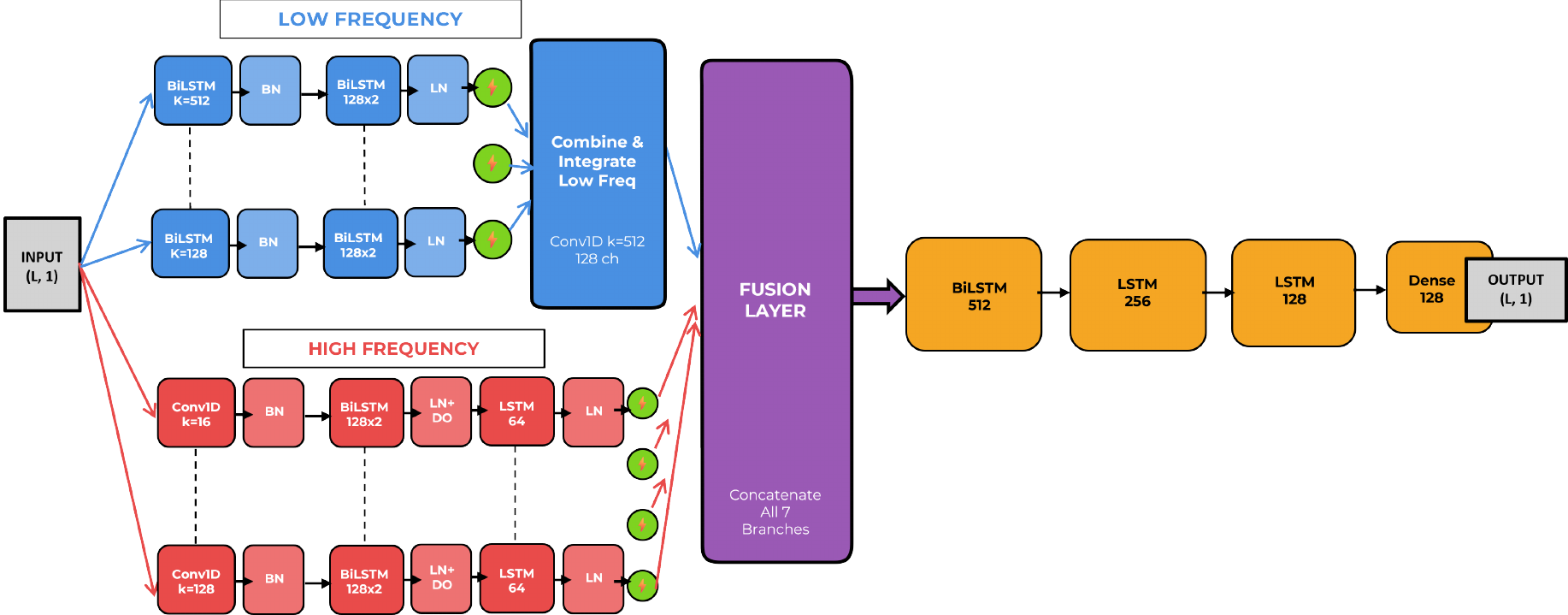}
\caption{Schematic diagram of the Multi-Frequency architecture. 
The input is processed by 7 parallel branches: 3 low-frequency branches 
(left, red/orange/yellow) with large kernels (512, 256, 128) targeting the 
inspiral band, and 4 high-frequency detail branches (right, 
green/blue/cyan/yellow) with small kernels (16, 32, 64, 128) capturing 
merger transients and ringdown. Each branch applies convolution + batch 
normalization, bidirectional LSTM (128 units), and self-attention (sigmoid 
gating). The 3 low-frequency branches are concatenated, integrated with a 
512-kernel convolution, and combined with the 4 high-frequency branches. 
Reconstruction proceeds through 3 LSTM layers (bidirectional 512, 
unidirectional 256 and 128) and a time-distributed dense layer (128 
units), concluding with $1 \times 1$ convolution. This explicit frequency 
decomposition enables specialized processing tailored to different spectral 
components. Total parameters: $\sim$59.1M.}
\label{fig:arch_multifreq_diagram}
\end{figure*}

The architecture processes the input through 3 low-frequency branches 
(large kernels targeting 20--300~Hz), 4 high-frequency detail branches 
(small kernels targeting 300--2000~Hz), cross-frequency integration, and 
multi-stage reconstruction, as illustrated in 
Figure~\ref{fig:arch_multifreq_diagram}.

\textit{Low-Frequency Processing Branches:}
Three parallel branches target progressively higher sub-bands using 
different kernel sizes:
\begingroup
\small
\begin{equation}
\begin{aligned}
\text{Branch}_{\text{lowest}} &: k = 512 
  \text{ (}\sim 125\text{ ms, }\sim 8\text{--}80\text{ Hz)} \\
\text{Branch}_{\text{mid-low}} &: k = 256 
  \text{ (}\sim 62\text{ ms, }\sim 16\text{--}160\text{ Hz)} \\
\text{Branch}_{\text{high-low}} &: k = 128 
  \text{ (}\sim 31\text{ ms, }\sim 32\text{--}320\text{ Hz)}
\end{aligned}
\label{eq:multifreq_lowfreq_branches}
\end{equation}
\endgroup
where frequency ranges are estimated via $f_{\rm min} \approx f_s/k$ and 
$f_{\rm max} \approx 10 f_{\rm min}$ at $f_s = 4096$~Hz. Each branch 
processes the input through convolution, bidirectional LSTM, and 
self-attention:
\begingroup
\small
\begin{equation}
\begin{aligned}
\mathbf{h}^{\text{low}}_i &= \text{BatchNorm}(\text{ReLU}(
  \text{Conv1D}_{k_i, 256}(\mathbf{x}))) \\
\mathbf{h}^{\text{low}}_i &\leftarrow \text{LayerNorm}(
  \text{BiLSTM}_{128}(\mathbf{h}^{\text{low}}_i)) \\
\alpha_i &= \text{sigmoid}(\text{Conv1D}_{1,1}(\text{ReLU}(
  \text{Conv1D}_{1,256}(\mathbf{h}^{\text{low}}_i)))) \\
\mathbf{h}^{\text{low, att}}_i &= \alpha_i \odot \mathbf{h}^{\text{low}}_i
\end{aligned}
\label{eq:multifreq_lowfreq_processing}
\end{equation}
\endgroup

\textit{High-Frequency Detail Processing Branches:}
Four parallel branches target different high-frequency scales:
\begingroup
\small
\begin{equation}
\begin{aligned}
\text{Level 1} &: k = 16 
  \text{ (}\sim 4\text{ ms, }\sim 250\text{--}2000\text{ Hz)} \\
\text{Level 2} &: k = 32 
  \text{ (}\sim 8\text{ ms, }\sim 125\text{--}1000\text{ Hz)} \\
\text{Level 3} &: k = 64 
  \text{ (}\sim 16\text{ ms, }\sim 60\text{--}500\text{ Hz)}  \\
\text{Level 4} &: k = 128 
  \text{ (}\sim 31\text{ ms, }\sim 32\text{--}320\text{ Hz)}
\end{aligned}
\label{eq:multifreq_highfreq_branches}
\end{equation}
\endgroup
Each branch applies convolution, batch normalization, bidirectional LSTM 
(128 units), unidirectional LSTM (64 units), and self-attention:
\begingroup
\small
\begin{equation}
\begin{aligned}
\mathbf{h}^{\text{high}}_j &=
  \text{BatchNorm}\!\left(
    \text{ReLU}\!\left(
      \text{Conv1D}_{k_j, 256}(\mathbf{x})
    \right)
  \right), \\[4pt]
\mathbf{h}^{\text{high}}_j &\leftarrow
  \text{LayerNorm}\!\Big(
    \text{Dropout}\!\Big(
      \text{LayerNorm}\!\big(\\[-2pt]
        &\text{BiLSTM}_{128}(\mathbf{h}^{\text{high}}_j)
      \big),
  \hspace{7.7em}0.2
    \Big)
  \Big), \\[4pt]
\mathbf{h}^{\text{high}}_j &\leftarrow
  \text{LayerNorm}\!\left(
    \text{LSTM}_{64}(\mathbf{h}^{\text{high}}_j)
  \right), \\[4pt]
\alpha_j &=
  \text{sigmoid}\!\left(
    \text{Conv1D}_{1,1}(\mathbf{h}^{\text{high}}_j)
  \right), \\[4pt]
\mathbf{h}^{\text{high, att}}_j &=
  \alpha_j \odot \mathbf{h}^{\text{high}}_j
\end{aligned}
\label{eq:multifreq_highfreq_processing}
\end{equation}
\endgroup

\textit{Cross-Frequency Integration:}
The three low-frequency outputs are concatenated and integrated via a 
large-kernel convolution to model inter-band interactions:
\begingroup
\small
\begin{equation}
\begin{aligned}
\mathbf{h}^{\text{low, integrated}} &=
  \text{LayerNorm}\!\Big(
    \text{ReLU}\!\Big(
      \text{Conv1D}_{512, 128}\!\big(\\[-2pt]
        &[\mathbf{h}^{\text{low, att}}_1,\, \mathbf{h}^{\text{low, att}}_2,
  \mathbf{h}^{\text{low, att}}_3]
      \big)
    \Big)
  \Big)
\end{aligned}
\label{eq:multifreq_low_integration}
\end{equation}
\endgroup
All branches are then concatenated:
\begingroup
\small
\begin{equation}
\begin{aligned}
\mathbf{h}^{\text{concat}} = 
  [&\mathbf{h}^{\text{low, integrated}},\,
   \mathbf{h}^{\text{high, att}}_1,\,
   \mathbf{h}^{\text{high, att}}_2,\\[-2pt]
   &\mathbf{h}^{\text{high, att}}_3,\,
   \mathbf{h}^{\text{high, att}}_4]
\end{aligned}
\label{eq:multifreq_concat}
\end{equation}
\endgroup

\textit{Multi-Stage Reconstruction:}
Three LSTM layers with decreasing dimensionalities integrate the 
multi-frequency features:
\begingroup
\small
\begin{equation}
\begin{aligned}
\mathbf{r}_1 &= \text{Dropout}(\text{LayerNorm}(
  \text{BiLSTM}_{512}(\mathbf{h}^{\text{concat}})), 0.2) \\
\mathbf{r}_2 &= \text{Dropout}(\text{LayerNorm}(
  \text{LSTM}_{256}(\mathbf{r}_1)), 0.2) \\
\mathbf{r}_3 &= \text{LayerNorm}(\text{LSTM}_{128}(\mathbf{r}_2))
\end{aligned}
\label{eq:multifreq_reconstruction}
\end{equation}
\endgroup
A time-distributed dense layer and final $1\times1$ convolution produce 
the output:
\begingroup
\small
\begin{equation}
\mathbf{r}_4 = \text{LayerNorm}(\text{ReLU}(
  \text{TimeDistributedDense}_{128}(\mathbf{r}_3)))
\label{eq:multifreq_dense}
\end{equation}
\begin{equation}
\hat{\mathbf{y}} = \text{Conv1D}_{1,1}(\mathbf{r}_4)
\label{eq:multifreq_output}
\end{equation}
\endgroup

The architecture contains $\sim$59.1M trainable parameters across the 7 
parallel branches and 3 reconstruction LSTM layers, with training time 
$\sim$1.8$\times$ that of the baseline LSTM encoder-decoder --- a 
favorable compute-to-parameter ratio representing explicit inductive bias 
alignment with the spectral structure of gravitational wave signals.

\section{\label{sec:training}Training Configuration and Optimization}

This appendix details the training methodology summarized in
\cref{sec:summary_training}, which is applied uniformly across all
five architectures to enable direct performance comparisons under
controlled conditions.

\paragraph{\label{sec:loss}Loss Function and Optimization Objective:}
We employ Mean Squared Error (MSE) as the primary loss function. For a 
training batch of $N$ samples with ground-truth waveforms 
$\mathbf{y}_i \in \mathbb{R}^T$ and model predictions 
$\hat{\mathbf{y}}_i(\boldsymbol{\theta})$:
\begin{equation}
\mathcal{L}(\boldsymbol{\theta}) = \frac{1}{N} \sum_{i=1}^{N} \frac{1}{T} 
\sum_{t=1}^{T} \left(y_i^{(t)} - \hat{y}_i^{(t)}(\boldsymbol{\theta})
\right)^2
\label{eq:mse_loss}
\end{equation}
where $T = 8192$ is the sequence length (2 seconds at 4096~Hz). MSE 
directly minimizes the $L^2$ distance between predicted and true waveforms, 
closely related to the matched-filter overlap metric~\citep{Allen:2005fk}, 
while its quadratic penalty prioritizes accurate reconstruction of the 
high-amplitude merger and ringdown phases that dominate the 
SNR~\citep{Goodfellow-et-al-2016}.

\paragraph{\label{sec:optimizer}Optimizer Configuration:}
All models are trained using the Adam optimizer~\citep{Kingma:2014vow} 
with an initial learning rate $LR_0 = 10^{-3}$. The 
\texttt{ReduceLROnPlateau} scheduler~\citep{Abadi:2016kic} halves the 
learning rate if validation loss fails to improve for 2 consecutive epochs:
\begin{equation}
LR_{k+1} = \max(\gamma \cdot LR_k, LR_{\text{min}})
\label{eq:lr_schedule}
\end{equation}
where $\gamma = 0.5$ and $LR_{\rm min} = 10^{-6}$, enabling aggressive 
updates during early training and finer steps near 
convergence~\citep{Smith:2018xbl}. Training runs for a maximum of 
$E_{\rm max} = 50$ epochs, with early stopping typically terminating 
training before this limit (see the early-stopping discussion below).

\paragraph{\label{sec:noise_augmentation}Pure Noise Augmentation Strategy:}
Models trained exclusively on signal-plus-noise samples may hallucinate 
spurious features in noise-dominated regions~\citep{Shen:2017jkj, 
Wei:2019zlc}. We address this by augmenting the training data with pure 
noise samples --- time series with $\mathbf{h}(t) = \mathbf{0}$ in 
Equation~(\ref{eq:noisy_data}) --- teaching the network to output zero 
in the absence of a gravitational wave signal. Ablation experiments on 
the baseline LSTM encoder-decoder comparing 0\%, 30\%, and 50\% pure 
noise fractions (detailed in ~\cref{sec:ablation}) establish 30\% 
as the optimal configuration, balancing noise-rejection behavior against 
waveform morphology learning. This fraction is adopted uniformly across 
all five architectures.

\paragraph{\label{sec:regularization}Regularization and Convergence Control:}
\textit{Early Stopping}
Training terminates when validation loss fails to improve for 5 
consecutive epochs~\citep{Prechelt2000}:
\begin{equation}
\text{Stop if } \mathcal{L}_{\text{val}}^{(e)} \geq \min_{e' < e} 
\mathcal{L}_{\text{val}}^{(e')} \text{ for } 5 \text{ consecutive epochs}
\label{eq:early_stopping}
\end{equation}
with model weights restored to the epoch achieving minimum validation loss.

\textit{Model Checkpointing}
Model weights are saved whenever a new validation loss minimum is achieved, 
ensuring the best-performing model is preserved regardless of when early 
stopping triggers.

\paragraph{\label{sec:infrastructure}Computational Infrastructure:}
All experiments were conducted on a single NVIDIA A40 GPU (46~GB VRAM, 
Driver 535.104.05, CUDA 12.2). All computations use double precision 
(float64) rather than the standard float32~\citep{micikevicius2018}, 
motivated by the whitening procedure's PSD division 
(~\cref{sec:whitening}), which can amplify numerical errors near 
cutoff frequencies, and the several-orders-of-magnitude dynamic range of 
normalized waveform amplitudes. TensorFlow's dynamic memory growth, XLA 
JIT compilation, and \texttt{tf.data} \texttt{AUTOTUNE} 
prefetching~\citep{Abadi:2016kic} are enabled to optimize GPU utilization 
and minimize idle time.

\paragraph{\label{sec:training_protocol}Training Protocol and Monitoring:}
Each epoch iterates through the training dataset in batches, performing 
forward propagation, MSE loss computation (Equation~\ref{eq:mse_loss}), 
backpropagation, and Adam parameter updates, followed by validation set 
evaluation. Training and validation MSE loss and Mean Absolute Error (MAE) 
are logged at the end of each epoch to monitor convergence and detect 
overfitting.

\section{\label{sec:pure_noise_appendix}Representative Denoised Outputs on Real Noise Segments}

Figure~\ref{fig:pure_noise_examples} shows four representative
segments from the pure-noise background study of
Section~\ref{sec:pure_noise_real}: two near the population median of
$\rho_{\rm out}$, representative of the network's typical behavior on
stationary Gaussian noise, and two from the extreme high-$\rho_{\rm out}$
tail, both visibly containing real noise transients (glitches) in the input
spectrogram. In the median segments, the network output is suppressed to
a negligible fraction of the input power throughout the analysis window.
In the glitch segments, the network instead removes the surrounding
Gaussian background while preserving the transient's time-frequency
structure, consistent with the interpretation in
Section~\ref{sec:pure_noise_real} that these outliers reflect real
non-Gaussian noise features rather than network-generated artifacts.

\begin{figure*}[htbp]
\centering
\subfloat[Median, GPS 1385941686\label{fig:median1}]{%
  \includegraphics[width=\columnwidth]{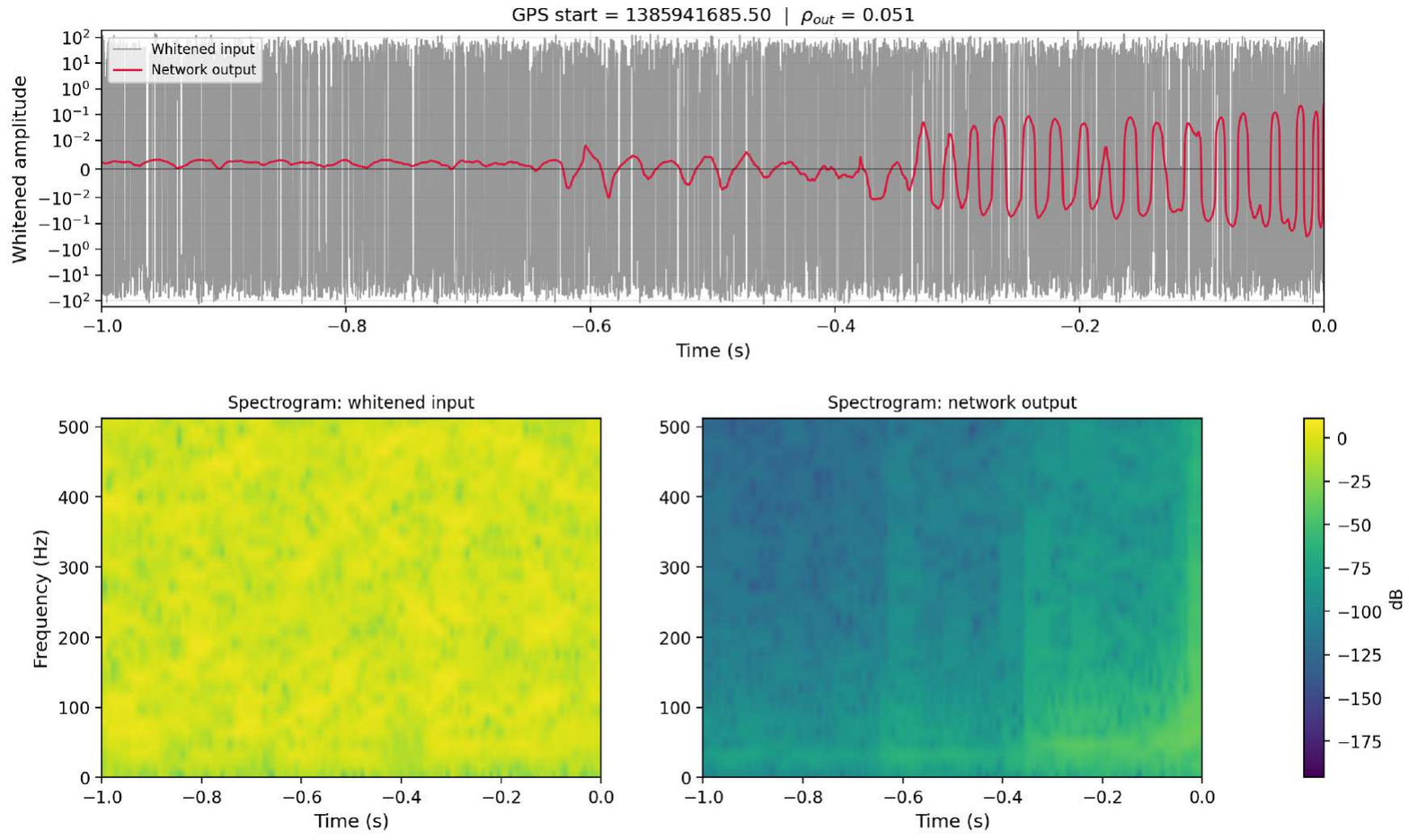}}\hfill
\subfloat[Median, GPS 1368333904\label{fig:median2}]{%
  \includegraphics[width=\columnwidth]{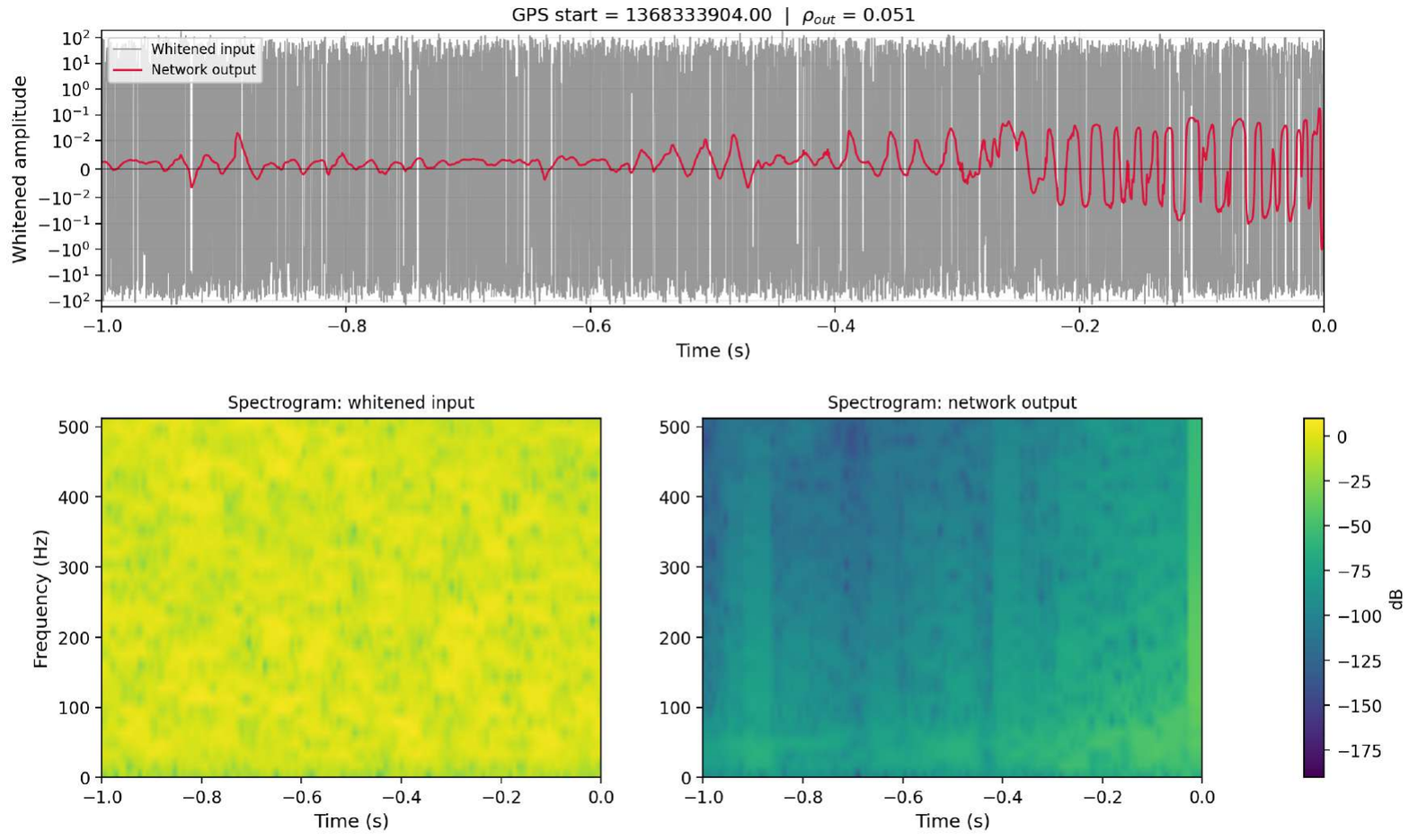}}
\caption{Representative pure-noise segments from the real H1 background
study (Section~\ref{sec:pure_noise_real}). Each panel shows the whitened
input (gray) and network output (red) over the $[-1.0,0.0]$\,s analysis
window (top), with spectrograms of the input and output below (shared
colorbar). \textbf{(a,b)} Segments near the population median of
$\rho_{\rm out}$ ($\rho_{\rm out} \approx 0.051$): the network suppresses the
output to a negligible level throughout, the expected behavior on
stationary Gaussian noise.}
\label{fig:pure_noise_examples}
\end{figure*}

\begin{figure*}[htbp]
\ContinuedFloat
\centering
\subfloat[Glitch (top-$\rho_{\rm out}$), GPS 1385942114\label{fig:top1}]{%
  \includegraphics[width=\columnwidth]{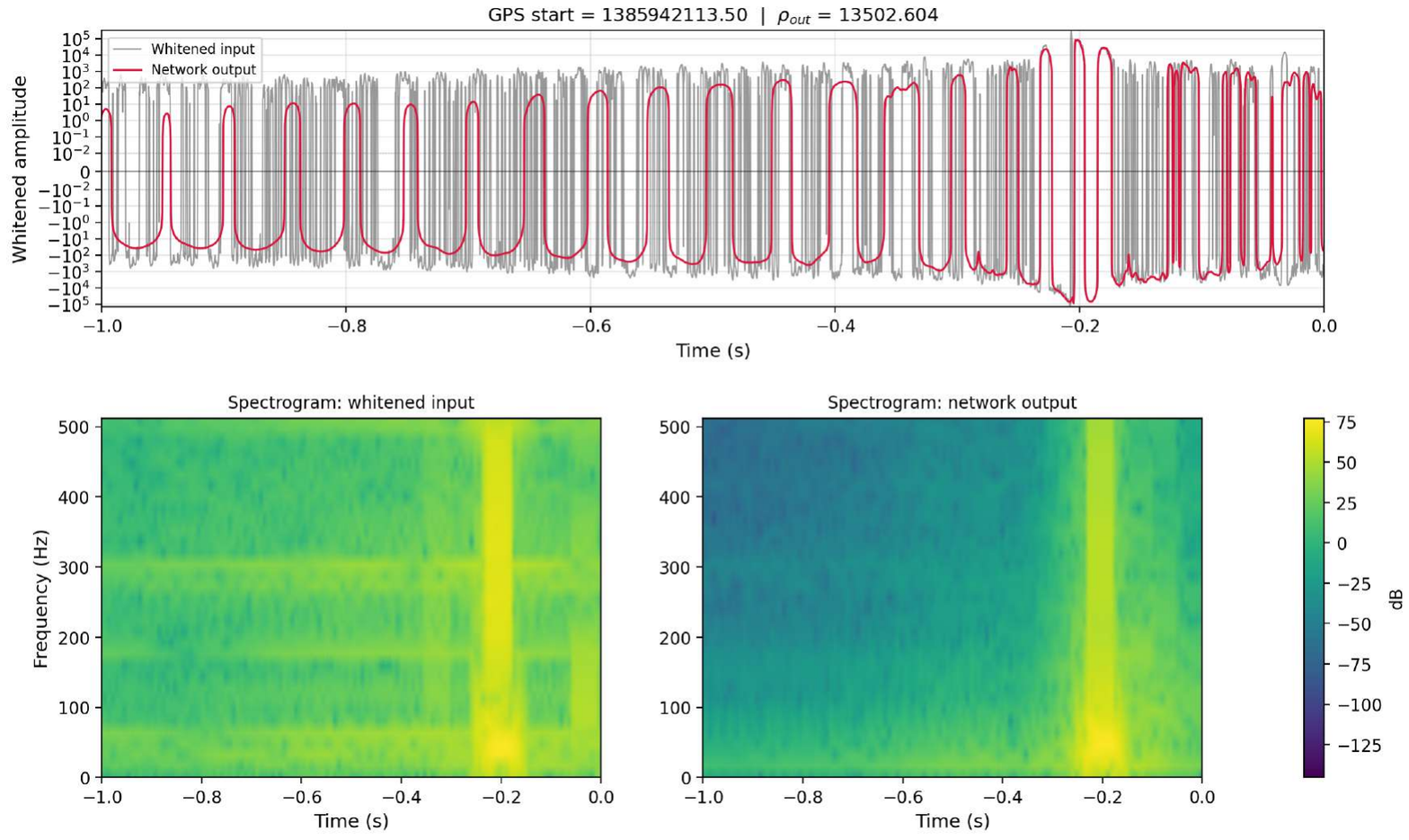}}\hfill
\subfloat[Glitch (top-$\rho_{\rm out}$), GPS 1371792945\label{fig:top2}]{%
  \includegraphics[width=\columnwidth]{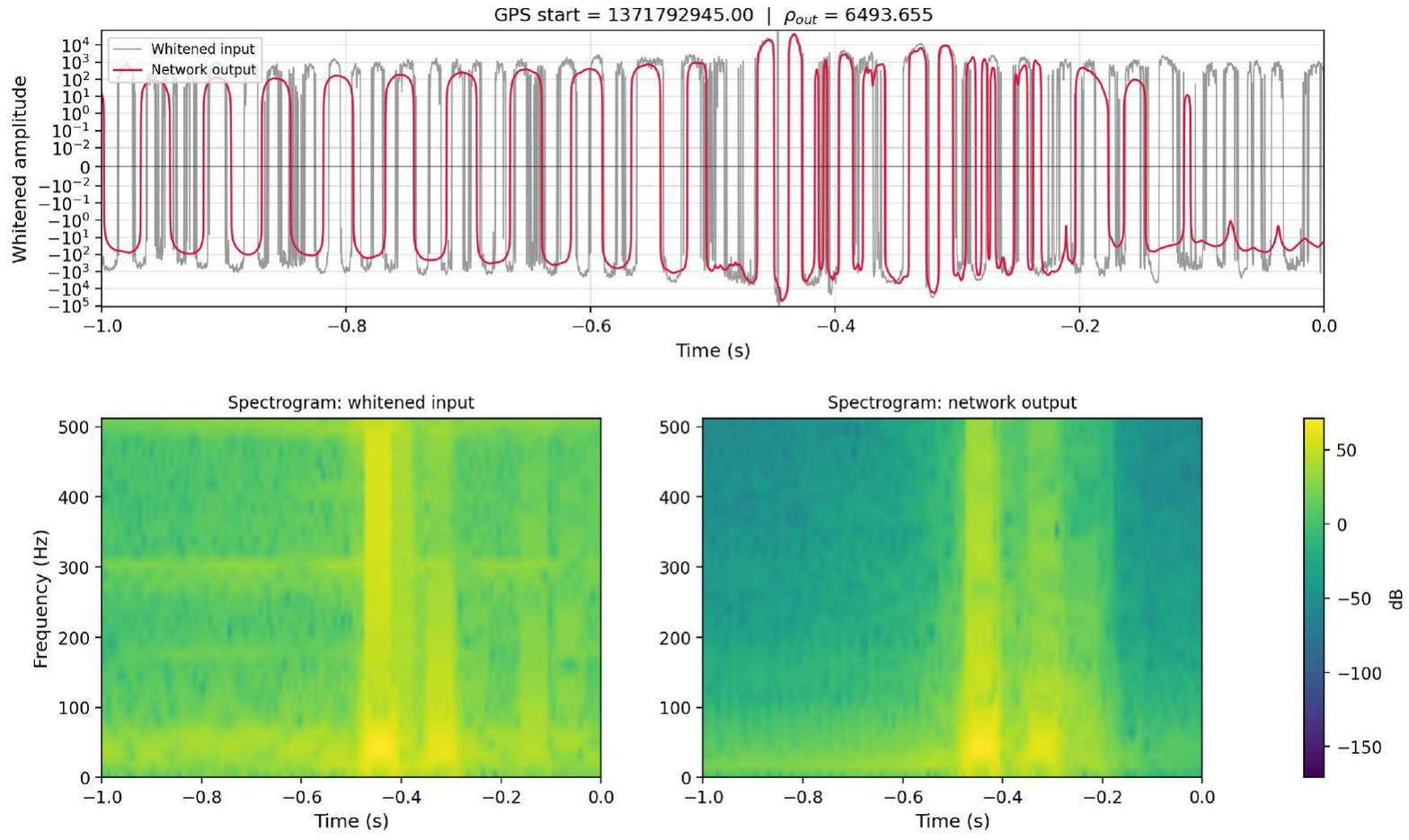}}
\caption{\textbf{(c,d)} Segments from the extreme
high-$\rho_{\rm out}$ tail ($\rho_{\rm out} = 13{,}502.6$ and $6493.7$
respectively), both visibly containing a real noise transient in the input
spectrogram: the network suppresses the surrounding Gaussian background
while retaining the transient's structure, consistent with these outliers
being driven by glitches rather than network-generated hallucination.}
\end{figure*}

\bibliographystyle{iopart-num}
\bibliography{apssamp}

\end{document}